\documentclass[aps,reprint,groupedaddress,floatfix,longbibliography,prx]{revtex4-1}
\usepackage[final]{graphicx}
\usepackage{natbib}
\usepackage{hyperref}
\usepackage{amssymb}
\usepackage{amsmath}
\usepackage{mathrsfs}
\usepackage{xcolor}
\usepackage{bm}
 
\makeatletter
\providecommand{\U}[1]{\protect\rule{.1in}{.1in}}

\begin{document}
\title{Quantum FFLO State in Clean Layered Superconductors}
	\author{Kok Wee Song}
\affiliation{
	Materials Science Division,
	Argonne National Laboratory,
	9700 South Cass Avenue, Lemont, Illinois 60639, USA
}
\author{Alexei E. Koshelev}
\affiliation{
	Materials Science Division,
	Argonne National Laboratory,
	9700 South Cass Avenue, Lemont, Illinois 60639, USA
}

\date{\today}

\begin{abstract}
	We investigate the influence of Landau quantization on the superconducting instability for a pure layered superconductor in the magnetic field directed perpendicular to the layers. We demonstrate that the quantization corrections to the Cooper-pairing kernel with finite Zeeman spin splitting promote the formation of the nonuniform state in which the order parameter is periodically modulated along the magnetic field, i.e., between the layers (Fulde-Ferrell-Larkin-Ovchinnikov [FFLO] state).  The conventional uniform state experiences such a quantization-induced FFLO instability at low temperatures even in a common case of predominantly orbital suppression of superconductivity when the Zeeman spin splitting is expected to have a relatively weak effect. The maximum relative FFLO temperature is given by the ratio of the superconducting transition temperature and the Fermi energy. This maximum is realized when the ratio of the spin-spitting energy and the Landau-level separation is half-integer. These results imply that the FFLO states may exist not only in the Pauli-limited superconductors but also in very clean materials with small Zeeman spin-splitting energy. 
	We expect that the described quantization-promoted FFLO instability is a general phenomenon, which may be found in materials with different electronic spectra and order-parameter symmetries.  
\end{abstract} 

\maketitle

\section{Introduction}

Superconductors exhibit a rich set of phenomena in a magnetic field due to the interplay of the electron orbital and spin degrees of freedoms.
One of the most intriguing phenomena due to the strong Zeeman spin-splitting effect is the emergence of Fulde-Ferrel-Larkin-Ovchinnikov (FFLO) states\cite{Fulde:PRev135.1964,Larkin:JETP20.1965}, in which the Cooper pairing occurs with nonzero total momentum. In the resulted state, the order parameter is modulated along the total momentum direction. This modulation allows the system to regain a part of the  Zeeman energy at the expense of the kinetic-energy loss.  
Although the existence of such states in clean Pauli-limited superconductors has been theoretically predicted a half-century ago, only recently indications of their experimental realization have been reported in the organic and heavy-fermion superconductors, see reviews \cite{Matsuda:JPSJ76.2007,BeyerLTP13,*WosnitzaAnnPhys18}.

In most materials, the orbital effect dominates meaning that it destroys superconductivity
before reaching the strong spin-splitting regime and 
FFLO states have no chance to develop. 
The relative role of the spin and orbital pair-breaking effect is standardly characterized by the Maki's parameter, $\alpha_M=\sqrt{2}H^O_{C2}/H^P_{C2}$, where $H^O_{C2}$ and $H^P_{C2}$ are the upper critical fields for the orbital and spin pair-breaking mechanism, respectively. The FFLO states may emerge only if  $\alpha_M>1$. %
The orbital effect is weak or absent and the Zeeman effect dominates in special situations of either quasi-one-dimensional materials or quasi-two-dimensional materials in the magnetic field applied parallel to the conducting layers. Naturally, most experimental search for the FFLO states\cite{Matsuda:JPSJ76.2007,BeyerLTP13} as well as theoretical studies of them\cite{TakadaPrThPhys1969,MachidaPhysRevB.30.122,ShimaharaPhysRevB.50.12760,BurkhardtAnnPhys94,Dupuis:PhysRevLett.70.2613,*Dupuis:PhysRevB.51.9074,*Dupuis:PhysRevB.49.8993,CroitoruPhysRevB.89.224506} have been focused on these favorable cases. %

Alternatively, the conditions for the FFLO instability in the presence of the orbital effect have been investigated by Gruenberg and Gunther \cite{Gruenberg:PRL16.1966} for a clean \emph{isotropic} superconductor within \emph{the quasiclassical approach}. In this case, the emerging FFLO state is the Abrikosov vortex lattice with additional periodic modulation of the order parameter along the magnetic field. Such a state appears only for very large Maki's parameter, $\alpha_M>1.8$, corresponding to huge Zeeman energy and/or very shallow band, conditions unlikely to be realized in any isotropic single-band material\footnote{In this paper, we generalize this consideration for a \emph{layered superconductor} in the magnetic field perpendicular to the layers and found that in this case, the critical Maki's parameter is even larger, for strong anisotropy it is $4.76$.}. In spite of this limitation, rich properties of the emerging modulated vortex state have been theoretically investigated in detail %
\cite{HouzetPhysRevB01,HouzetPhysRevB2006,ManivPhysRevB2008,Zhuravlev:PRB80.2009,Shimahara:PRB80.2009}.

The most recent development in the field has been motivated by the discovery of iron-based superconductors \cite{Paglione:NatPhys6.2010,*StewartRevModPhys.83.1589,*Hosono:PhysC514.2015,*SiNatRevMat16}.
These materials are characterized by several electron and hole bands with rather small Fermi energies which can be tuned by doping or pressure. In addition, these compounds have high transition temperatures and very high upper critical fields $H_{C2}$, up to 70 T, likely limited by the paramagnetic effect. 
These properties make iron-based superconductors plausible candidates for the realization of the FFLO state, which motivated generalization of the theory of this state for multiple-band materials \cite{Gurevich:PRB82.2010,*Gurevich:RPP74.2011,AdachiJPSJ15,Ptok:JLowT172.2013,*Ptok:EPJB87.2014,*Ptok:JPhysCM27.2015,*Ptok:NJPhys19.2017}.

Practically all investigations of the orbital-effect influence on the FFLO transition have been done so far within the quasiclassical approximation. In pure materials, however, the superconducting instability in the magnetic field may be influenced by the orbital quantization%
\cite{Rajagopal:PLett23.1966,Gruenberg:PRev176.1968,Tesanovic:PRB39.1989,Reick:PhysicaC170.1990,MacDonald:PRB45.1992,Maniv:PRB46.1992}. This influence is most pronounced in superconductors with shallow bands and high upper critical fields, i.e., for the conditions also favoring the FFLO instability. It is not widely recognized that the quantization actually may profoundly promote this instability due to the one-dimensional nature of the quasiparticle's spectrum at the Landau levels.
Such quantization-induced FFLO states have been recently demonstrated in a special situation motivated by the physics of multiple-band iron-based superconductors, when one of the shallow bands is close to the extreme quantum limit so that the cyclotron frequency $\omega_c$  near $H_{C2}$ is comparable with the band's Fermi energy $ \epsilon_F$ \cite{SongHc2LifLay}. 
\begin{figure}[htbp]
	\centering
	\includegraphics[width=3.in]{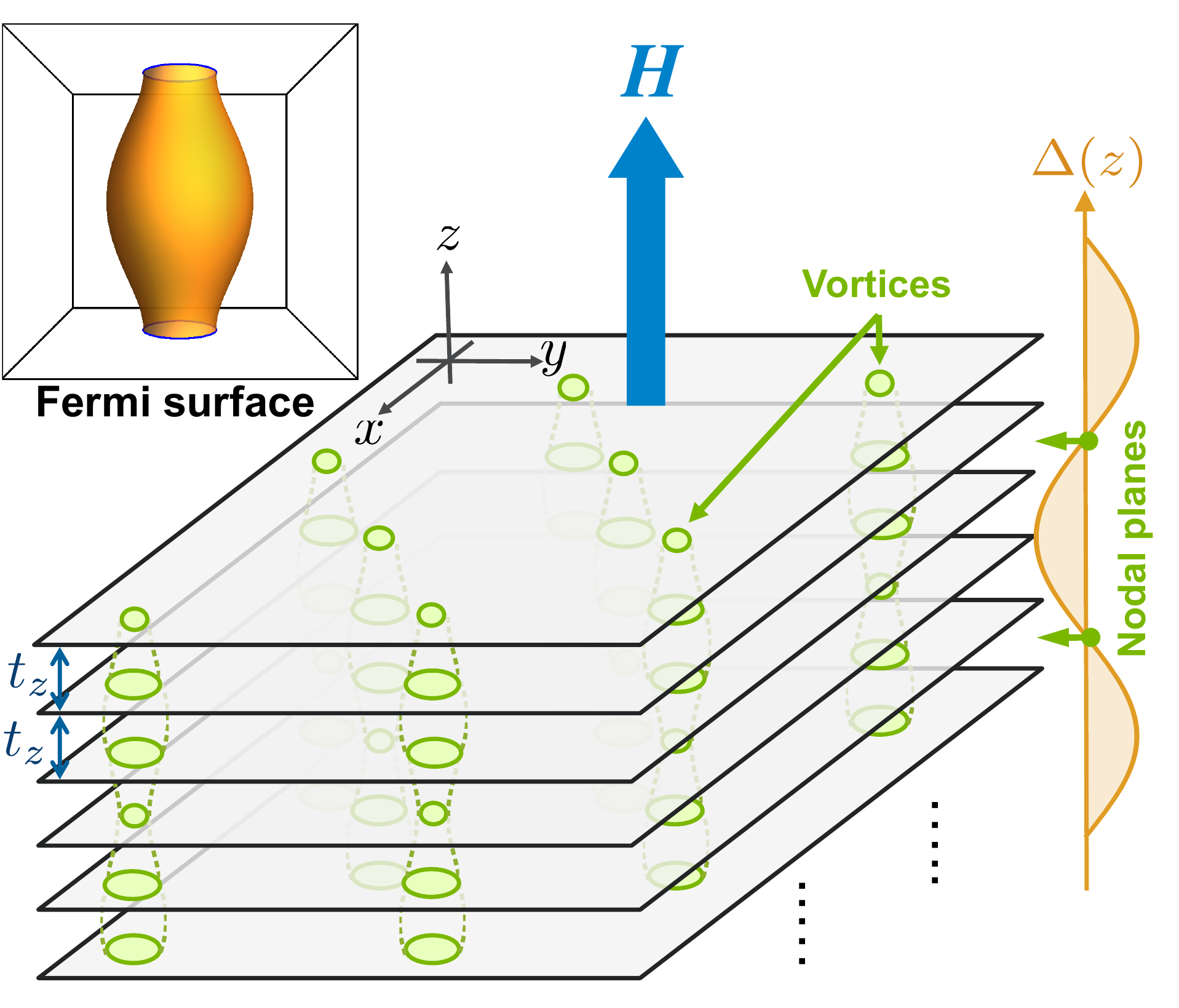}
	\caption{Schematic figure illustrating a layered superconductor in the out-of-plane magnetic field, open Fermi surface, and the emerging vortex state with interlayer Larkin-Ovchinnikov modulation.}
	\label{fig:Model}
\end{figure}

In this paper, we investigate the impact of Landau quantization on the FFLO instability in a generic and common case of an s-wave single-band layered material in the magnetic field applied perpendicular to the layers, see Fig.\ \ref{fig:Model}.  Evaluating the quantum-oscillating correction to the Cooper-pairing kernel with the finite spin splitting, we demonstrate, that, surprisingly, the quantum effects persistently promote the FFLO states in pure materials even in the limit of large Fermi energies, where the quasiclassical approximation is supposed to work well. Even though the quantum correction is smaller than the quasiclassical pairing kernel in this limit, at low temperatures, it acquires strong oscillating dependence on the FFLO modulation wave vector.  As a result, the optimal pairing in the low-temperature limit typically occurs at a finite wave vector, and the uniform-along-the-field state becomes unstable below the FFLO temperature, $T_{\mathrm{FFLO}}$. 
Because of the quantization, the electronic spectrum is composed of one-dimensional Landau-level branches depending on the c-axis momentum. The immediate reason for the emergence of the nonuniform state is the mismatch between the c-axis Fermi momenta for the branches with opposite spin orientation. One-dimensionality of the spectrum further enhances the instability. Contrary to the case of quasi-one-dimensional superconductors \cite{MachidaPhysRevB.30.122,Dupuis:PhysRevLett.70.2613,*Dupuis:PhysRevB.51.9074,*Dupuis:PhysRevB.49.8993,CroitoruPhysRevB.89.224506}, the optimal modulation vector is a result of an interplay between multiple branches. As for the classical FFLO states, the modulation allows to gain the Zeeman energy exceeding the loss of the condensation energy caused by a nonuniform order parameter. %

The specific behavior is sensitive to the relation between the spin-spitting energy $2\mu_zH$ and the Landau-level separation $\hbar \omega_c$, where $\mu_z$ is the band-electron magnetic moment. The FFLO temperature has the oscillating dependence on 
the field-independent ratio $2\gamma_z=2\mu_z H /(\hbar \omega_c)$, and its maximal value is given by the superconducting transition temperature squared divided by the Fermi energy, $T_{\mathrm{FFLO}}^{\mathrm{max}}\sim T_C^2/\epsilon_F$. This maximum is achieved when the ratio of the spin-spitting energy and the Landau-level separation is half-integer, $2\gamma_z\!=\!n\!+\!\tfrac{1}{2}$. On the other hand, $T_{\mathrm{FFLO}}$ vanishes, and the uniform state remains stable down to zero temperature only in the exceptional cases when this ratio is integer $2\gamma_z=n$. 
The modulation wave vector of the emerging FFLO state continuously grows from zero at the transition point to the low-temperature value which is proportional to the ratio of the cyclotron frequency and the interlayer hopping integral. The modulation period remains much larger than the interlayer separation. These results imply that the conditions for the onset of the FFLO state are much milder than it is generally anticipated. \emph{This state may actually appear in materials with small Zeeman energy, and, correspondingly, small Maki's parameter.} The only demanding requirement is the material's purity. The natural experimental indication of the required purity level is noticeable quantum oscillations in the normal state  near the superconducting instability.  

We focus here on the case of the magnetic field applied perpendicular to the layers, along the c axis. In this case $\gamma_z$ is a material's parameter. It is important to note, however, that this parameter can be effectively tuned by tilting the magnetic field away from the c axis\cite{WosnitzaFSLowDSC1996}, because the Zeeman energy is determined by the total magnetic field while the Landau-level separation is mostly determined by the c-axis field component. Therefore, the FFLO-instability temperature should have strongly oscillating dependence on the tilting angle.

This paper is organized as follows: In Sec.\ \ref{sec:model}, we describe our model of a layered superconductor and derive the equation describing superconducting instability in the out-of-plane magnetic field taking into account the quantum contribution to the pairing kernel and assuming a possibility of the FFLO modulation along the field. The derivation details of the quantum correction to the kernel are presented in Appendix \ref{app:OscCorr}.
For completeness, we derive in Appendix \ref{app:QCFFLO} the criterion for the emergence of the FFLO state in the \emph{quasiclassical approximation} generalizing previous consideration \cite{Gruenberg:PRL16.1966} to the case of layered superconductors. %
In Sec.\ \ref{sec:HT}, based on the derived equations, we investigate the influence of the quantum contributions to pairing on  
the interlayer FFLO transition 
using both analytical estimates and numerical calculations for representative parameters. 
In Appendix \ref{app:disorder}, we consider suppression of the quantum FFLO state by impurity scattering.
Finally, the summary and discussion are presented in Sec.\ \ref{Sec:Summary}.

\section{Superconducting instability in a layered superconductor}\label{sec:model}

We investigate the influence of the orbital-quantization effects on the onset of superconductivity for layered materials in the out-of-plane magnetic field. We use the tight-binding model with the nearest-layer  hopping term described by the Hamiltonian
\begin{align}
	&\mathcal{H}=\sum_{ j}\int\mathrm{d}^2\mathbf{r}\Big[c^\dagger_{ j s}(\mathbf{r})\left(\xi (\hat{\mathbf{k}}) \delta_{ss'}-\mu_zH\sigma^z_{ss'}\right)c_{js'}(\mathbf{r})\notag\\
	&-\!t_z c^\dagger_{j s}\!(\mathbf{r})c_{j+\!1,s}(\mathbf{r})\!+\!\text{H.c.}\!-\!Uc^\dagger_{ j\downarrow}\!(\mathbf{r})c^\dagger_{ j\uparrow}\!(\mathbf{r})
				c_{ j\downarrow}\!(\mathbf{r})c_{ j\uparrow}\!(\mathbf{r})\Big],
	\label{eqn:modelH}
\end{align}
where $\mathbf{r}=(x, y)$ is the in-plane coordinate, $j$ is the layer index, and $s$ represents spin (summation over $s$ and $s'$ is assumed). Furthermore, $t _z$ is the interlayer hopping energy, $\xi(\hat{\mathbf{k}})\!=\!\hat{\mathbf{k}}^2/(2m)\!-\!\mu$  is the intralayer energy dispersion with the band mass $m$, the Fermi energy $\mu$, and the momentum operator $\hat{\mathbf{k}}\! =\!-\mathrm{i}\nabla_{\mathbf{r}}\!-e\mathbf{A}/c$ \footnote{In the technical part, we use a natural system of units in which $k_B=1$ and $\hbar=1$.}.
We use the symmetric gauge for the vector potential, $\mathbf{A}=\frac{H}{2}(-y,x,0)$. We also include in the model the Zeeman spin splitting which is determined by the band electron's magnetic moment $\mu_z\!=\!g\mu_{B}/2$, where $\mu_{B}$ is the Bohr magneton and $g$ is the $g$-factor.    The full three-dimensional normal-state spectrum of the model is $\xi_{3D}(\mathbf{k},k_z)=\xi(\mathbf{k})-2t_z\cos k_z$. The corresponding open Fermi surface for $\mu>2t_z$ is illustrated in Fig.\ \ref{fig:Model}.

\begin{figure}[htbp]
	\centering
	\includegraphics[width=3.3in]{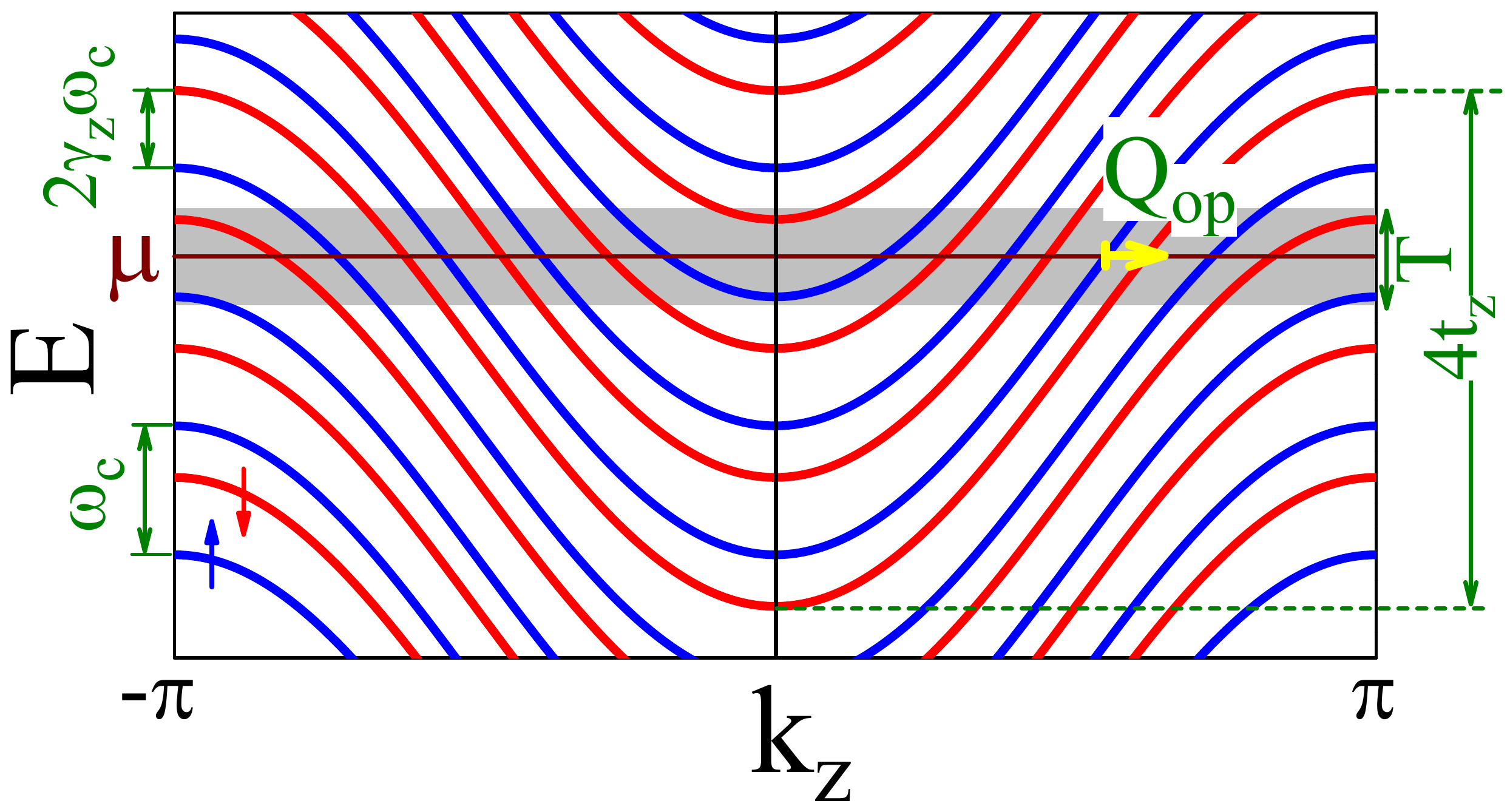}
	\caption{Illustrative plot of the spin-split Landau-level branches crossing the Fermi level. The plot also shows the relevant energy scales and the optimal modulation wave vector $Q_{\mathrm{op}}$.}
	\label{fig:LLBranches}
\end{figure}

To study the superconducting pairing instabilities for the model in Eq.\ \eqref{eqn:modelH}, we follow the standard approach and write the linearized gap equation as
\begin{equation}\label{eqn:gap-eqn}
\Delta_{j}(\mathbf{r})\!=\!UT\sum_{\omega_n}\sum_{j' } \!\int_{\mathbf{r}'}K_{\omega_n}(\mathbf{r} j,\mathbf{r}' j')\Delta_{j'}(\mathbf{r}'),
\end{equation}
where $\Delta_{j}(\mathbf{r})\!=U \langle c_{ j\downarrow}(\mathbf{r})c_{j\uparrow}(\mathbf{r})\rangle$
is the gap function, 
we used the notation $\int_{\mathbf{r}}=\int\mathrm{d}\mathbf{r}$, $\omega_n=2\pi T(n+1/2)$ are the Matsubara frequencies, and the kernel 
\begin{equation}
K_{\omega_n}\!(\mathbf{r} j,\mathbf{r}' j')
=G^+_{\omega_n}(\mathbf{r} j,\mathbf{r}' j')\bar{G}^- _{\omega_n}(\mathbf{r}' j',\mathbf{r} j),
\label{KernelDef}
\end{equation}
is determined by the one-particle Green's functions in the normal phase,  $G^\pm_{\omega_n}(\mathbf{r} j,\mathbf{r}' j')$, in which the superscripts $+$ or $-$ describe spin orientation and the overbar of the Green's function represents the complex conjugate.
These functions can be presented in the form of expansion over the exact Landau-level eigenstates as\cite{Rajagopal:PRB44.1991,Maniv:RMP73.2001}
\begin{equation}\label{eqn:GreenFH}
G^\pm_{\omega_n}\!=\!\frac{\exp\left(\mathrm{i}\frac{\mathrm{[\mathbf{r}\times\mathbf{r}']_z}}{2l_H^2}\right)}{2\pi l_H^2}\sum^\infty_{\ell=0}\left\langle\!\frac{\mathrm{e}^{-x/2}\mathrm{e}^{-\mathrm{i}k_z(j-j')}L_\ell(x)}{\mathrm{i}\omega_n-\xi_{\pm}(\ell\!+\frac{1}{2},k_{z})}\!\right\rangle_z,
\end{equation}	
where $\langle\dots\rangle_z\equiv \int_{-\pi}^\pi\dots\mathrm{d}k_z/(2\pi)$, $l_H=\sqrt{c/(eH)}$ is the magnetic length, $x=|\mathbf{r}-\mathbf{r}'|^2/2l_H^2$,  and $L_{\ell}(x)$ are the Laguerre polynomials. Furthermore, 
\begin{equation}
\xi_{\pm}(\ell\!+\tfrac{1}{2},k_{z})\equiv\omega_{c}\left(\ell\!+\tfrac{1}{2}\pm\gamma_{z}\right)\!-2t_{z}\cos k_{z}\!-\mu
\end{equation}
with $\omega_c=eH/(mc)$ being the cyclotron frequency and $\gamma_z\!=\!\mu_zmc/e\!=\!gm/4m_0$ being the reduced spin-splitting parameter, where $m_0$ is the free electron mass. The electronic spectrum is composed of the one-dimensional Landau-level branches, see Fig.\ \ref{fig:LLBranches}. In the limit $t_z\gg \omega_c$ roughly $4t_z/\omega_c$ of these branches cross the Fermi level for each spin orientation. 

In finite out-of-plane magnetic field, the gap parameter $\Delta_{j}(\mathbf{r})$ in the form of the \textit{lowest Landau-level} eigenfunction typically yields the leading instability.
In addition, in the presence of the Zeeman splitting, the order parameter may be periodically modulated between the layers, i.e., along the magnetic field\cite{Gruenberg:PRL16.1966}. Such a modulation is the realization of the nonuniform FFLO state \cite{Fulde:PRev135.1964,Larkin:JETP20.1965}
Therefore, we assume the solution in the form
\begin{equation}\label{eqn:ansatz}
\Delta_{j}(\mathbf{r})=\Delta _0\exp\left(-\frac{r^2}{2l^2_H}+\mathrm{i}Q_z j\right).
\end{equation}
The solution with the modulation vector $Q_z$ giving the maximal transition temperature, $T_{C2}$, is realized. Below this instability temperature, one has to compose a proper combination of the lowest Landau-level wavefunctions corresponding to the Abrikosov vortex lattice. The order parameter with the phase modulation along the field in Eq.\ \ \eqref{eqn:ansatz} is usually called the Fulde-Ferrel state. Alternatively, the state with the amplitude modulation, $\Delta_{j}\propto \cos (Q_z j)$, known as the Larkin-Ovchinnikov state, may emerge, see Fig.\ \ref{fig:Model}. We only investigate the instability location here, which is identical for both of these states. 
The gap function in Eq.\ \eqref{eqn:ansatz} is the exact eigenfunction of the kernel,
\[
\sum_{j'}\int_{\mathbf{r}'}K_{\omega_{n}}(\mathbf{r}j,\mathbf{r}'j')\Delta_{j'}(\mathbf{r}')\!=\!\pi\nu\lambda_{\omega_{n},Q_{z}}\Delta_{j}(\mathbf{r}),
\]
where $\nu=m/2\pi$ is the density of state per layer. This allows us to reduce the gap equation, Eq.\ \eqref{eqn:gap-eqn}, to
\begin{equation}
\Lambda^{-1}=2\pi T\, \text{Re}\sum^\Omega_{\omega_n>0} \lambda_{\omega_n,Q_z},
\end{equation}
where $\Lambda=\nu U$ is the coupling constant and $\Omega$ is the cutoff energy.

Using the expansion of the one-particle Green's function over the exact Landau-level basis, Eq.\ \eqref{eqn:GreenFH}, one can derive the exact presentation for the kernel eigenvalue\cite{Gruenberg:PRev176.1968,SongHc2LifLay},
\begin{align}\label{lh}
&\lambda_{\omega_n,Q_z}\!=\! -\frac{1}{2\pi\omega_c}\\
\times\!&
\sum_{\ell_{1},\ell_{2}}\!\left\langle \frac{(\ell_{1}+\ell_{2})!/(2^{\ell_{1}+\ell_{2}}\ell_{1}!\ell_{2}!)}{(\mathrm{i}\bar{\omega}_{n}\!-\!\ell_{1}\!-\frac{1}{2}-\!\tilde{\gamma}_{z}\!+\!\tilde{\mu})(\mathrm{i}\bar{\omega}_{n}\!+\!\ell_{2}\!+\frac{1}{2}\!-\!\tilde{\gamma}_{z}\!+\!\tilde{\mu})}\right\rangle _z\!,\nonumber
\end{align}
where we introduced the following notations
\begin{subequations}
\begin{align}
\tilde{\mu}(k_z,Q_z)&=\bar{\mu}+2\bar{t}_{z}\cos k_{z}\cos\tfrac{Q_{z}}{2}\label{tmu},\\
\tilde{\gamma}_{z}(k_z,Q_z)&=\gamma_{z}-2\bar{t}_{z}\sin k_{z}\sin\tfrac{Q_{z}}{2}\label{tg}.
\end{align}
\end{subequations}
Here all normalized quantities marked by bars are defined as $\bar{a}\equiv a/\omega_c$ (with $a=\omega_n,\mu,t_z$). 

The Matsubara-frequency sums are logarithmically-divergent and have to be cut at $\omega_n\sim \Omega$.  This divergence can be eliminated using the zero-field gap equation giving
\begin{equation}\label{eqn:gap-regularized}
\tfrac{1}{2}\ln(T/T_C)-\Upsilon_T+\Upsilon_{T_C}-\mathcal{J}(H,T,Q_z)
=0,
\end{equation}
where
$
\Upsilon_T
=-\int^{\infty}_0\!\frac{\mathrm{d}s}{\pi s}\ln\tanh(\pi T s)\sin(2\mu s)J_0(4t_z s),
$ \cite{SongHc2LifLay}
and the field-dependent parts of the pairing-kernel eigenvalue is
\begin{equation}
\mathcal{J}(H,T,Q_z)\!=\!2\pi T\!\sum^\Omega_{\omega_n>0}
\!\text{Re}\Big(\lambda_{\omega_n,Q_z}\!-\frac{1}{2\omega_n}\Big)\!-\!\Upsilon_T\label{Jh},
\end{equation}
with $\mathcal{J}(0,T,0)\!=\!0$. 
Therefore, the UV cutoffs are explicitly removed and the Matsubara-frequency sum on the left-hand side in Eqs.\ \eqref{Jh} converges now in the limit of $\Omega\to\infty$.
We can represent the functions $\mathcal{J} $ in this limit as \cite{SongHc2LifLay}
\begin{widetext}
	\begin{align}
\mathcal{J}(H,T,Q_{z})\! & =\frac{1}{4}\!\sum_{m=0}^{\infty}\sum_{\ell=0}^{m}\frac{m!}{2^{m}\left(m\!-\!\ell\right)!\ell!}\left\langle \frac{\mathcal{T}(\ell+\tilde{\gamma}_{z}-\!\tilde{\mu})+\mathcal{T}(m\!-\!\ell-\tilde{\gamma}_{z}-\!\tilde{\mu})-2\mathcal{T}(\frac{m}{2}-\tilde{\mu})}{m+1-2\tilde{\mu}}\right\rangle _{\!z}\nonumber \\
& -\frac{1}{2}\left\langle \int\limits _{0}^{1/2}dz\frac{\mathcal{T}(\frac{z-1}{2}-\tilde{\mu}_{0})}{z-2\tilde{\mu}_{0}}+\sum_{m=0}^{\infty}\int\limits _{-1/2}^{1/2}dz\left[\frac{\mathcal{T}(\frac{m+z}{2}-\tilde{\mu}_{0})}{m+1+z-2\tilde{\mu}_{0}}-\frac{\mathcal{T}(\frac{m}{2}-\tilde{\mu})}{m+1-2\tilde{\mu}}\right]\right\rangle _{\!z},\label{J1llSumPres}
	\end{align}
\end{widetext}
where $\tilde{\mu}_{0}(k_{z}) \!=\!\tilde{\mu}(k_{z},0) \!=\!\bar{\mu}\!+\!2\bar{t}_{z}\cos k_{z}$ and  $\mathcal{T}(x)\!\equiv\!\tanh[\omega_c(x\!+\!1/2)/2T]$. We remind that the parameters $\tilde{\mu}$ and $\tilde{\gamma}_{z}$ depend on $k_z$ and $Q_z$, see Eqs.\ \eqref{tmu} and \eqref{tg}. 
Different terms in the above sum describe the contributions to pairing from two quasiparticle states with opposite spin orientations located at the Landau-level branches with indices $\ell$ and $m-\ell$.  The c-axis momenta of the pairing states at these branches $\pm k_z+Q_z/2$ are mismatched by the modulation wave vector $Q_z$.
For the fixed magnetic field, we have to obtain the transition temperature $T_{C2}(H,Q_z)$ by solving Eq.\ \eqref{eqn:gap-regularized} and then find $Q_z$ which gives its maximum.

We are mostly interested in the quasiclassical limit set by the related conditions $\mu-2t_z\gg \omega_c,T_{C}$, that are satisfied in an overwhelming majority of materials. In this case, the problem can be significantly simplified. First,  in the limit $\mu\gg T_{C}$, we have  $-\Upsilon_{T}+\Upsilon_{T_{C}}\approx\tfrac{1}{2}\ln(T/T_{C})$ meaning that Eq.\ \eqref{eqn:gap-regularized} simplifies as $\ln(T/T_{C})-\mathcal{J}(H,T,Q_{z})=0$. 
Furthermore, in the limit  $\mu\!-\!2t_z\!\gg \! \omega_c$ high Landau levels $\ell_{1}, \ell_{2}\gg 1$ give the dominating contribution to the sum in Eq.\ \eqref{lh}. Therefore, the main term is obtained by neglecting discreteness of the spectrum and replacing the summation over these indices by integration over in-plane energies of the pairing states, $\epsilon_{1,2}=\omega_c(\ell_{1,2}+1/2)$, which gives the quasiclassical kernel, $\mathcal{J}_{\mathrm{cl}}$, 
\begin{align}
&\mathcal{J}_{\mathrm{cl}}(H,T,Q_z)=2\int^\infty_0\mathrm{d}s\ln\tanh\Big(\frac{\pi T}{\omega_c}s\Big)\nonumber\\
&\times\left\langle\exp(-\tilde{\mu}s^2)
[\tilde{\mu}s\cos(2\tilde{\gamma}_{z}s)\!+\!\tilde{\gamma}_{z}\sin(2\tilde{\gamma}_{z}s)]\right\rangle_z,
\label{eqn:J1qc}
\end{align}
This contribution is the famous quasiclassical Werthamer-Helfand-Hohenberg (WHH) result\cite{Helfand:PRev.1966,*Werthamer:PRev147.1966,Kogan:RPP75.2012}, which is widely used to describe the temperature dependence of the upper critical field in clean superconducting materials. This quasiclassical term usually favors the uniform state, $Q_z=0$, unless the Maki's parameter set by the Zeeman energy is anomalously large. We analyze this issue in Appendix \ref{app:QCFFLO}. 

The discreteness of the Landau-level spectrum leads to the quantum correction to the quasiclassical kernel,  $\mathcal{J}_{\mathrm{q}}(H,T,Q_z)$,  which we derive in Appendix \ref{app:OscCorr}.
This correction is the sum of terms that are (i) oscillating functions of the in-plane energies of two pairing states, $\epsilon_{1,2}$, with the period equal to the cyclotron frequency $\omega_c$, $\propto \exp [i (m_1\epsilon_{1}-m_2\epsilon_{2})/\omega_c]$, and (ii) rapidly decrease with separation between the average in-plane energy $(\epsilon_1+\epsilon_2)/2$ and the average in-plane Fermi energy for the pairing states with the c-axis wave vectors $\pm k_z+Q_z/2$. Therefore, the sum over the two Landau-level indices in Eq.\ \eqref{lh} is replaced by the sum over two harmonic indices $m_{1,2}$, in which all terms have to be integrated over the two energies and averaged over $k_z$.  This double sum can be further split into two contributions with qualitatively different behavior. The terms with  mismatched harmonic indices $m_{1}\!\neq\! m_2$ rapidly oscillate with the ratio $\mu/\omega_c$ but weakly depend on the modulation wave vector $Q_z$. On  the other hand, in the same-harmonic terms with $m_{1}\!=\! m_2$ the strong magnetic oscillations cancel but these terms have instead a pronounced dependence on $Q_z$ with typical scale given by average separation between Landau-level branches, as illustrated in Fig.\ \ref{fig:LLBranches}.  
This dependence appears because the modulation partially compensates the momentum mismatch at the branches caused by the spin splitting.  This key property is the origin of the effects discussed in this paper. All contributions together can be presented in the following concise form
\begin{align}
&\mathcal{J}_{\mathrm{q}}(H,T,Q_z)=\frac{2\pi^{3/2}T}{\omega_{c}}\!\sum_{k\!=\!1}^{\infty}\!(-1)^{k}\frac{\cos\!\left(2\pi k\gamma_{z}\right)}{\sinh\left(\frac{2\pi^{2}kT}{\omega_{c}}\right)}\nonumber\\
&\times\!\left\langle \!\frac{\cos\!\left(4\pi k\bar{t}_z\sin k_{z}\!\sin\frac{Q_{z}}{2}\right)
	\!\sin\!\left[2\pi k\tilde{\mu}(k_{z},Q_{z})\right]}
{\sqrt{\tilde{\mu}(k_{z},Q_{z})}\tan\left[2\pi\tilde{\mu}(k_{z},Q_{z})\right]}\!\right\rangle_{\!z}\!.
\label{eqn:J1q}
\end{align}
Here the oscillating part of the factor $\sin( k \phi)/\tan \phi$ with $\phi\!=\!2\pi\tilde{\mu}(k_{z},Q_{z})$ is coming from mismatched-harmonics terms, while its average part equal to 1 for even $k$ originates from the same-harmonics terms. The oscillating contribution has a structure resembling other quantum-oscillation quantities such as the de Haas-van Alphen oscillating magnetization, see, e.g., Ref.\ \cite{ShoenbergMagnOscBook}.  Namely, it is the sum of terms which are periodic functions of $1/H$ (since $\tilde{\mu}(k_{z},Q_{z}),\bar{t}_z\propto 1/H$) and exponentially decay with the temperature for $T>\omega_c$. Also, these terms contain the familiar factors $\cos\!\left(2\pi k\gamma_{z}\right)$ due to the spin splitting. The analogy, however, is not complete because, in contrast to single-electron normal-state quantities, the quantum pairing kernel is a two-electron property. In particular, the same-harmonics contribution to the pairing kernel does not have an analogue in the quantum correction to the normal-state magnetization.

Therefore, in the standard quasiclassical limit $\mu-2t_z\gg \omega_c, T_C$, the total pairing kernel can be split into classical and quantum contributions $\mathcal{J}(H,T,Q_z)\approx\mathcal{J}_{\mathrm{cl}}(H,T,Q_z)\!+\!\mathcal{J}_{\mathrm{q}}(H,T,Q_z)$ and Eq.\ \eqref{eqn:gap-regularized} for the upper critical field $H_{C2}$ can be approximated as
\begin{equation}
\ln(T/T_C)-\mathcal{J}_{\mathrm{cl}}(H,T,Q_z)\!-\!\mathcal{J}_{\mathrm{q}}(H,T,Q_z)=0.
\label{eqHc2-approx}
\end{equation}
The quantum contribution is expected to be small in the quasiclassical limit. We will demonstrate, however, that, while weakly influencing the absolute value of $H_{C2}$, this correction strongly promotes the formation of the FFLO state at low temperatures.

\section{Interlayer FFLO transitions}\label{sec:HT}

In this section, we address the problem of interlayer FFLO instability. 
It is well established that \emph{within the quasiclassical approximation} the FFLO state emerges only when the Maki's parameter of the material exceeds a certain critical value. In particular, for an isotropic 3D material this value was evaluated as $\sim 1.8$  in Ref.\ \cite{Gruenberg:PRL16.1966}. For completeness, in Appendix \ref{app:QCFFLO} we generalize this quasiclassical consideration to the system we analyze here, a quasi-two-dimensional layered superconductor in the magnetic field applied perpendicular to the layers. 
The Maki's parameter for this system is expressed via the electronic parameters as
\begin{equation}\label{eqn:alpha}
\alpha_{M}=\frac{\pi T_{C}\gamma_{z}}{\mu}\frac{4}{1+\sqrt{1-4t_z^2/\mu^2}}.
\end{equation} 
We can see that this parameter may be large only if the band is not too deep and the spin-splitting factor is very large. We found that in the open-Fermi surface regime, $\mu> 2t_z$, the critical Maki's parameter is   $\approx 4.76$.  This result suggests that the formation of the FFLO state in layered materials requires even higher Zeeman energy than in the isotropic case.

We argue, however, that this established quasiclassical picture is incomplete
and only provides the correct criterion for the FFLO instability if the temperature is not too low, $T>\omega_c$.
The conditions for FFLO instability at very low temperatures dramatically change when the orbital-quantization correction ($\mathcal{J}_{\mathrm{q}}$) in the pairing kernel is taken into account. To see this, we investigate the influence of this correction on the onset of the FFLO state for the case when the Zeeman spin-splitting parameter is not near the resonant values. We start with approximate analytical analysis for the common particular case
$\omega_{c}\ll t_{z}\ll\mu$. The quantum correction in Eq.\ \eqref{eqn:J1q} is a sum of the oscillating terms exponentially decaying with the temperature. In the range  $T\gtrsim\omega_{c}$ the dominating contribution to
$\mathcal{J}_{\mathrm{q}}$ is coming from the first several terms. The first two terms can be evaluated as (see Appendix \ref{app:OscCorr} for details)
\begin{subequations}
\begin{align}
\mathcal{J}_{\mathrm{q}}^{(1)}\! & \approx\!-\sqrt{\frac{2\pi}{t_{z}}}\!\frac{T\cos\!\left(2\pi\gamma_{z}\right)}{\sinh\left(\frac{2\pi^{2}T}{\omega_{c}}\right)}\!\sum_{ \delta_{t}=\pm 1}\! \frac{\cos\!\left(2\pi\frac{\mu+2\delta_{t}t_{z}}{\omega_{c}}\!-\!\frac{\delta_{t}\pi}{4}\right)}
{\sqrt{\mu\!+\!2\delta_{t}t_{z}\cos^{2}\frac{Q_{z}}{2}}}
,\label{eq:JQ1}\\
\mathcal{J}_{\mathrm{q}}^{(2)} & \!\approx\!2\pi^{3/2}\!\frac{T\cos\!\left(4\pi\gamma_{z}\right)}{\sinh\left(\frac{4\pi^{2}T}{\omega_{c}}\right)}\Bigg[\!\frac{1}{\sqrt{\omega_{c}\mu}}J_{0}\left(4\pi\frac{2t_{z}}{\omega_{c}}\sin\frac{Q_{z}}{2}\right)\nonumber\\
&+\sum_{ \delta_{t}=\pm 1}\frac{\cos\left(4\pi\frac{\mu\!+\!2\delta_{t}t_{z}}{\omega_{c}}-\frac{\delta_{t}\pi}{4}\right)}{4\pi\sqrt{t_{z}\left(\mu\!+\!2\delta_{t}t_{z}\cos^{2}\frac{Q_{z}}{2}\right)}}
\Bigg],
\label{eq:JQ2}
\end{align}
\end{subequations}
where $J_{0}(x)$ is the Bessel function. We see that the first term oscillates with the magnetic field in the same way as the normal-state magnetization (de Haas-van Alphen effect) and conductivity (Shubnikov-de Haas effect) and the second term also has such magnetic-oscillating contribution given by the second line in Eq.\ \eqref{eq:JQ2}
\footnote{The ratio $(\mu\pm 2t_z)/\omega_c$ in Eqs.\ \eqref{eq:JQ1} and \eqref{eq:JQ2} can be rewritten in a more common form as  $F_\pm/H$, where $F_\pm=(c/2\pi e)A_\pm$ is the de Haas-van Alphen frequency and $A_\pm$ is the area of the corresponding extremal Fermi-surface cross section, see, e.g.,  Ref.\ \cite{ShoenbergMagnOscBook}}. 
As discussed in the previous section, such terms appear from the oscillating contributions of the two pairing electronic states with opposite spins, which have mismatched periodicities in the in-plane energy dependence.
In addition, the second term $\mathcal{J}_{\mathrm{q}}^{(2)}$ has qualitatively different contribution 
described by the first line in Eq.\ \eqref{eq:JQ2}
that does not oscillate with $(\mu\pm 2t_z)/\omega_c$ but has a pronounced oscillating dependence on $Q_{z}$.
It originates from the same-harmonics contributions of the two pairing electronic states, also discussed in the previous section. 
This crucial part of the kernel is essentially a two-particle property which does not have analogues in  single-electron normal properties.
For large spin splitting between the Landau-level branches, the pairing at finite $Q_z$ allows the system to mitigate this split. This enhancement of pairing is quantitatively described by this contribution, which dominates the $Q_{z}$ dependence of the \emph{total} quantum correction at temperatures $T\gtrsim\omega_{c}$. Moreover, at low temperatures this enhancement occurs to be much stronger than the suppression of the quasiclassical kernel with $Q_z$, Eq.\ \eqref{eqn:J1qc}, in the usual regime of the dominating orbital effect.  

Stability of the uniform state is determined by the second derivative
of $\mathcal{J}_{\mathrm{q}}$ with respect to $Q_{z}$ at $Q_z=0$, $\mathcal{J}_{\mathrm{q}}^{\prime\prime}\equiv\partial^{2}\mathcal{J}_{\mathrm{q}}/\partial Q_{z}^{2}|_{Q_z=0},$
\begin{equation}
\mathcal{J}_{\mathrm{q}}^{\prime\prime}\approx-\pi^{3/2}T\!\frac{\cos\!\left(4\pi\gamma_{z}\right)}{\sinh\left(\frac{4\pi^{2}T}{\omega_{c}}\right)}\!\frac{1}{\sqrt{\omega_{c}\mu}}\left(4\pi\frac{t_{z}}{\omega_{c}}\right)^{2}.\label{eq:Sec-derQuant}
\end{equation}
It is positive for the spin-splitting factors in the range $|2\gamma_{z}-n/2|<1/4$
with $n=1,2\ldots$ meaning that the quantum correction strongly favors the
modulated state  within this range. Adding the quasiclassical term $\mathcal{J}_{\mathrm{cl}}^{\prime\prime}\approx-t_{z}^{2}/\left(\omega_{c}\mu\right)$
valid in the limit $\gamma_{z}\ll\sqrt{\mu\!/\!\omega_{c}}$, we obtain
the total second derivative
\begin{align}
&\mathcal{J}^{\prime\prime}\!\approx\!-\frac{t_{z}^{2}}{\omega_{c}\mu}\left(1\!+\!4\pi^{3/2}\cos\!\left(4\pi\gamma_{z}\right)\!\frac{\frac{4\pi^{2}T}{\omega_{c}}}{\sinh\left(\frac{4\pi^{2}T}{\omega_{c}}\right)}\sqrt{\frac{\mu}{\omega_{c}}}\right)\nonumber\\
&\approx\!-\frac{t_{z}^{2}}{\omega_{c}\mu}\left(1\!+\!
\cos\!\left(4\pi\gamma_{z}\right)\!\frac{16\sqrt{\pi}\mathtt{C}_{\mathrm{E}}^{3/2}\mu^{2}T/T_{C}^{3}}{\sinh\left(4\mathtt{C}_{\mathrm{E}}\mu T/T_{C}^{2}\right)}\right),\label{eq:SecDerTot2}
\end{align}
where in the second line, assuming $T\ll T_{C}$, we substituted the low-temperature limit
for $\omega_{c}$ at the upper critical field for $t_{z}\!\ll\!\mu$,
$\omega_{c}\approx \pi^{2}T_{C}^{2}\!/\!(\mathtt{C}_{\mathrm{E}}\mu)$ with  $\mathtt{C}_{\mathrm{E}}\!\approx\!\exp(0.5772)\!\approx\!1.781$ being the exponential of the Euler–-Mascheroni constant.
We can see that the quantum correction exponentially decays above
the temperature scale $T_{q}=\omega_{c}/4\pi^{2}=0.14T_{C}^{2}/\mu$.
However, at $T\sim T_{q}$ it is already larger than the quasiclassical
term by the factor $\mu/T_{C}\gg 1$.  In the case $\cos\!\left(4\pi\gamma_{z}\right)<0$,
this property allows us to evaluate the FFLO instability temperature from the equation $\mathcal{J}^{\prime\prime}\!=\!0$ with logarithmic accuracy,  
\begin{equation}
T_{\mathrm{FFLO}}\approx\!\frac{T_{C}^{2}}{4\mathtt{C}_{\mathrm{E}}\mu}\ln
\left(\left|\cos\!\left(4\pi\gamma_{z}\right)\right|\frac{C_{\mu}\mu}{T_{C}}\right).
\label{eq:TFFLO}
\end{equation}
where $C_{\mu}$ is a numerical factor $\approx 100\!-\!150$. This estimate is valid until the expression under the logarithm is large, i.e., it breaks near the points $|2\gamma_{z}\!-n/2|\!=\!1/4$ where $\cos\!\left(4\pi\gamma_{z}\right)$ vanishes.  
For spin-splitting factors outside the ranges $|2\gamma_{z}-n/2|<1/4$, the system may still have the FFLO instability, but it realizes at temperatures much smaller than $T_q$, meaning that its evaluation requires accounting for multiple terms in the sum in Eq.\ \eqref{eqn:J1q} and lacks a simple description.
In addition, even though the uniform state becomes unstable, the transition
to the modulated state takes place at a noticeable fraction of the zero-field
transition temperature only if the ratio $T_{C}/\mu$ is not too small.

At temperatures smaller than $T_{\mathrm{FFLO}}$, the $Q_{z}$ dependence
of the pairing kernel is dominated by the quantum term, Eq.\ \eqref{eq:JQ2},
and the optimal modulation vector has to be close to its maximum.
As the first minimum of the Bessel function $J_{0}\left(x\right)$ is located
at $x_{\mathrm{min}}\!=\!3.832$, the optimal $Q_{z}$ can be evaluated as
\begin{equation}
Q_{\mathrm{op}}\!\approx\!3.832\omega_{c}/(4\pi t_{z})\!\approx\!0.305\omega_{c}/t_{z}.
\label{eq:Qz-Hth}
\end{equation}
This wave vector is close to the average separation between the neighboring opposite-spin Landau-level branches near $k_z\!=\!\pi/2$, see Fig. \ref{fig:LLBranches}. Note that, in contrast to \emph{minimum} spacing between the branches, the \emph{average} separation does not depend on $\gamma_z$. 
Weak dependence of the modulation wave vector on the spin splitting is an unusual feature distinguishing our state from conventional FFLO states. We remind that the above result is obtained under the assumption $\omega_{c}\!\ll\!t_{z}$,  meaning that the modulation period in this regime is much larger than the distance between the layers.
Strictly speaking, the result in Eq.\ \eqref{eq:Qz-Hth} is derived assuming that the $Q_z$ dependence is mostly determined by the term $\mathcal{J}_{\mathrm{q}}^{(2)}$ in Eq.\ \eqref{eq:JQ2}, which is justified for $T\gtrsim T_{q}/2$. For lower temperatures, higher-$k$ terms in Eq. \eqref{eqn:J1q} become important, which may influence the value of $Q_{\mathrm{op}}$.  Further numerical checks, however, show that Eq.\ \eqref{eq:Qz-Hth} gives a good approximation for  $Q_{\mathrm{op}}$ within a rather wide temperature range.

\begin{figure}[htbp]
	\centering
	\includegraphics[width=3.4in]{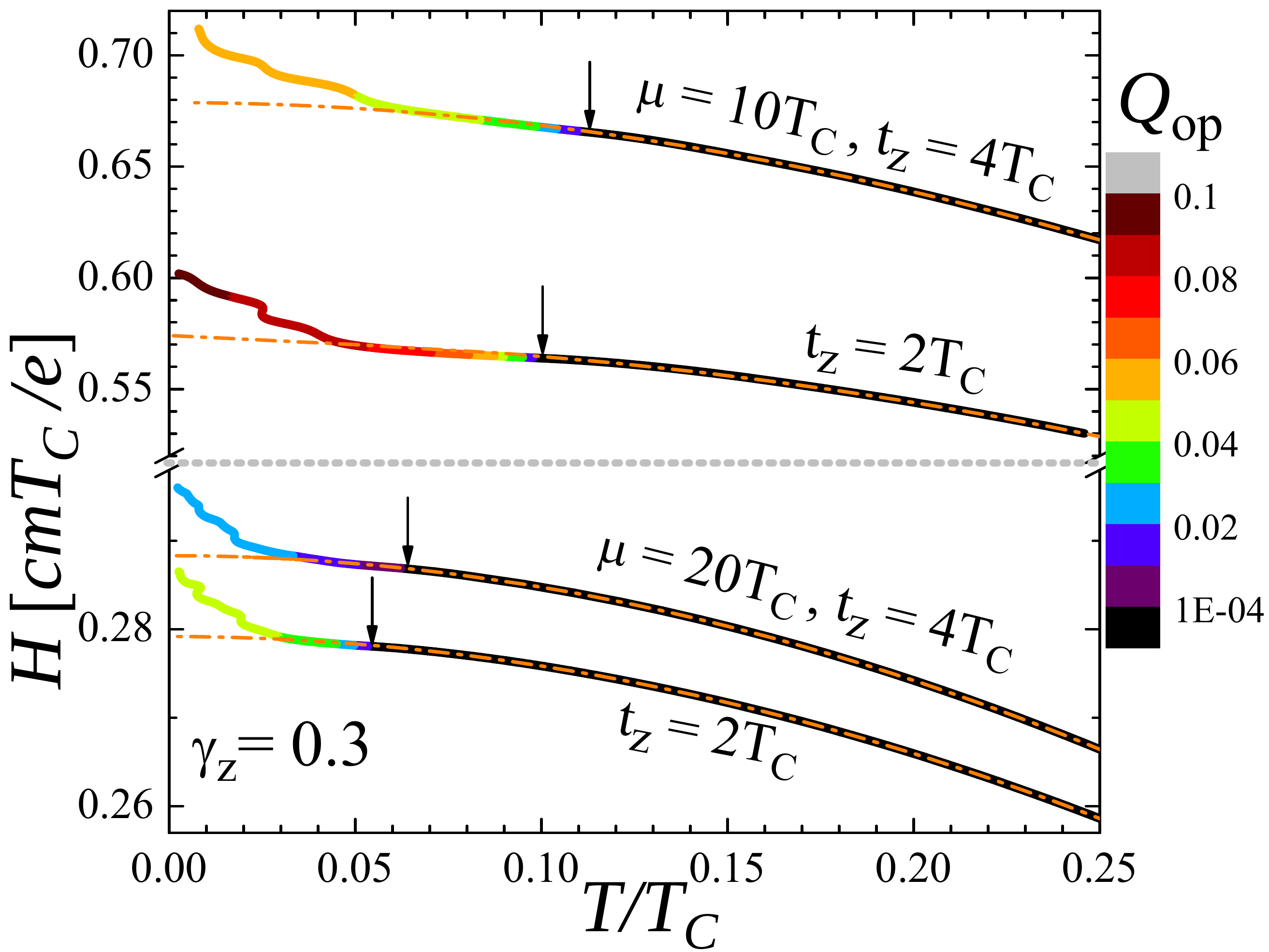}
	\caption{Examples of the temperature dependences of the upper critical field for the spin-splitting factor $\gamma_z=0.3$, two Fermi energies, $\mu=10T_C$ (upper part) and $20T_C$(lower part), and two hopping energies, $t_z=2T_C$ and $4T_C$. The boundaries are color-coded by the optimal modulation wave vector $Q_{\mathrm{op}}$. The arrows mark the location of the FFLO transition temperature, $T_{\mathrm{FFLO}}$. The orange dash-dot lines show quasiclassical results.}
	\label{fig:HC2Tmu10mu20}
\end{figure}
\begin{figure}[htbp] 
	\centering
	\includegraphics[width=3.2in]{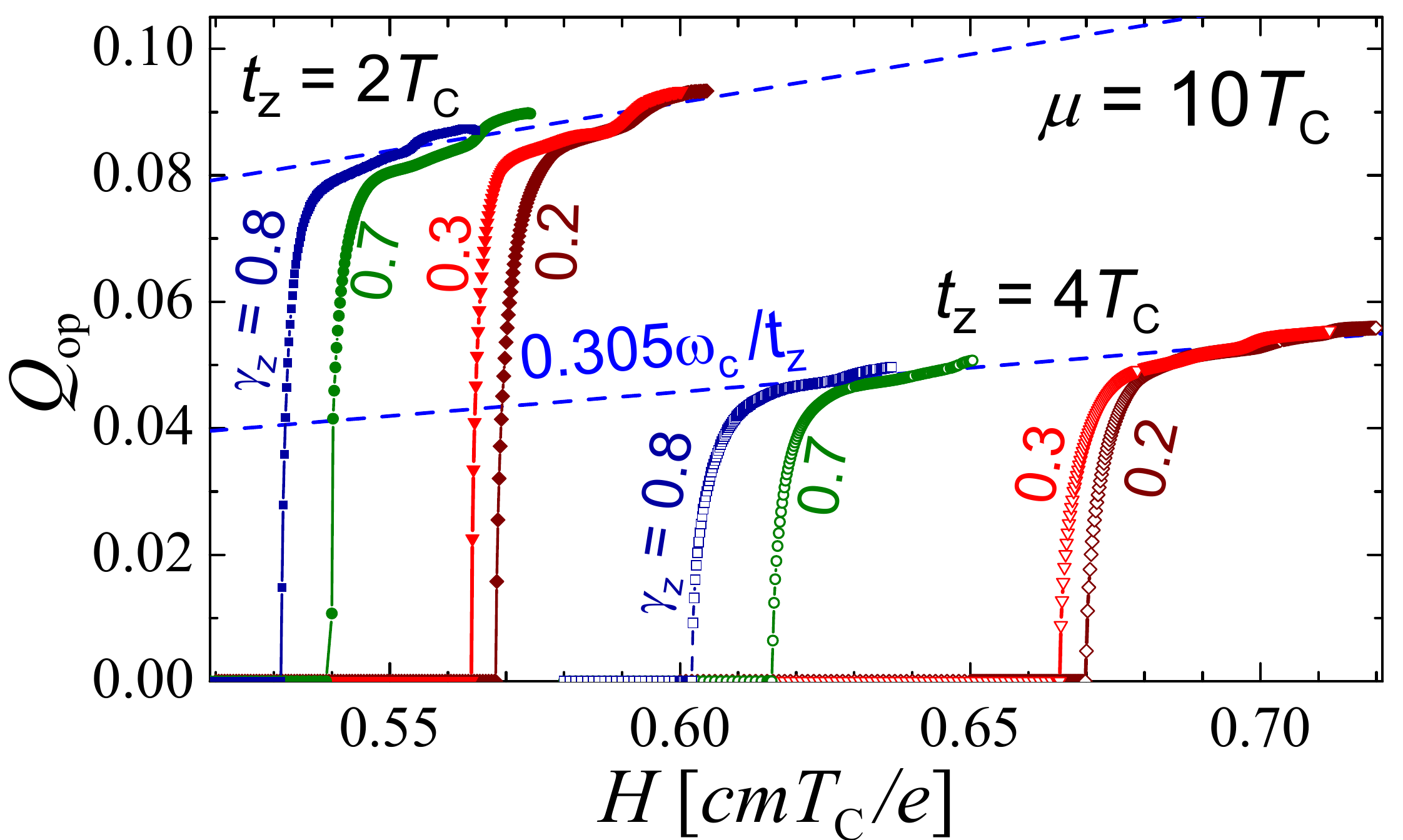}
	\caption{The field dependences of the optimal modulation wave vector $Q_z$ for $\mu=10T_C$, two values of $t_z$, $2T_C$ and $4T_C$, and several values of the spin-splitting parameter $\gamma_z$ near the optimal values $0.25$ and $0.75$.  The curves are marked by the values of $\gamma_z$. The dashed lines show the expected low-temperature behavior, Eq.\ \eqref{eq:Qz-Hth}}.
	\label{fig:QzHTmu10}
\end{figure}
To support and verify these analytical results, we proceed with the discussion of the numerically-computed phase diagrams. We remind that the transition temperature $T_{C2}$ at fixed $H$ and $Q_z$ can be computed
using exact equation, Eq.\ \eqref{eqn:gap-regularized}, with the exact result for the kernel $\mathcal{J}(H,T,Q_z)$, Eq.\ \eqref{J1llSumPres}. In the quasiclassical limit, $\mu\!-\!2t_z\gg T_C,\omega_c$, however,  calculations are much easier with the  approximate equation, Eq.\ \eqref{eqHc2-approx}, in which the classical and quantum contributions to the kernel are given by Eqs.\ \eqref{eqn:J1qc}, and \eqref{eqn:J1q}, respectively. The modulation vector maximizing $T_{C2}$ has to be selected.
Figure \ref{fig:HC2Tmu10mu20} shows the representative upper critical field lines at low temperatures for  the spin-splitting factor $\gamma_z=0.3$, two Fermi energies, $\mu=10T_C$  and $20T_C$, and two hopping energies, $t_z=2T_C$ and $4T_C$. 
This choice of electronic parameters corresponds to small values of the  Maki's parameter. From Eq.\ \eqref{eqn:alpha}, we estimate $\alpha_M\!\sim\! 0.2$ for $\mu\!=\!10T_C$ and $\alpha_M\!\sim\! 0.1$ for $\mu\!=\!20T_C$.
Nevertheless, in all shown cases the FFLO instability develops below the critical temperature, $T_\text{FFLO}$. 
For $\mu=10T_C$ this critical temperature is slightly above $0.1T_C$ and for $\mu=20T_C$ it is slightly above $0.05T_C$, in accidental agreement with a simple estimate $T_\text{FFLO}/T_C\sim T_C/\mu$. At somewhat lower temperature, $\sim 0.5T_\text{FFLO}$, the oscillatory upturn of $H_{C2}(T)$ develops. 
The optimal modulation wave vector, $Q_{\mathrm{op}}$, continuously increases below $T_\text{FFLO}$. 
Figure \ref{fig:QzHTmu10} shows the field dependences of $Q_{\mathrm{op}}$ for $\mu=10T_C$, two values of $t_z$, $2T_C$ and $4T_C$, and several values of the spin-splitting parameter located near the optimal values $0.25$ and $0.75$, including $\gamma_z\!=\!0.3$ used in Fig.\ \ref{fig:HC2Tmu10mu20}. The last points at these curves are typically at temperatures $0.005T_C$ and $0.01T_C$ for $t_z\!=\!2T_C$ and $4T_C$, respectively. We see that the modulation wave vector sharply increases below  $T_\text{FFLO}$ and at low temperatures it starts to approximately follow the linear dependence on the magnetic field predicted by Eq.\ \eqref{eq:Qz-Hth}. The value of $Q_{\mathrm{op}}$ in this regime weakly depends on  the spin-splitting parameter which only determines the field range where such a behavior is realized. 

\begin{figure}[htbp] 
	\centering
	\includegraphics[width=3.2in]{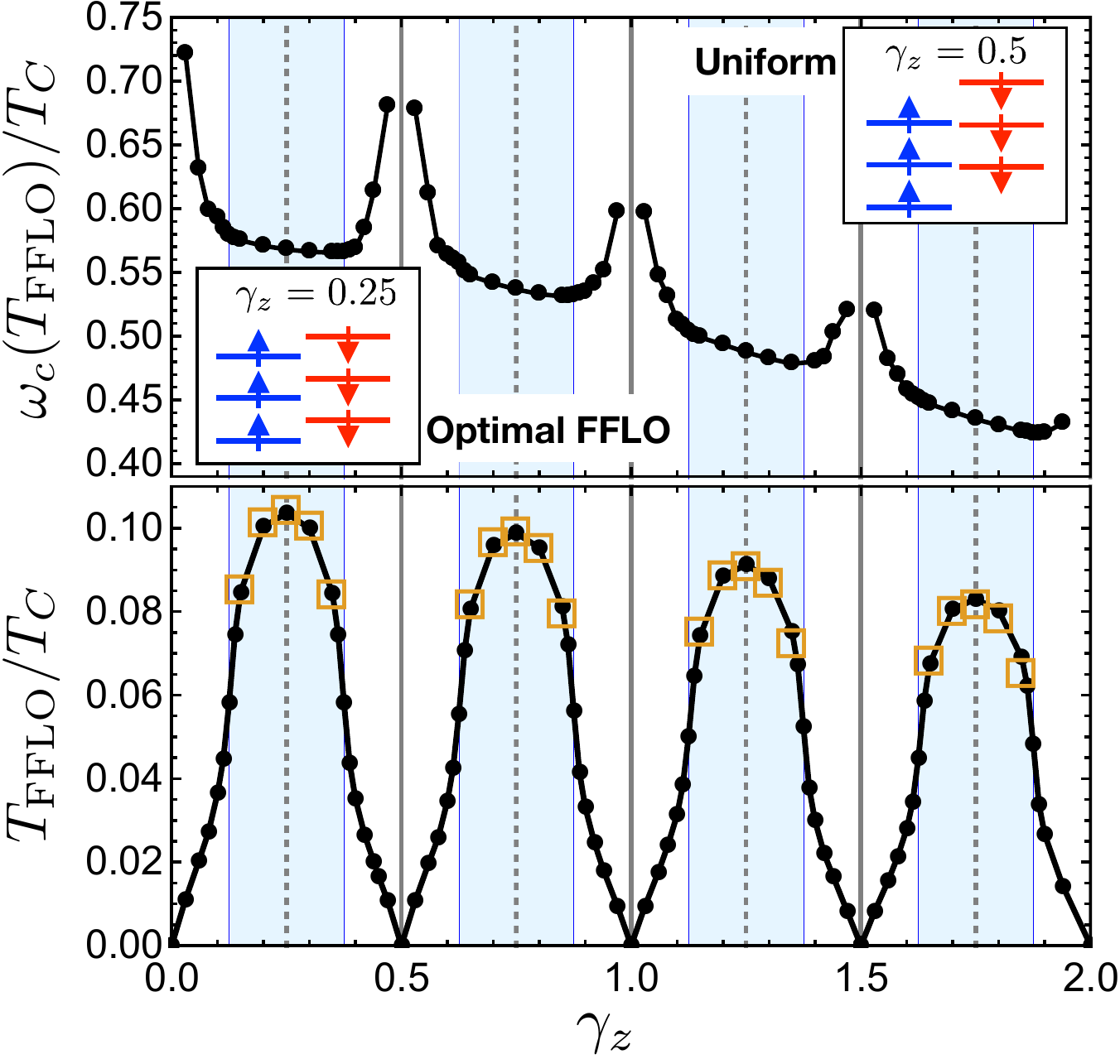}
	\caption{The lower panel shows the dependence of the FFLO onset temperature $T_\text{FFLO}$ on the spin-splitting factor $\gamma_z$. The parameters in this plot are $t_z/\mu=0.2$ and $\mu/T_C=10$. In contrast to the quasiclassical case, the FFLO states emerge at finite $T_\text{FFLO}$ even when the $\gamma_z$ is very small. Furthermore, $T_\text{FFLO}(\gamma_z)$ oscillates and its maxima are located at $2\gamma_z=n+1/2$ corresponding to the largest minimum separation between the opposite-spin Landau levels, as illustrated in the inset for $\gamma_z=0.25$. Open squares show analytical estimates using Eq.\ \eqref{eq:SecDerTot2}. The upper panel shows $\gamma_z$ dependence of the cyclotron frequency at the upper critical field for $T=T_\text{FFLO}$. We omit the points at the resonances ($2\gamma_z=1,2,\dots$), where the spin degeneracy of the Landau levels is restored, see inset for $\gamma_z=0.5$. The FFLO transition is absent at these points.
	}
	\label{fig:TFFLO}
\end{figure}
The analytical estimate for the instability temperature in Eq.\ \eqref{eq:TFFLO}  is valid only for the Zeeman spin-splitting parameters within certain ranges, $|2\gamma_{z}-n/2|<1/4$. Outside these ranges, no simple analytical results are available.
The lower panel of Fig.\ \ref{fig:TFFLO} shows the numerically-computed $\gamma_z$ dependence of the FFLO instability temperature. We see that this dependence is oscillatory with the slowly decaying amplitude. As follows from Eq.\ \eqref{eq:SecDerTot2}, $T_\text{FFLO} \propto \omega_c$ suggesting that this slow decay is caused by an overall suppression of  $\omega_c \propto H_{C2}(T_\text{FFLO})$ with $\gamma_z$. Such a suppression is indeed seen in the upper panel of Fig.\ \ref{fig:TFFLO}, which presents the $\gamma_z$ dependence of the cyclotron frequency at the instability point. We note that the ratio $T_\text{FFLO}/\omega_c$ displays much better $\gamma_z$ periodicity with slightly increasing amplitude (not shown). The dependence  $T_\text{FFLO}(\gamma_z)$ within the ranges $|2\gamma_{z}-n/2|<1/4$ is in excellent agreement with 
the evaluation based on the analytical result, Eq.\ \eqref{eq:SecDerTot2}. 
When $\gamma_{z}$ is shifted outside this range, the instability temperature sharply drops and moves in the region of oscillatory $H_{C2}(T)$ behavior. Nevertheless, we see that the instability is always present unless $2\gamma_{z}$ exactly equals an integer.  
Therefore, both analytical and numerical analyses of this section consistently demonstrate that the quantum-oscillation contribution to the Cooper pairing favors the FFLO instability, especially in the case of strong spin splitting between the Landau-level branches. Our theory provides a quantitative description of this instability.

\section{Summary and discussion}\label{Sec:Summary}

In this paper, we investigate the FFLO instabilities in a clean single-band layered superconductor in the out-of-plane magnetic field taking into account the orbital-quantization effects. The quasiclassical analysis predicts the emergence of the FFLO state only at very large Maki's parameters, $\alpha_M> 4.76$. We found, however, that the quantum effects promote the formation of this state at low temperature even in the range of parameters where the quasiclassical approximation is expected to work well. Contrary to the quasiclassical predictions, the FFLO state in a clean system can emerge even in the weak Zeeman-effect regime ($\alpha_M\ll 4.76$). 
The instability of the uniform state is caused by the mismatch between the c-axis Fermi momenta for the one-dimensional Landau-level branches with opposite spin orientation.  Correspondingly, the optimal modulation vector at low temperatures is given by the typical separation between the branches. The condensation-energy loss in the modulated state is compensated by the higher gain in the Zeeman energy. Therefore, this state is expected to have higher electronic spin polarization in comparison with the uniform state. 
Note that in the case of a very strong Zeeman effect, the FFLO state may emerge via the first-order phase transition \cite{HouzetPhysRevB2006,Zhuravlev:PRB80.2009}. In our case of weak Zeeman energy, the transition is known to be continuous in the quasiclassical regime, and incorporating the quantization corrections will not change this scenario,
because the quantum correction to the quartic coefficient in the Ginzburg-Landau expansion is small and cannot change its sign.

We mostly focused on the typical situation when many Landau-level branches cross the Fermi level near the upper critical field, which also may realize in materials with different electronic spectra. In layered superconductors, this situation corresponds to the condition $4t_z\gg \omega_c \sim T_C^2/\mu$.   
We note, however, that some key results also hold for extremely anisotropic layered materials, in which the miniband width $4t_z$ is comparable to or smaller than the cyclotron frequency $\omega_c$.  In particular, the results for the classical and quantum contributions to the pairing kernel, Eqs.\ \eqref{eqn:J1qc} and \eqref{eqn:J1q}, are valid for arbitrary $t_z$ within the open Fermi surface regime. More importantly, the crucial contribution to the quantum kernel given by the first term in Eq.\ \eqref{eq:JQ2} remains valid for $t_z<\omega_c$. As a consequence, the FFLO instability persists for arbitrarily small interlayer hopping, and the instability temperature for the case $\cos\!\left(4\pi\gamma_{z}\right)\!<\!0$  can be still estimated from  Eq.\ \eqref{eq:TFFLO}.  The nature of the FFLO state developing at low temperatures, however, changes qualitatively when $4t_z$ becomes comparable with $\omega_c$. In this regime, only a few Landau-level branches cross the Fermi level. Correspondingly, the optimal modulation emerges as the result of competition between a few favorable wave vectors connecting the Fermi wave vectors of the opposite-spin branches. In this case, the low-temperature modulation period is expected to be comparable with the interlayer spacing and will have complicated magnetic-field dependence.     

The most demanding requirement for the observation of the quantum FFLO instability is the material's purity.  
The impurity scattering has detrimental effects on the quantum contributions, and strong disorder restores the quasiclassical behavior.  
As demonstrated in the Appendix \ref{app:OscCorr}, within the simplest lifetime approximation, the impurity scattering leads to appearance of the additional ``Dingle factors'' $\exp[-2\pi k\Gamma/\omega_c]$ in the sum for the quantum correction in Eq.\ \eqref{eqn:J1q}.
Here the impurity broadening $\Gamma$ is related to the scattering time as $\Gamma=1/2\tau$.  In addition to the reduction of the Landau-quantization corrections, the impurity factors make the sum in Eq.\ \eqref{eqn:J1q} convergent in zero-temperature limit for an arbitrary modulation wave vector.  
In Appendix \ref{app:disorder}, we consider suppression of the FFLO instability by impurities. We evaluate the critical impurity broadening $\Gamma_{\mathrm{cr}}$ above which the FFLO state is suppressed, $\Gamma_{\mathrm{cr}}  \!=\!\left({\omega_{c}}/{4\pi}\right)\ln\left(4\pi^{3/2}\left|\cos\!\left(4\pi\gamma_{z}\right)\right|\!\sqrt{{\mu}/{\omega_{c}}}\right)$ for $\cos\!\left(4\pi\gamma_{z}\right)\!<\!0$. We also illustrate the evolution of superconducting instability boundaries with increasing $\Gamma$.   

In our consideration, we assumed fixed chemical potential and neglected its quantum magnetic oscillation $\delta\mu(H)$, which should be present if the system is not coupled to a charge reservoir.
In layered metals in the limit $4 t_z/\omega_c\gg1$, the low-temperature oscillating amplitude of $\delta\mu(H)$ scales as $\omega_c\sqrt{\omega_c/t_z}$\cite{Gvozdikov:PRB68.2003}. To check if these chemical-potential oscillations have a noticeable influence on our results, we computed $\delta\mu(H)/\omega_c$ using the precise formula provided in Ref.\ \cite{Gvozdikov:PRB68.2003}. We found that for the parameters $\mu/T_C\!=\!10$ and $t_z/T_C\!=\!2$ in Fig.\ \ref{fig:HC2Tmu10mu20}, the maximum amplitude of $\delta\mu(H)/\omega_c$ is only $\sim 0.01$ near the FFLO transition. Moreover, even for the points with small $T_\text{FFLO}$  near the resonance spin splittings in Fig. \ref{fig:TFFLO}, $\delta\mu(H)/\omega_c$ does not exceed $0.05$. We conclude that the oscillating contribution to the chemical potential has a minor influence on the FFLO instability in the studied parameter range.

The FFLO transition temperature ($T_\text{FFLO}$) is an oscillating function of the spin-splitting parameter $\gamma_z=\mu_zH/\omega_c$. It has maxima at $2\gamma_z=n+\frac{1}{2}$ corresponding to the largest splitting between the Landau levels with opposite spin orientations, 
which is the least favorable situation for the uniform state. On the other hand, the uniform state remains stable down to zero temperature at the resonances $2\gamma_z=n$ corresponding to spin-degenerate Landau levels.  
As pointed out in the Introduction, the effective $\gamma_z$ can be tuned by tilting the magnetic field \cite{WosnitzaFSLowDSC1996} and we expect that $T_\text{FFLO}$ will also be an oscillating function of the tilting angle. 
This consideration is very general meaning that, in principle, such a quantum FFLO instability may appear in any layered superconducting material provided it can be prepared sufficiently pure. 
Moreover, we believe that specific assumptions for the electronic spectrum and s-wave symmetry of the order parameter made in derivation are not really essential, and we expect that the predicted promotion of the FFLO state by the quantization is a general phenomenon which also takes place in more complicated situations. 
We note, however, that in a typical good metal with very large Fermi energy $\mu > 100 T_C$ the FFLO transition temperature becomes vanishingly small, and, correspondingly, purity requirements may be unrealistic. Therefore, the best materials for observation of the predicted behavior are superconductors with strong pairing and not-too-deep bands so that  $\mu/T_{C}<30$ and $\mu/\omega_c<50$. Clearly, the predicted quantum effects become more pronounced with decreasing these ratios.

Among the known materials, the possible candidates for the predicted behavior may be found in organic and iron-based superconductors. 
Consider, for example, the well-studied organic superconducting material,
$\kappa\text{-(BEDT-TTF)}_2\text{Cu(NCS)}_2$ with $T_C=10.4$K\cite{Singleton:ContemPhys43.2002,Wosnitza:JLowTPhys146.2007,BeyerLTP13,*WosnitzaAnnPhys18} and $H_{C2}(0)\approx7$T\cite{Murata:SynMetals27.1988}.
As other organic materials, it can be made exceptionally clean
so that the quantum oscillations may be observed even inside the superconducting state\cite{SasakiPhysRevB.67.144521}. 
In addition, strong experimental support for the classical FFLO state caused by the large Zeeman energy already exists for the magnetic field \emph{oriented along the layers} \cite{BeyerLTP13,*WosnitzaAnnPhys18} and this state may realize only if impurity scattering is very weak. 
The band structure of this material is composed of one holelike corrugated cylindrical Fermi surface and two electronlike Fermi planar sheets. 
The Landau-quantization effect is relevant only for the holelike Fermi surface which is characterized by the effective mass $\sim 3.2 m_e$ \cite{MeyerEPL95} and interlayer hopping energy $t_z\approx0.04$ meV \cite{Singleton:PRL88.2002}.
The ratio $\mu/\omega_c$ is equal to the ratio of the de Haas-van Alphen frequency $\sim 599$T and $H_{C2}$ giving $\approx 86$. With above effective mass this yields $\omega_c\approx 0.25$ meV,  $\mu\approx 22$ meV, and $\mu/T_C=25$.
The spin-splitting parameter, $\gamma_z=1.3$ has been extracted using the `spin-zero' effect in the de Haas-van Alphen oscillations \cite{MeyerEPL95} and it is actually close to the optimal value of $1.25$ for the quantum FFLO scenario. 
The pairing in this and similar molecular crystals may be mediated by spin fluctuations leading to the $d$-wave symmetry of the order parameter \cite{SchmalianPhysRevLett.81.4232, KurokiJPSJ06}. Several material’s properties are consistent with the $d$-wave symmetry including, NMR\cite{DeSotoPhysRevB.52.10364}, low-temperature behavior of the London penetration depth\cite{CarringtonPhysRevLett.83.4172, MilbradtPhysRevB.88.064501} and specific heat\cite{TaylorPhysRevLett.99.057001}, dependences of specific heat\cite{MalonePhysRevB.82.014522}, and thermal conductivity\cite{IzawaPhysRevLett.88.027002} on the magnetic-field direction.  The consideration of this paper can be straightforwardly generalized to the d-wave case and we expect a very similar behavior.
The instability temperature for this material can be estimated from Eq.\ \eqref{eq:TFFLO} as $T_\text{FFLO}\sim0.045T_C\sim 0.45 K$. 
From the material's parameters, we estimate $4t_z/\omega_c\approx 0.7$ meaning that the Fermi level typically crosses only one Landau-level branch for every spin direction. In this case, one can expect a large modulation wave vector which is determined by the two Fermi momenta of the opposite-spin branches. 
We can conclude that this organic superconductor has almost ideal electronic parameters and is a very feasible candidate for the realization of the quantum FFLO state. It is, however, a challenge to demonstrate it experimentally. An additional complicating factor is that the transition may not be described by the mean-field theory due to strong quantum fluctuations \cite{UjiPhysRevB2018} which may smear the static configuration. The described FFLO instability actually enhances these fluctuations due to the reduction of the vortex-lattice tilt stiffness.

High values of transition temperatures and upper critical fields as well as small Fermi energies make \emph{iron-based superconductors} natural candidates for observing the predicted phenomenon.  The weak impurity scattering limit probably cannot be achieved in compounds obtained by doping from nonsuperconducting parent materials. Fortunately, there are also several stoichiometric compounds, such as FeSe, LiFeAs, and CaKFe$_4$As$_4$, which, at least in principle, can be made pure. 
For example, the compound FeSe has a transition temperature $\sim 8$K\cite{Medvedev:NatMat8.2009} and a rather high low-temperature upper critical field $\sim17$T\cite{Kasahara:PNAS111.2014,Terashima:PRB90.2014}. The material can be made clean allowing for the observation of quantum oscillations down to fields $\sim 20$T \cite{Terashima:PRB90.2014,Watson:PRB91.2015,AudouardEPL2015}, only slightly above $H_{c2}$. Its band structure is composed of hole and electron pockets with very small Fermi surfaces. An analysis of the Shubnikov-de Haas oscillations \cite{Terashima:PRB90.2014} gives the smallest Fermi energy for the electron and hole bands of only 3.9 and 5.4 meV, respectively, and ARPES measurements\cite{Shimojima:PRB90.2014,*Nakayama:PRL113.2014,*Watson:PRB91.2015,*FedorovSciRep2016} are consistent with these estimates. This means that the ratios $\epsilon_F/\omega_c$ are in the range of $4-6$, clearly indicating the relevance of quantum effects. 
Moreover, experimental indications of a possible phase transition inside the superconducting state have been reported recently.  It was demonstrated that the diagonal and Hall thermal conductivities have kinklike features near the magnetic field $H^\ast\sim15$T at $T< 1.5$K below $H_{C2}(T)$\cite{Kasahara:PNAS111.2014,Watashige:JPSJ86.2017}. It is feasible that this transition corresponds to the quantization-induced FFLO state with modulation along the magnetic field. 
Even if this interpretation is correct, the simple model used in this paper probably does not quantitatively describe this transition, because it is likely influenced by multiple-band effects. 

We conclude that the generally accepted picture of the true superconducting ground state in high magnetic fields is incomplete for clean materials. The quantization effects promote the formation of the FFLO state in which the order parameter is periodically modulated along the magnetic field. Such a state may actually realize in several existing pure materials, even though a direct experimental proof for it may be quite challenging.

\begin{acknowledgements}
The authors would like to thank Alexander Buzdin and Yakov Kopelevich for useful discussions. 
This work is supported by the U.S. Department of Energy, Office of Science, Basic Energy Sciences, Materials Sciences and Engineering Division. K.W.S. is supported by the Center for Emergent Superconductivity, an Energy Frontier Research Center funded by the U.S. Department of Energy, Office of Science, under Award No. DEAC0298CH1088.
\end{acknowledgements}

\appendix

\begin{widetext}
\section{Derivation of the oscillating correction to the pairing kernel eigenvalue $\mathcal{J}$}
\label{app:OscCorr}

The starting point of the derivation is the exact result for the eigenvalue
$\lambda_{\omega_{n},Q_{z}}$, Eq.\ \eqref{lh}, which we rewrite in
real variables as
\begin{equation}
\lambda_{\omega_{n},Q_{z}}\!=\!-\frac{\omega_{c}}{2\pi}\!\sum_{\ell_{1},\ell_{2}=0}^{\infty}\!\frac{\left(\ell_{1}+\ell_{2}\right)!}{2^{\ell_{1}\!+\!\ell_{2}}\ell_{1}!\ell_{2}!}\left\langle \frac{1}{\left[\mathrm{i}(\omega_{n}+\zeta_\omega\Gamma) \!-\!\xi_{+}(\ell_{1}\!+\frac{1}{2},k_{z}\!+\!Q_{z}/2)\right]\left[\mathrm{i}(\omega_{n}+\zeta_\omega\Gamma) \!+\!\xi_{-}(\ell_{2}\!+\frac{1}{2},k_{z}\!-\!Q_{z}/2)\right]}\right\rangle _{z}\label{eq:lhApp}
\end{equation}
with $\zeta_\omega=\text{sign}(\omega_n)$ and $\xi_{\pm}(\ell\!+\frac{1}{2},k_{z})\equiv\omega_{c}\left(\ell\!+\frac{1}{2}\pm\gamma_{z}\right)\!-2t_{z}\cos k_{z}\!-\mu$ being the quasiparticle energies in a finite out-of-plane magnetic field. 
While in most part of the paper we consider clean case, here
we also include a finite broadening  
$\Gamma$ related to the scattering time by nonmagnetic impurities as $\Gamma=1/2\tau$. We use the simplest lifetime approximation neglecting the
vertex impurity corrections in the pairing kernel, which is justified
at high magnetic fields for $\Gamma<\omega_{c}$ \cite{Gruenberg:PRev176.1968,Mineev:PhilMagB80.2000,Champel:PhilMagB81.2001}.
This simple model is sufficient for us to understand the qualitative
behavior of the impurities effects. We expect that a more accurate
treatment will not change the qualitative features of the result. 
The initial steps of derivation are similar to ones in Refs.\ \cite{Gruenberg:PRL16.1966,Champel:PhilMagB81.2001}.
In the quasiclassical limit $\mu\!\gg\!\omega_{c}$ the main contribution
is coming from large Landau-level indices $\ell_{\alpha}\gg1$. In
this limit, using the Stirling formula $x!\approx\sqrt{2\pi}x^{x+1/2}\exp(-x)$,
the combinatorial factor can be approximated as
\[
\frac{\left(\ell_{1}+\ell_{2}\right)!}{2^{\ell_{1}+\ell_{2}}\ell_{1}!\ell_{2}!}  \approx\frac{\exp\left[-\left(\ell_{1}-\ell_{2}\right)^{2}/4\ell_{1}\right]}{\sqrt{\pi\ell_{1}}}.
\]
Using also the Poisson summation formula $\sum_{\ell=0}^{\infty}f(\ell\!+\!\frac{1}{2})\!=\!\int_{0}^{\infty}\!dx\,f(x)\sum_{m=-\infty}^{\infty}\!(-1)^{m}\exp\left(2\pi imx\right)$,
we obtain an approximate presentation for $\lambda_{\omega_{n},Q_{z}}$
\begin{align}
\lambda_{\omega_{n},Q_{z}}\!=\! & \sum_{m_1,m_2\!=\!-\!\infty}^{\infty}\lambda_{\omega_{n},Q_{z}}^{m_1m_2},\nonumber\\
\lambda_{\omega_{n},Q_{z}}^{m_1m_2}\!\approx\! & -(\!-\!1)^{m_1+m_2}\frac{\omega_{c}}{2\pi^{3/2}}\int\limits _{0}^{\infty}\frac{dx_{1}}{\sqrt{x_{1}}}\int\limits _{0}^{\infty}dx_{2}\!\left\langle \frac{\exp\left[2\pi i\left(m_1x_{1}-m_2x_{2}\right)-\left(x_{1}-x_{2}\right)^{2}\!/4x_{1}\right]}{\left[\mathrm{i}(\omega_{n}+\zeta_\omega\Gamma) \!-\!\xi_{+}(x_{1},k_{z}\!+\!\frac{Q_{z}}{2})\right]\left[\mathrm{i}(\omega_{n}+\zeta_\omega\Gamma) \!+\!\xi_{-}(x_{2},k_{z}\!-\frac{Q_{z}}{2})\right]}\right\rangle _{z}.
\label{eq:KernelOscExp}
\end{align}
Here, the variables $x_{1,2}$ are reduced in-plane energies $\epsilon_{1,2}$ of the pairing states, $x_{1,2}=\epsilon_{1,2}/\omega_c$. The terms with nonzero $m_\alpha$ give the oscillating contributions to the kernel with respect to these energies due to the discrete spectrum of the two pairing electronic states in the magnetic field.    
Making the variable change $\epsilon=\omega_{c}(x_{1}+x_{2})/2$ and $\epsilon_{-}=\omega_{c}(x_{2}-x_{1})$,
we obtain
\[
\lambda_{\omega_{n},Q_{z}}^{m_1m_2}\!\approx\!\frac{(-1)^{m_1+m_2}}{2\pi^{3/2}\sqrt{\omega_{c}}}\!\int\limits _{-\infty}^{\infty}\!d\epsilon_{-}\!\int\limits _{|\epsilon_{-}|/2}^{\infty}\!\frac{d\epsilon}{\sqrt{\epsilon}}\!\left\langle \frac{\exp\left[2\pi i\left(m_1-m_2\right)\epsilon/\omega_{c}\!-\pi i\left(m_1+m_2\right)\epsilon_{-}/\omega_{c}\!-\epsilon_{-}^{2}/4\omega_{c}\epsilon\right]}{\left(\epsilon\!-\!\mu(k_{z},Q_{z})\right)^{2}\!+\left(\omega_{n}+\zeta_\omega\Gamma\!-\mathrm{i}\frac{\epsilon_{-}}{2}-\!\mathrm{i}\omega_{c}\tilde{\gamma}_{z}(k_{z},Q_{z})\right)^{2}}\right\rangle _{z},
\]
where $\mu(k_{z},Q_{z})\!\equiv\!\mu\!+\!2t_{z}\cos k_{z}\cos\frac{Q_{z}}{2}$ is the average in-plane Fermi energy for two pairing states with c-axis wave vectors $\pm k_z+Q_z/2$
and $\tilde{\gamma}_{z}(k_{z},Q_{z})$ is defined in Eq.\ \eqref{tg}. Further derivation steps deviate from Refs.\ \cite{Gruenberg:PRL16.1966,Champel:PhilMagB81.2001} and lead to somewhat more physically transparent presentation for the kernel eigenvalue. 
Assuming $\mu(k_{z},Q_{z})\gg\omega_{n},\frac{\epsilon_{-}}{2},\omega_{c}\gamma_{z},2t_{z}\sin k_{z}\sin\frac{Q_{z}}{2}$,
we can approximately integrate over the mean in-plane energy $\epsilon$,
\begin{align*}
\lambda_{\omega_{n},Q_{z}}^{m_1m_2}&\!\approx\!\frac{(-1)^{m_1+m_2}}{2\sqrt{\pi\omega_{c}}}\\
\times&\int\limits _{-\infty}^{\infty}\!\!d\epsilon_{-}\!
\left\langle\! \frac{\exp\!
	\left\{\!2\pi\mathrm{i}\left(m_1\!-\!m_2\right)\!\frac{\mu(k_{z},Q_{z})}{\omega_{c}}\!-\!\pi\mathrm{i}\left(m_1\!+\!m_2\right)\!\frac{\epsilon_{-}}{\omega_{c}}\!-\!\frac{\epsilon_{-}^{2}/4}{\omega_{c}\mu(k_{z},Q_{z})}\!-\!\frac{2\pi|m_1\!-\!m_2|}{\omega_{c}}\!\left[|\omega_{n}|+\Gamma\!-\!\mathrm{i}\zeta_{\omega}\left(\frac{\epsilon_{-}}{2}\!+\!\omega_{c}\tilde{\gamma}_{z}(k_{z},Q_{z})\right)\right]\!\right\}}
{\sqrt{\mu(k_{z},Q_{z})}\left[|\omega_{n}|+\Gamma\!-\!\mathrm{i}\zeta_{\omega}\left(\frac{\epsilon_{-}}{2}+\!\omega_{c}\tilde{\gamma}_{z}(k_{z},Q_{z})\right)\right]}
\!\right\rangle _{z}
\end{align*}
with $\zeta_{\omega}\equiv\mathrm{sign}(\omega_{n})$. We use
the presentation
\[
\frac{1}{|\omega_{n}|+\Gamma\!-\!\mathrm{i}\zeta_{\omega}\left(\frac{\epsilon_{-}}{2}+\!\omega_{c}\tilde{\gamma}_{z}(k_{z},Q_{z})\right)}=\!\int\limits _{0}^{\infty}\!\frac{2ds}{\omega_{c}}\exp\left[-\frac{2\left(|\omega_{n}|+\Gamma\!-\!\mathrm{i}\zeta_{\omega}\left(\frac{\epsilon_{-}}{2}+\!\omega_{c}\tilde{\gamma}_{z}(k_{z},Q_{z})\right)\right)s}{\omega_{c}}\right],
\]
which allows us to integrate over $\epsilon_{-}$,
\begin{align*}
\lambda_{\omega_{n},Q_{z}}^{m_1m_2}\!\approx&2(-1)^{m_1+m_2}\!
\int\limits _{0}^{\infty}\!\frac{ds}{\omega_{c}}\Big\langle \!\exp\Big\{
\left(2\pi\mathrm{i}\left(m_1\!-\!m_2\right)\!-\left[\pi\left(m_1\!+\!m_2\!-\!\zeta_{\omega}|m_1\!-\!m_2|\right)\!-\!s\zeta_{\omega}\right]^{2}\right)\tilde{\mu}(k_{z},Q_{z})\\
&\!-\!\left(\pi|m_1\!-\!m_2|\!+\!s\right)\frac{2(|\omega_{n}|+\Gamma)}{\omega_{c}}\!+\!\left(\pi|m_1\!-\!m_2|\!+\!s\right)2\mathrm{i}\zeta_{\omega}\tilde{\gamma}_{z}(k_{z},Q_{z})
\Big\} \Big\rangle _{z}
\end{align*}
with $\tilde{\mu}(k_{z},Q_{z})\equiv\mu(k_{z},Q_{z})/\omega_{c}$, see Eq.\ \eqref{tmu}.
In the next step, we perform summation over the Matsubara frequencies
\begin{align}
& \pi T\!\sum_{\omega_{n}\!=\!-\infty}^{\infty}\!\lambda_{\omega_{n},Q_{z}}^{m_1m_2}\approx(-1)^{m_1+m_2}\!\int\limits _{0}^{\infty}\!ds\sum_{\zeta_{\omega}=\pm1}\\
&\times \left\langle \!\frac{\pi T\exp\left\{ \left(2\pi\mathrm{i}\left(m_1\!-\!m_2\right)\!-\!\left[\pi\left(m_1\!+\!m_2\!-\!\zeta_{\omega}|m_1\!-\!m_2|\right)\!-\!s\zeta_{\omega}\right]^{2}\right)\tilde{\mu}(k_{z},Q_{z})\!+\!2\left(\pi|m_1\!-\!m_2|\!+\!s\right)[\mathrm{i}\zeta_{\omega}\tilde{\gamma}_{z}(k_{z},Q_{z})-\frac{\Gamma}{\omega_c}]\right\} }{\omega_{c}\sinh\left(\left(\pi|m_1\!-\!m_2|\!+\!s\right)\frac{2\pi T}{\omega_{c}}\right)}\right\rangle _{z}.\nonumber
\end{align}
In the sum over $m_1$ and $m_2$, it is convenient to introduce new summation
indices $k\!=\!m_1\!-\!m_2$ and $r\!=\!\zeta_{\omega}\left(m_2\!+\!\frac{k\!-\!\zeta_{\omega}|k|}{2}\right)$,
which leads to the following presentation for $\mathsf{J}\equiv\pi T\!\sum_{\omega_{n}\!=\!-\infty}^{\infty}\lambda_{\omega_{n},Q_{z}}$:
\begin{align}
\mathsf{J}\approx & \sum_{r,k\!=\!-\!\infty}^{\infty}\mathsf{J}^{rk},\\
\mathsf{J}^{rk}\approx & (-1)^{k}\!\frac{\pi T}{\omega_{c}}
\int\limits _{0}^{\infty}\!ds
\left\langle \!
\frac{\exp\left\{ \left[2\pi\mathrm{i}k\!-\left(2\pi r\!-\!s\right)^{2}\right]\tilde{\mu}(k_{z},Q_{z})\right\} 
	\cos\left[2\left(\pi|k|\!+\!s\right)\tilde{\gamma}_{z}(k_{z},Q_{z})\right]}{\sinh\left(\left(\pi|k|\!+\!s\right)\frac{2\pi T}{\omega_{c}}\right)}\exp \left[-2(\pi|k|+s)\frac{\Gamma}{\omega_c}\right]
\right\rangle _{z}.
\label{eq:Jrk}
\end{align}
We remind again that the parameters $\tilde{\mu}(k_{z},Q_{z})$ and $\tilde{\gamma}_{z}(k_{z},Q_{z})$ are defined in Eqs.\ \eqref{tmu} and \eqref{tg}.
As we consider the regime $\tilde{\mu}(k_{z},Q_{z})\gg1$, the terms
with $r<0$ are exponentially small and can be neglected. For further
transformations, we split $\mathsf{J}=\mathsf{J}_{\mathrm{qc}}+\mathsf{J}_{\mathrm{q}}^{I}+\mathsf{J}_{\mathrm{q}}^{II}$
with
\[
\mathsf{J}_{\mathrm{qc}}=\mathsf{J}^{00},\:\mathsf{J}_{\mathrm{q}}^{I}=\sum_{r\!=\!1}^{\infty}\mathsf{J}^{r0},\:\mathsf{J}_{\mathrm{q}}^{II}=2\sum_{k\!=\!1}^{\infty}\sum_{r\!=\!0}^{\infty}\mathsf{J}^{rk}.
\]
Here the first term is the conventional quasiclassical result. It contains a logarithmic divergence which can be eliminated by subtracting its value at zero magnetic field and $Q_z=0$ leading to Eq.\ \eqref{eqn:J1qc}. The other two terms give a quantum-oscillation correction to the kernel eigenvalue. In these terms we can approximately perform $s$ integration in Eq.\ \eqref{eq:Jrk} assuming $|s\!-\!2\pi r|\sim\sqrt{\omega_{c}\!/\!\mu(k_{z},Q_{z})}\ll1$, which allows us to keep $s$ dependence only in the first exponential factor.   
For the term $\mathsf{J}_{\mathrm{q}}^{I}$ this gives 
\begin{equation}
\mathsf{J}_{\mathrm{q}}^{I}  \approx\frac{2\pi^{3/2}T}{\omega_{c}}\sum_{r\!=\!1}^{\infty}\!\left\langle \!\frac{\cos\left(4\pi r\gamma_{z}\right)\cos\left(4\pi r\frac{2t_{z}}{\omega_{c}}\sin k_{z}\sin\frac{Q_{z}}{2}\right)\exp(-4\pi r\Gamma/\omega_c)}
{\sqrt{\tilde{\mu}(k_{z},Q_{z})}\sinh\left({4\pi^{2}rT}/{\omega_{c}}\right)}\right\rangle _{z},
\label{JsameOsc}
\end{equation}
where we also substituted $\cos\left[4\pi r\tilde{\gamma}_{z}(k_{z},Q_{z})\right] \rightarrow \cos\left(4\pi r\gamma_{z}\right)\cos\left(4\pi r\frac{2t_{z}}{\omega_{c}}\sin k_{z}\sin\frac{Q_{z}}{2}\right)$.
This quantity represents the contribution to the kernel originating from the oscillating components of the two pairing electronic states \emph{with the same periodicity}, i.\ e., from terms with $m_1\!=\!m_2$ in Eq.\ \eqref{eq:KernelOscExp}.  We note that $\mathsf{J}_{\mathrm{q}}^{I}$ monotonically decreases with $\tilde{\mu}(k_{z},Q_{z})$ meaning that it does not contain terms periodically varying with the large ratio $\mu/\omega_c$ that are typical for quantum-oscillating corrections to normal-state quantities.  Other contributions to the pairing kernel considered below do contain such terms.  On the other hand, this kernel contribution has a pronounced oscillating dependence on the modulation wave vector $Q_z$. The latter property is very crucial for the consideration of the FFLO instability.

For the term $\mathsf{J}_{Q}^{II}$ we note that the $s$ integral 
for the $r=0$ term is from $0$ and it is approximately two times smaller than for 
the $r\neq0$ terms for which we can extend the lower integration limit to $-\infty$, i.e.,
\begin{align*}
\mathsf{J}_{\mathrm{q}}^{II} & \approx\frac{4\pi^{3/2}T}{\omega_{c}}\sum_{k\!=\!1}^{\infty}\sum_{r\!=\!0}^{\infty}(-1)^{k}\!\left\langle \!\frac{\cos\left[2\pi k\tilde{\mu}(k_{z},Q_{z})\right]\cos\left[2\pi\left(k\!+\!2r\right)\tilde{\gamma}_{z}(k_{z},Q_{z})\right]\exp[-2\pi (k+2r)\Gamma/\omega_c]}{\sqrt{\tilde{\mu}(k_{z},Q_{z})}\sinh\left[\left(k\!+\!2r\right)\frac{2\pi^{2}T}{\omega_{c}}\right]}\right\rangle _{z}\\
& -\frac{2\pi^{3/2}T}{\omega_{c}}\sum_{k\!=\!1}^{\infty}(-1)^{k}\!\left\langle \!\frac{\cos\left[2\pi k\tilde{\mu}(k_{z},Q_{z})\right]\cos\left[2\pi k\tilde{\gamma}_{z}(k_{z},Q_{z})\right]\exp(-2\pi k\Gamma/\omega_c)}
{\sqrt{\tilde{\mu}(k_{z},Q_{z})}\sinh\left(k\frac{2\pi^{2}T}{\omega_{c}}\right)}\right\rangle _{z}
\end{align*}
We see that, in contrast to $\mathsf{J}_{\mathrm{q}}^{I}$, Eq.\ \eqref{JsameOsc}, this contribution is composed of terms proportional to $\cos\left[2\pi k\tilde{\mu}(k_{z},Q_{z})\right]$ that oscillate with the ratio $\mu/\omega_c$. We can reduce the double summation in the first line to a single
sum using
\begin{align*}
\sum_{k\!=\!1}^{\infty}\sum_{r\!=\!0}^{\infty}(\!-\!1)^{k}\cos(kx)g(k\!+\!2r) & =\!\sum_{j\!=\!1}^{\infty}\sum_{n\!=\!1}^{j}\left[\cos(2nx)g(2j)-\cos\left[(2n-1)x\right]g(2j\!-\!1)\right]\\
=\! & \sum_{j\!=\!1}^{\infty}\left[\frac{\sin\left[\left(2j\!+\!1\right)x\right]\!-\!\sin x}{2\sin x}g(2j)\!-\!\frac{\sin\left(2jx\right)}{2\sin x}g(2j\!-\!1)\right]\\
= & \frac{1}{2}\!\sum_{j\!=\!1}^{\infty}(\!-\!1)^{j}\left[\sin\left(jx\right)\cot x\!+\!\cos\left(jx\right)\right]g(j)-\frac{1}{2}\sum_{j\!=\!1}^{\infty}g(2j),
\end{align*}
with $x\!=\!2\pi \tilde{\mu}(k_{z},Q_{z})$ and $g(j)\!=\!\cos\left[2\pi j\tilde{\gamma}_{z}(k_{z},Q_{z})\right]\exp[-2\pi j\Gamma/\omega_c]/\!
\sinh\left(2\pi^{2}jT/\omega_{c}\right)$, which leads to the presentation
\begin{align}
\mathsf{J}_{\mathrm{q}}^{II} & \approx\frac{2\pi^{3/2}T}{\omega_{c}}\sum_{j\!=\!1}^{\infty}
(-1)^{j}\!\left\langle 
\!\frac{\sin\left[2\pi j\tilde{\mu}(k_{z},Q_{z})\right]
	\cos\left[2\pi j\tilde{\gamma}_{z}(k_{z},Q_{z})\right]\exp(-2\pi j\Gamma/\omega_c)}
{\sqrt{\tilde{\mu}(k_{z},Q_{z})}\tan\left[2\pi\tilde{\mu}(k_{z},Q_{z})\right]
	\sinh\left({2\pi^{2}jT}/{\omega_{c}}\right)}
\right\rangle _{z}\nonumber\\
- & \frac{2\pi^{3/2}T}{\omega_{c}}\sum_{j\!=\!1}^{\infty}\!\left\langle \!\frac{\cos\left[4\pi j\tilde{\gamma}_{z}(k_{z},Q_{z})\right]\exp(-4\pi j\Gamma/\omega_c)}
{\sqrt{\tilde{\mu}(k_{z},Q_{z})}\sinh\left({4\pi^{2}jT}/{\omega_{c}}\right)}\right\rangle _{z}.
\label{OscContrKernel}
\end{align}
We can see that the second term exactly cancels $\mathsf{J}_{\mathrm{q}}^{I}$, Eq.\ \eqref{JsameOsc}, meaning that the first line gives the final result for the full quantum correction to the pairing kernel 
$\mathcal{J}_{\mathrm{q}}=\mathsf{J}_{\mathrm{q}}^{I}+\mathsf{J}_{\mathrm{q}}^{II}$,
\begin{equation}
\mathcal{J}_{\mathrm{q}}(H,T,Q_z)\approx\frac{2\pi^{3/2}T}{\omega_{c}}
\sum_{j\!=\!1}^{\infty}(-1)^{j}
\frac{\cos\left[2\pi j\tilde{\gamma}_{z}(k_{z},Q_{z})\right]\exp(-2\pi j\Gamma/\omega_c)}{\sinh\left({2\pi^{2}jT}/{\omega_{c}}\right)}
\left\langle 
\!\frac{\sin\left[2\pi j\tilde{\mu}(k_{z},Q_{z})\right]}
{\sqrt{\tilde{\mu}(k_{z},Q_{z})}\tan\left[2\pi\tilde{\mu}(k_{z},Q_{z})\right]}
\right\rangle _{z}.
\label{eq:JqDis}
\end{equation}
\end{widetext}
In the clean case, $\Gamma\!=0$, this gives Eq.\ \eqref{eqn:J1q} of the main text. 
We note that the oscillating factor $\sin\left[2\pi j\tilde{\mu}(k_{z},Q_{z})\right]/\tan\left[2\pi\tilde{\mu}(k_{z},Q_{z})\right]$ in this result
containing the large parameter $\tilde{\mu}(k_{z},Q_{z})$ has zero average over the period for odd $j$,
while for even $j$ its average equals to one.  The oscillating part of this factor for all $j$ originates from $\mathsf{J}_{\mathrm{q}}^{II}$ and the average part for even $j$ originates from $\mathsf{J}_{\mathrm{q}}^{I}$.
\begin{figure}[htbp] 
	\centering
	\includegraphics[width=3.2in]{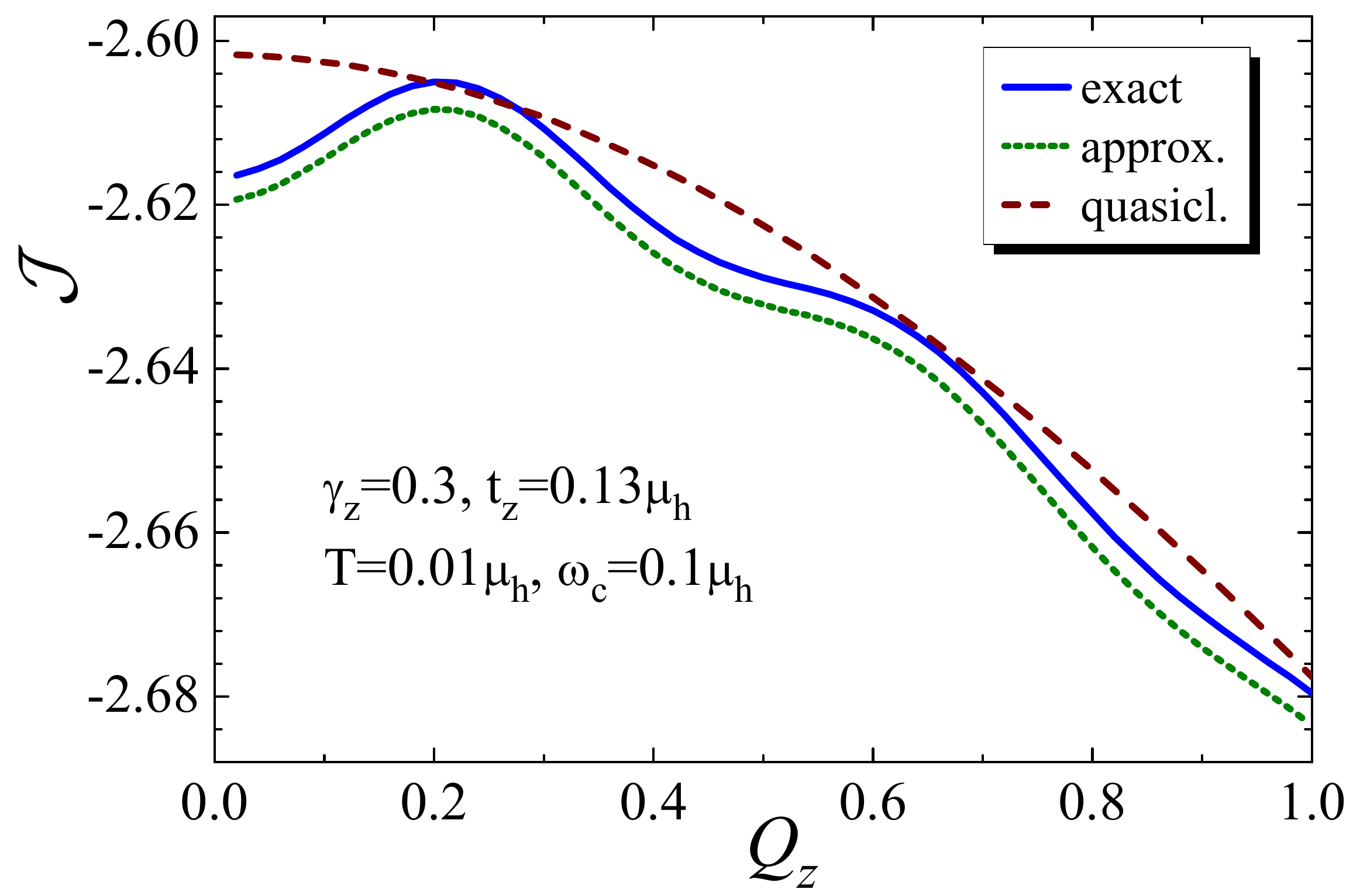}
	\caption{Comparison of the representative $Q_z$ dependence of the exact total kernel eigenvalue $\mathcal{J}$, Eq.\ \eqref{J1llSumPres}, with quasiclassical formula, Eq.\ \eqref{eqn:J1qc}, and with the approximation which accounts for the oscillating correction, Eq.\ \eqref{eqn:J1q}, to the quasiclassical result. The used parameters are shown in the plot.
	}
	\label{fig:Jplot}
\end{figure} 
Figure \ref{fig:Jplot} compares the $Q_z$ dependence of the exact total kernel eigenvalue $\mathcal{J}$ given by Eq.\ \eqref{J1llSumPres}, with the quasiclassical result from Eq.\ \eqref{eqn:J1qc} and with the more accurate approximation which also takes into account the oscillating correction, Eq.\ \eqref{eqn:J1q}. We can see that this correction accurately reproduces the oscillating behavior. A small difference with the exact result is only present in the smooth part and is obviously related to the inaccuracy of the quasiclassical contribution of the order of $\omega_c/\mu$.

Further simplification can be achieved in the common limit $\omega_{c}\ll t_{z}$.
In this limit, as usual in the physics of quantum-oscillating corrections,
the dominating contributions are coming from the extremal cross sections
of the Fermi surface. We will demonstrate this for the first two terms
in the sum giving the quantum correction in Eq.\ \eqref{eqn:J1q}. 
These terms provide the main contribution to the total sum almost everywhere except very low temperatures, $T<\omega_{c}/4\pi^2$.
The first term, $k=1$, we rewrite as
\begin{align*}
&\mathcal{J}_{\mathrm{q}}^{(1)}=-\pi^{3/2}T\frac{\cos\!\left(2\pi\gamma_{z}\right)\exp(-2\pi \Gamma/\omega_c)}{\sinh\left({2\pi^{2}T}/{\omega_{c}}\right)}\!\\
\times&\int\limits_{-\pi}^{\pi}\frac{dk_{z}}{2\pi}\frac{\sum_{\delta_{\textsc{q}}=\pm1}
\cos\!\left\{ 2\pi\left[\frac{\mu}{\omega_{c}}\!+\!\frac{2t_{z}}{\omega_{c}}\cos\left(k_{z}\!-\!\delta_{\textsc{q}}\frac{Q_{z}}{2}\right)\right]\right\}}{\sqrt{\omega_{c}\left(\mu\!+\!2t_{z}\cos k_{z}\cos\frac{Q_{z}}{2}\right)}}.
\end{align*}
In the limit $\omega_{c}\ll t_{z}$, the dominating contributions
to the $k_{z}$ integration for the two rapidly-oscillating terms
in the nominator are coming from the regions near $k_{z}-\delta_{\textsc{q}}Q_{z}/2=0$
and $\pi$, where we can expand $\cos\left(k_{z}\!-\delta_{\textsc{q}}Q_{z}\!/2\right)\!\approx\!\pm\left[1\!-\!\left(k_{z}\!-\delta_{\textsc{q}}Q_{z}\!/2\right)^{2}/2\right]$.
Also, we can neglect the $k_{z}$ dependence in the denominator substituting
$\cos k_{z}\rightarrow\pm\cos\frac{Q_{z}}{2}$. These approximations give
\begin{align*}
\mathcal{J}_{\mathrm{q}}^{(1)}\approx&-\sqrt{\pi}T\frac{\cos\!\left(2\pi\gamma_{z}\right)\exp(-2\pi \Gamma/\omega_c)}{\sinh\left({2\pi^{2}T}/{\omega_{c}}\right)}\!\\
\times&\sum_{\delta_{t}=\pm1}\int\limits_{-\pi}^{\pi}dk_{z}\frac{\cos\!\left[2\pi\frac{\mu}{\omega_{c}}\!+\!2\pi\delta_{t}\frac{2t_{z}}{\omega_{c}}\left(1-k_{z}^{2}/2\right)\right]}{\sqrt{\omega_{c}\left(\mu\!+\!2\delta_{t}t_{z}\cos^{2}\frac{Q_{z}}{2}\right)}}.
\end{align*}
Using $\int_{-\infty}^{\infty}\cos(a\pm bx^{2})dx=\sqrt{\pi/b}\cos\left(a\pm\pi/4\right),$
we can approximately perform the $k_{z}$ integration giving Eq.\ \eqref{eq:JQ1}. 

For the similar evaluation of the second term in Eq.\ \eqref{eqn:J1q}, we represent it in
the form
\begin{widetext}
\[
\mathcal{J}_{\mathrm{q}}^{(2)}\!=\!
\pi^{3/2}T\frac{\cos\!\left(4\pi\gamma_{z}\right)\exp(-4\pi \Gamma/\omega_c)}
{\sinh\left(4\pi^{2}T/\omega_{c}\right)}
\int\limits_{-\pi}^{\pi}\frac{dk_{z}}{2\pi}\!
\frac{2\cos\!\left(4\pi\frac{2t_{z}}{\omega_{c}}\sin k_{z}\sin\frac{Q_{z}}{2}\right)+\sum_{\delta_{\textsc{q}}=\pm1}
	\cos\!\left\{4\pi\left[\frac{\mu}{\omega_{c}}
	+\frac{2t_{z}}{\omega_{c}}\cos\left(k_{z}\!-\!\delta_{\textsc{q}}\frac{Q_{z}}{2}\right)\right]\right\}}
{\sqrt{\omega_{c}\left(\mu\!+\!2t_{z}\cos k_{z}\cos\frac{Q_{z}}{2}\right)}}.
\]
\end{widetext}
The qualitative difference from $\mathcal{J}_{\mathrm{q}}^{(1)}$ is the presence of
the first term in the nominator, which gives the rapidly oscillating
with $Q_{z}$ contribution. In the limit $t_{z}\ll\mu$, we can neglect
the $k_{z}$-dependent term in the denominator allowing us to compute
this contribution using $\int_{-\pi}^{\pi}\frac{dx}{2\pi}\cos\!\left(a\sin x\right)=J_{0}(x)$ with $J_{0}(x)$ being the Bessel function.
The last two terms can be evaluated similarly to $\mathcal{J}_{\mathrm{q}}^{(1)}$ leading to the result in Eq.\ \eqref{eq:JQ2}. The first term in this result having strong oscillating dependence on $Q_z$ originates from the kernel part given by Eq.\ \eqref{JsameOsc} describing the oscillating contributions of the two pairing states with the same periodicity. 

\section{Critical Maki's parameter of layered superconductors within quasiclassical approximation}\label{app:QCFFLO}

\begin{figure}[htbp] 
	\centering
	\includegraphics[width=3.3in]{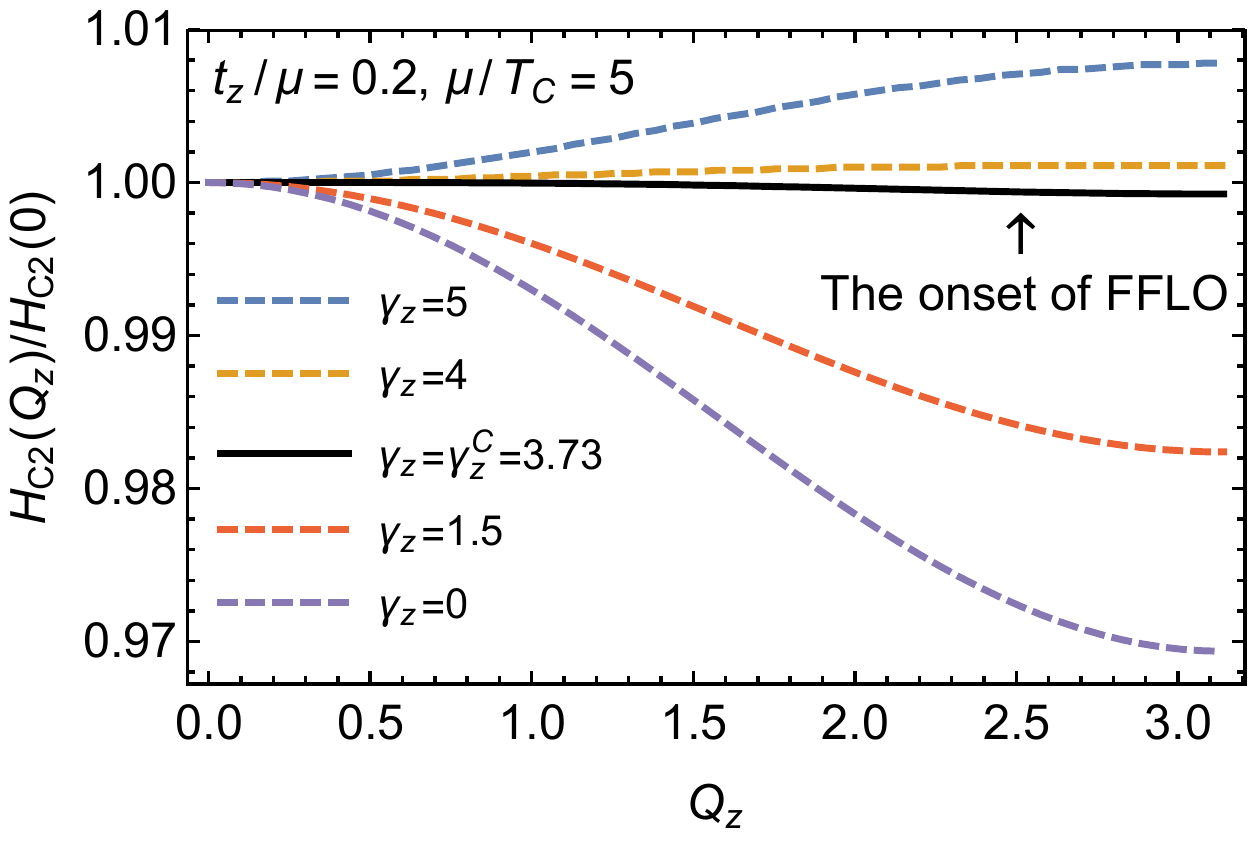}
	\caption{The dependence $H_{C2}$ on $Q_z$ for single-band layered superconductor with different $\gamma_z$. The parameters that are used in the plot: $t_z/\mu=0.3$. For $\gamma_z<\gamma^C_z$, the maximum $H_{C2}$ is always at $Q_z=0$. If $\gamma_z>\gamma^C_z$, the maximum $H_{C2}$ moves  to finite $Q_z$.}
	\label{fig:qcvsLL}
\end{figure}
In this appendix, we investigate
the onset of the interlayer FFLO state in the magnetic field applied perpendicular to the
layers for the quasiclassical case. This problem provides a natural reference for the
case in which the quantum effects are taken into account. Note that
in most theoretical papers, only the case of the magnetic field applied \emph{along} the layers
has been considered because orbital effects are weak in this geometry which is favorable
for the FFLO modulation. %
For isotropic materials with large Zeeman effect, a similar problem of the FFLO state
along the magnetic field was considered in the seminal paper of Gruenberg and Gunther
\cite{Gruenberg:PRL16.1966}. It was demonstrated that the FFLO modulation appears when the
Maki's parameter exceeds $1.8$. Surprisingly, this consideration was never generalized
to the case of layered superconductors with open Fermi surfaces.

Our consideration is based on the result for the field-dependent pairing kernel, Eq.\ \eqref{eqn:J1qc}. To investigate superconducting instability at zero temperature, we approximate
$\tanh x\approx x$. The single-band $H_{C2}$ equation $\ln(T/T_C)=\mathcal{J}(H,T,Q_z)$
can be transformed into the following form using the substitution $z=\sqrt{\bar{\mu}}\bar{s}$
\begin{align}\label{eqn:Rsq}
&R(\omega_c,Q_z)\equiv \ln\frac{\pi T_C}{\sqrt{\mu\omega_c}}+2\int^\pi_{-\pi}\frac{\mathrm{d}k_z}{2\pi}\int^\infty_0\mathrm{d} z \mathrm{e}^{-\beta_{Q} z ^2}\ln z \notag\\
&\times\Big[\beta_{Q} z \cos\Big(2\gamma_{Q} z \Big)+\gamma_{Q}\sin\Big(2\gamma_{Q} z \Big)\Big]=0
\end{align}
with $Q_z$-dependent parameters
\begin{align*}
\beta_{Q}(k_z,Q_z)&=\frac{\tilde{\mu}}{\bar{\mu}}=1+2\frac{t_z}{\mu}\cos k_{z}\cos\frac{Q_{z}}{2},\\
\gamma_{Q}(k_z,Q_z)&=\frac{\tilde{\gamma}_{z}}{\sqrt{\bar{\mu}}}=\frac{\gamma_{z}-2(t_{z}/\omega_c)\sin k_{z}\sin\frac{Q_{z}}{2}}{\sqrt{\mu/\omega_c}}.
\end{align*}
This equation determines the reduced upper critical field $\omega_c/T_C=eH_{C2}/(cmT_C)$ as a function of the modulation wave vector $Q_z$ and the reduced electronic parameters $\mu/T_C$, $t_z/\mu$, and $\gamma_z$. The value of $Q_z$ giving the largest $H_{C2}$ is realized. For the uniform case, the value $\omega_c^\ast\equiv \omega_c(Q_z\!=\!0)$ is determined by a simpler equation
\begin{align}
&\ln\!\left(\!\frac{\pi T_{c}}{\sqrt{\mu\omega_{c}^\ast}}\!\right)\!+\!f_R\!=\!0,\label{eqn:snglHc2Unif}\\
&f_R(a_\gamma,\tfrac{t_z}{\mu })
\!=\!2\!\int\limits^\infty_0\mathrm{d} z \exp(\!-\!z^2)\ln z \Big\{I_0(2\tfrac{t_z}{\mu }z^2)a_\gamma\sin(2a_\gamma z )\notag\\
&+[I_0(2\tfrac{t_z}{\mu }z^2)-2\tfrac{t_z}{\mu }I_1(2\tfrac{t_z}{\mu }z^2)] z \cos(2a_\gamma z )\Big\},\label{eqn:fR}
\end{align}
where  $a_\gamma\equiv \gamma_z\sqrt{\omega^\ast_c/\mu }$ and $I_n(x)$ is the modified Bessel function.

Figure  \ref{fig:qcvsLL} shows representative $Q_z$  dependences of $H_{C2}$ for $t_z/\mu=0.2$, $\mu/T_C=5$, and different $\gamma_z$.
We can see that for small $\gamma_z$, the maximum $H_{C2}$ is located at $Q_z=0$. When $\gamma_z$ exceeds the critical value, $\gamma^C_z$, the maximum of $H_{C2}$
moves to finite $Q_z$. For large $\gamma_z$ the maximum is realized at $Q_z=\pi$.

The critical spin splitting $\gamma_z^C $ can be expressed via the Maki's parameter, $\alpha_{M}=\sqrt{2}H^O_{C2}/H^P_{C2}$, where
\begin{equation}
H^P_{C2}=\frac{\pi T_C}{\sqrt{2}\mathtt{C}_{\mathrm{E}}\mu_z}=\frac{\pi cT_Cm}{\sqrt{2}\mathtt{C}_{\mathrm{E}}e\gamma_z}
\end{equation}
is the Pauli-limiting field at $T=0$
and $H^O_{C2}$ is the orbital upper critical field at $T=0$.
In our case we can obtain it from Eq. \eqref{eqn:snglHc2Unif} with $\gamma_z=0$,
\begin{equation}
H^O_{C2}=\frac{\pi^2 c T_C^2 m }{\mathtt{C}_{\mathrm{E}} e \mu}\frac{2}{1+\sqrt{1-4t_z^2/\mu^2 }}.
\end{equation}
Combining these results for the critical fields, we obtain the presentation of the Maki's parameter via the electronic parameters given by Eq.\ \eqref{eqn:alpha} of the main text. 
We also derive the presentation
\begin{equation}\label{eqn:alpha1}
\alpha_{M}=a_\gamma\mathrm{e}^{-f_R
}\frac{4}{1+\sqrt{1-4t_z^2/\mu^2}},
\end{equation}
which is convenient for the numerical evaluation of the critical Maki's parameter.

\begin{figure}[htbp] 
	\centering
	\includegraphics[width=3.in]{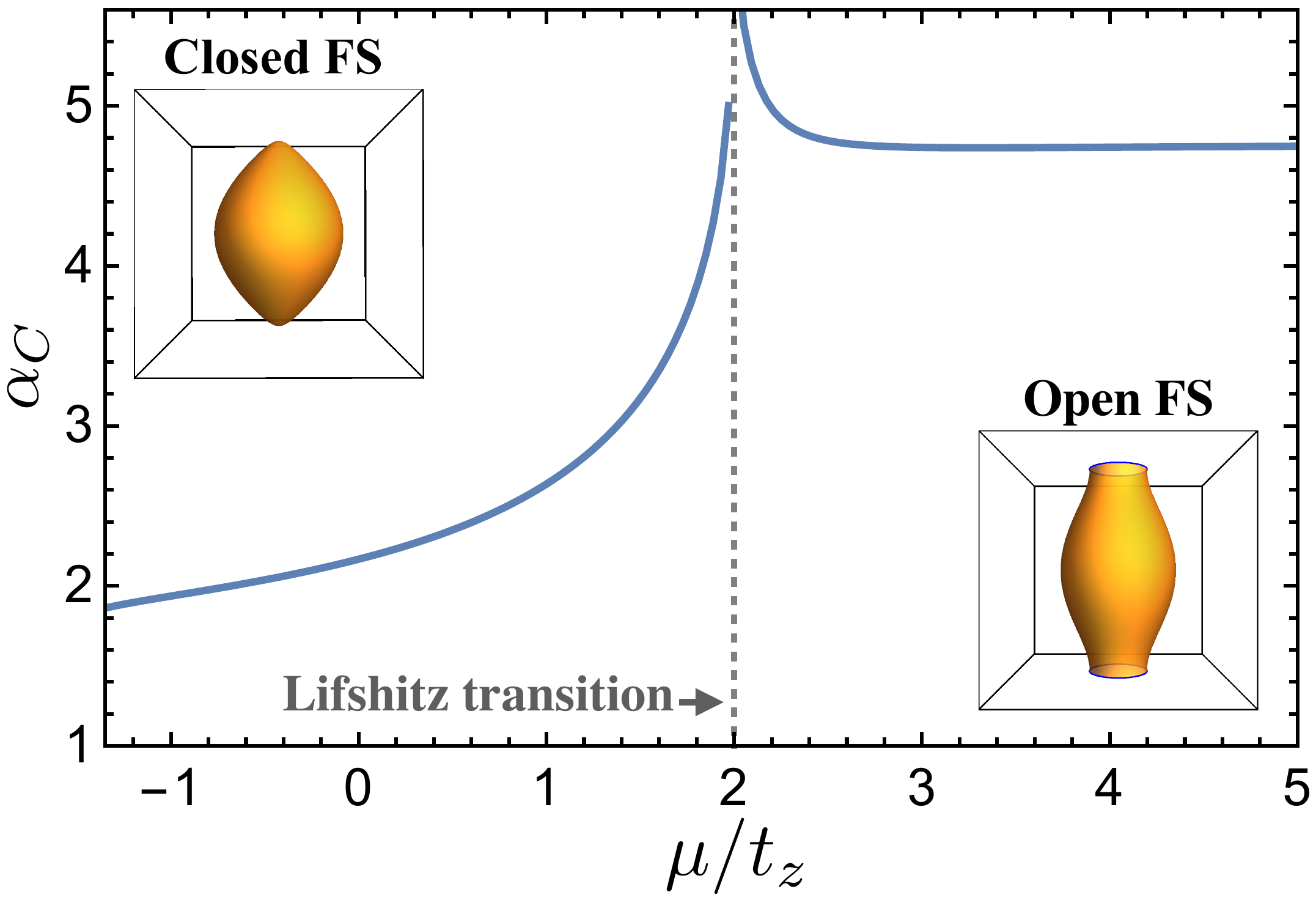}
	\caption{The dependence of the critical Maki's parameter $\alpha_C$ on $\mu/t_z$. Note that
		at negative $\mu$ this parameter approaches
		the known value for the isotropic case $\alpha_C\approx 1.8$ \cite{Gruenberg:PRL16.1966}.
	}
	\label{fig:Maki}
\end{figure}
We proceed with calculation of the critical spin splitting $\gamma^C_z$ as function of the
parameters $\mu/T_C$ and $t_z/\mu$ and will relate it with the critical Maki's parameter using Eq.\ \eqref{eqn:alpha}. At
$\gamma_z=\gamma^C_z$  the second derivative
$\mathrm{d}^2\omega_c/\mathrm{d}Q_z^2|_{Q_z=0}$ vanishes, which coincides with the
condition
\begin{equation}
R''\equiv\frac{\partial^2R(\omega_c^\ast,0)}{\partial Q_z^2}=0.
\end{equation}
From Eq.\ \eqref{eqn:Rsq} in the limit $\mu\gg \omega_c$, we derive the presentation
\begin{equation}\label{eqn:R2result}
R''\!=\!-\frac{2t_{z}}{\omega_c}\!\int\limits_{0}^{\infty}\frac{\mathrm{d}z}{z}\exp\left(\!-z^{2}\right)
\cos\left(2a_{\gamma}s\right)I_{1}\left(2\tfrac{t_{z}}{\mu}z^{2}\right).
\end{equation}
In the limit $t_z\!\ll \!\mu$, the second derivative $R''$ changes sign at $a_{\gamma}\!=\!a_{\gamma}^C\!\approx \!0.9241$. This means that $\gamma_z^C$ scales as $\sqrt{\mu/\omega_c^\ast}$. From Eq.\ \eqref{eqn:fR} we evaluate $f_R^C=f_R(a_\gamma^C,0)\approx\! -1.8922$ and $\omega_{c}^\ast=\pi^2\exp(2f_R^C)T_C^2/\mu$ giving $\gamma_z^C\approx 0.758\mu/T_C$. Substituting the evaluated parameters into Eq.\ \eqref{eqn:alpha1}, we find the critical Maki's parameter $\alpha_C\!\approx \!4.761$ in the limit $t_z\!\ll \!\mu$.

At finite $t_z$, we find the dependence $a_\gamma^C(t_z/\mu)$ by numerically solving  equation $R''=0$ using the presentation in Eq.\ \eqref{eqn:R2result}, then compute the function $f_R[a_\gamma^C(t_z/\mu),t_z/\mu]$, Eq.\ \eqref{eqn:fR}, and, finally, evaluate the critical Maki's parameter from Eq.\ \eqref{eqn:alpha1}. In Fig. \ref{fig:Maki}, we plot the resulting dependence of the critical Maki's parameter $\alpha_C$ on $\mu/t_z$. We can see that $\alpha_C$ has a sharp increase at the neck-interruption transition $t_z=\mu/2$ and monotonically decreases with $\mu/t_z$ approaching the value $1.8$ for the isotropic case.

\section{Suppression of the FFLO state by impurity scattering}\label{app:disorder}

In this appendix, we investigate the influence of the impurity scattering on the FFLO transition. The quantum correction to the pairing kernel $\mathcal{J}_{\mathrm{q}}(H,T,Q_z)$ taking into account impurity broadening $\Gamma$ has been derived in Appendix \ref{app:OscCorr} and it is given by  Eq.\ \eqref{eq:JqDis}. Impurities lead to the appearance of the ``Dingle factors''  $\exp(-2\pi j\Gamma/\omega_c)$, well known in the theory of quantum oscillations \cite{ShoenbergMagnOscBook}.  We consider here the case of relatively weak impurity scattering $\Gamma < \omega_c$ and neglect the scattering correction to the quasiclassical kernel, Eq.\ \eqref{eqn:J1qc}. This correction provides only a small and smooth contribution, which do not influence the location of the FFLO transition. In addition, it is not captured correctly by the used lifetime approximation.

The impurity factors suppress the higher-order terms in the sum for $\mathcal{J}_{\mathrm{q}}(H,T,Q_z)$ in the same way as the temperature. At noticeable scattering the main contribution is given by the two low-order terms, Eqs. \eqref{eq:JQ1} and \eqref{eq:JQ2}, where the first and second terms acquire the factors $\exp(-2\pi \Gamma/\omega_c)$ and $\exp(-4\pi \Gamma/\omega_c)$, respectively. One can straightforwardly generalize the criterion for the FFLO transition following from Eq.\ \eqref{eq:SecDerTot2} to the case of finite scattering rate. Adding the factor $\exp(-4\pi \Gamma/\omega_c)$ to Eq.\ \eqref{eq:Sec-derQuant}, we obtain the following equation for the FFLO temperature in the case $\cos\!\left(4\pi\gamma_{z}\right)<0$ and $t_z\ll \mu$
\begin{equation}
4\pi^{3/2}|\!\cos\!\left(4\pi\gamma_{z}\right)\!|\exp\!\left(\!-\frac{4\pi\Gamma}{\omega_{c}}\right)\!\frac{\frac{4\pi^{2}T_{\mathrm{FFLO}}}{\omega_{c}}}{\sinh\left(\!\frac{4\pi^{2}T_{\mathrm{FFLO}}}{\omega_{c}}\!\right)}\sqrt{\frac{\mu}{\omega_{c}}}\!=\!1.\label{eq:TFFLO-Dis}
\end{equation}
In particular, this equation gives the critical scattering broadening $\Gamma_{\mathrm{cr}}$ completely eliminating the FFLO state
\begin{align}
\Gamma_{\mathrm{cr}} & =\frac{\omega_{c}}{4\pi}\ln\left(4\pi^{3/2}\left|\cos\!\left(4\pi\gamma_{z}\right)\right|\!\sqrt{\frac{\mu}{\omega_{c}}}\right)\nonumber\\
& \approx\frac{\pi T_{C}^{2}}{4\mathtt{C}_{\mathrm{E}}\mu}\ln\left(4\sqrt{\pi\mathtt{C}_{\mathrm{E}}}\left|\cos\!\left(4\pi\gamma_{z}\right)\right|\!\frac{\mu}{T_{C}}\right)
\label{eq:CritScatRate}
\end{align}
One can see that a small numerical factor $1/4\pi$ in the ratio $\Gamma_{\mathrm{cr}}/\omega_{c}$ is   partially compensated by the large logarithm. In the regime $\cos\left(4\pi\gamma_{z}\right)\!>\!0$ the FFLO state is suppressed by much smaller scattering broadening.

\begin{figure}[htbp]
	\centering
	\includegraphics[width=3.3in]{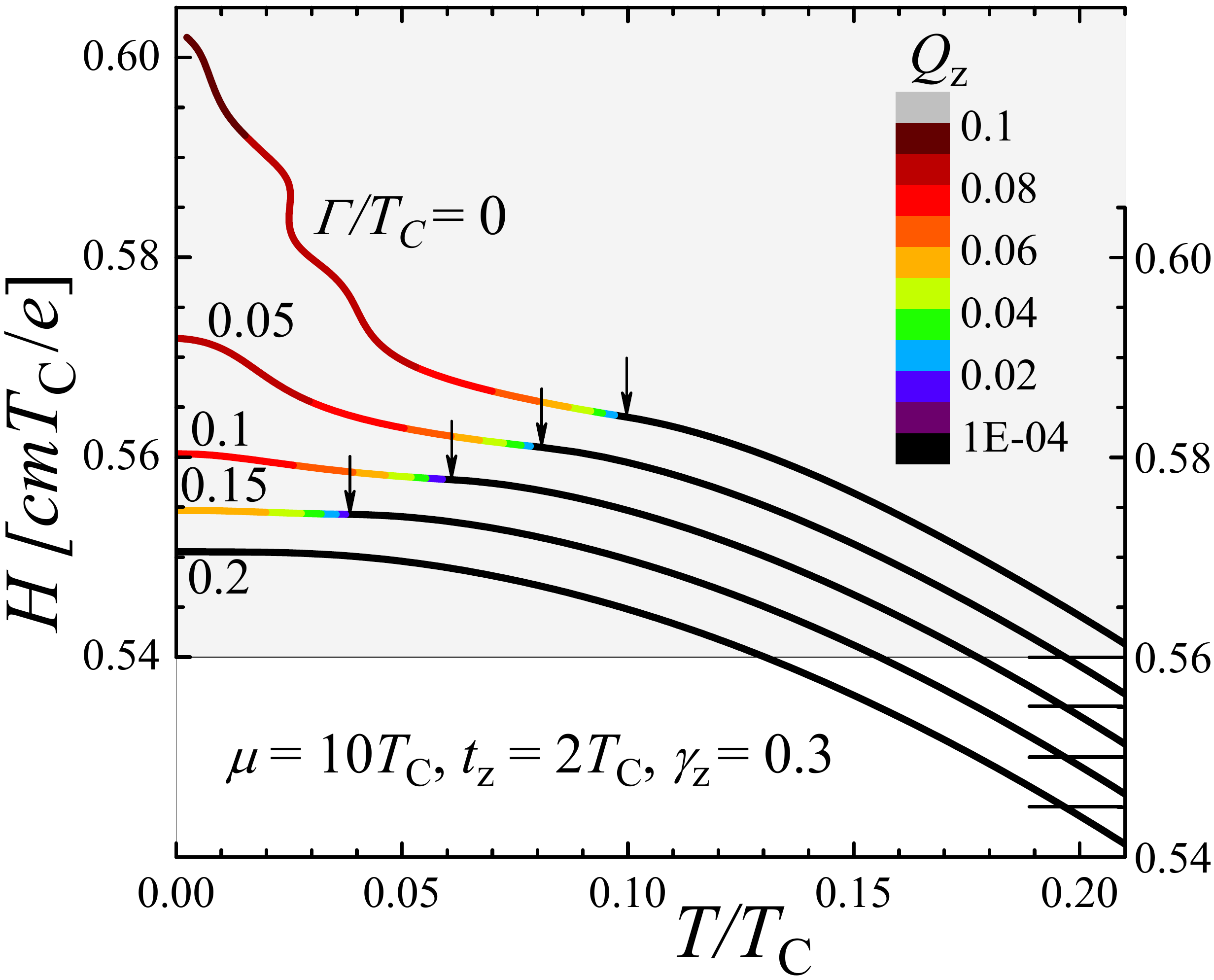}
	\caption{Evolution of the upper critical field line with increasing the impurity broadening $\Gamma$. 
		The plots are vertically displaced for clarity (the left-axis (right-axis) labels correspond to the upper (lower) curve). The used electronic parameters are shown in the plot. Arrows mark the locations of the FFLO transitions.
		For these parameters, at $\Gamma=0.2T_C$ the FFLO state is destroyed. }
	\label{fig:disorder}
\end{figure}
Figure \ref{fig:disorder} shows the evolution of the instability boundary with increasing scattering broadening $\Gamma$ for one of the parameter sets used in Fig.\ \ref{fig:HC2Tmu10mu20}, 
$\mu\!=\! 10T_{C}$, $t_z\!=\! 2T_{C}$, and $\gamma_z\!=\!0.3$. 
One can see that the impurities scattering reduces the FFLO transition temperature and  at $\Gamma/T_C\!=\!0.2$ the FFLO state is completely suppressed. This is consistent with Eq.\ \eqref{eq:CritScatRate} giving $\Gamma\!\approx\!0.19T_C\! \approx\! 0.35\omega_c$ for $\mu\!=\! 10T_{C}$ and $\gamma_z\!=\!0.3$. 
It is well known that impurity scattering increases the upper critical field within the quasiclassical approximation. We see that suppression  of the quantum term has the opposite effect: the pronounced low-temperature upturn of the instability curve existing in the clean case rapidly diminishes with increasing scattering.
On the other hand, we can observe that the absence of this upturn does not exclude the FFLO instability. For example, for  $\Gamma/T_C\!=\!0.15$ the FFLO state still exists, even though the shape of the upper-critical-field curve does not suggest any anomalies. 

\begin{figure}[htbp]
	\centering
	\includegraphics[width=3.3in]{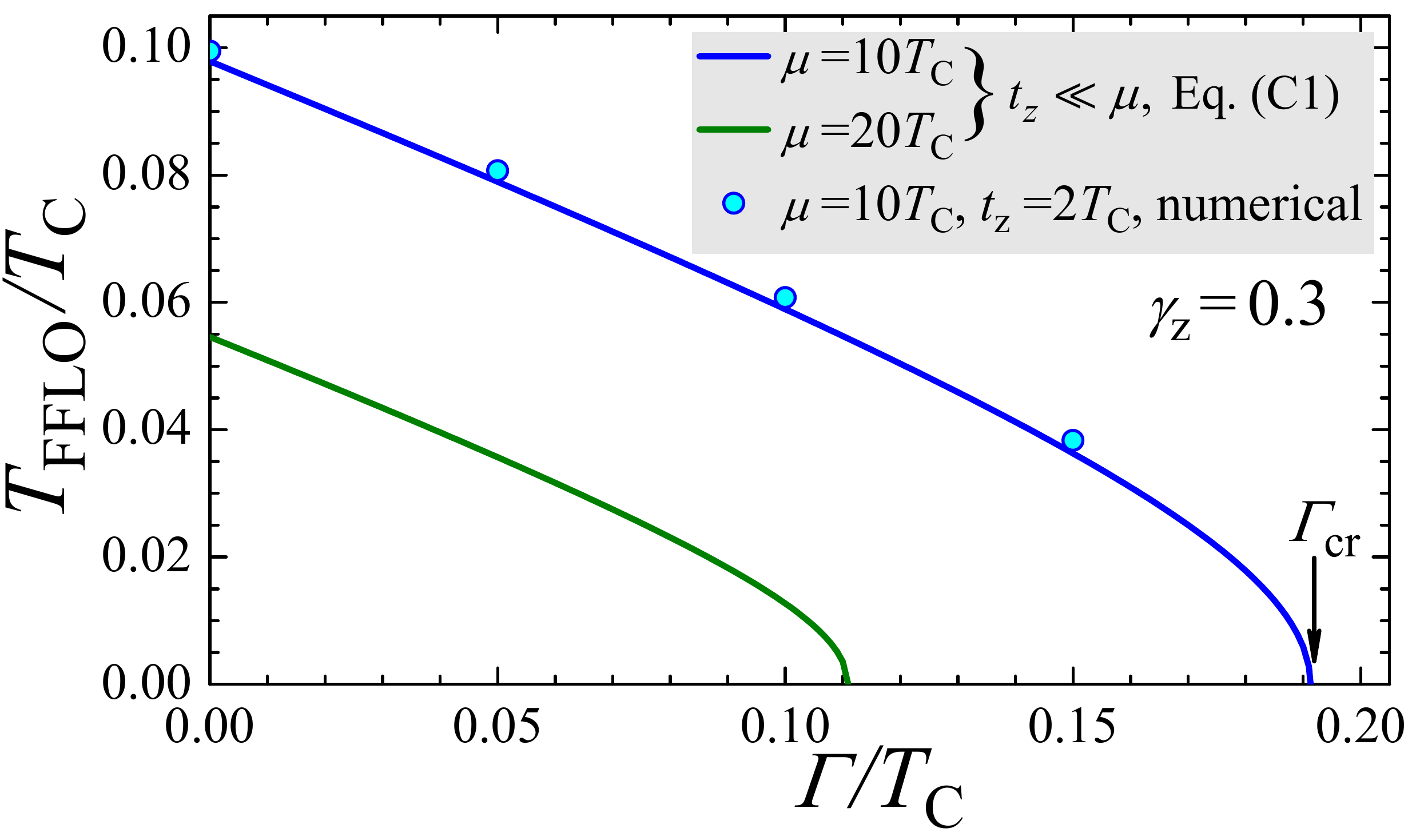}
	\caption{The representative dependences of the FFLO-instability temperature on the scattering broadening $\Gamma$. The solid lines are obtained from Eq.\ \eqref{eq:TFFLO-Dis} valid in the limit $t_z\ll \mu$. The solid symbols are the transition points for plots in Fig.\ \ref{fig:disorder}.}
	\label{fig:TFFLO-Gamma}
\end{figure} 
Figure \ref{fig:TFFLO-Gamma} shows the dependences of the FFLO transition temperature on the impurity broadening. The lines show plots computed using Eq.\ \eqref{eq:TFFLO-Dis} for $\gamma_z=0.3$ and two values of $\mu/T_C$, 10 and 20. We remind that this equation is valid in the limit $t_z\ll \mu$. The solid symbols show the transition points obtained from the computed instability boundaries shown in Fig.\ \ref{fig:disorder}. One can see that the two calculations give consistent results. A small deviation is caused by correction from the finite value of $t_z\!=\!0.2\mu$. As expected, the critical impurity broadening decreases with increasing the ratio $\mu/T_{C}$.

We can conclude that the fragile quantum FFLO state is destroyed by a quite small impurity broadening.
This is the most obvious reason why such a state is difficult to realize in existing superconducting materials.

\bibliography{2Lifshitz}

\begin{thebibliography}{74}%
\makeatletter
\providecommand \@ifxundefined [1]{%
 \@ifx{#1\undefined}
}%
\providecommand \@ifnum [1]{%
 \ifnum #1\expandafter \@firstoftwo
 \else \expandafter \@secondoftwo
 \fi
}%
\providecommand \@ifx [1]{%
 \ifx #1\expandafter \@firstoftwo
 \else \expandafter \@secondoftwo
 \fi
}%
\providecommand \natexlab [1]{#1}%
\providecommand \enquote  [1]{``#1''}%
\providecommand \bibnamefont  [1]{#1}%
\providecommand \bibfnamefont [1]{#1}%
\providecommand \citenamefont [1]{#1}%
\providecommand \href@noop [0]{\@secondoftwo}%
\providecommand \href [0]{\begingroup \@sanitize@url \@href}%
\providecommand \@href[1]{\@@startlink{#1}\@@href}%
\providecommand \@@href[1]{\endgroup#1\@@endlink}%
\providecommand \@sanitize@url [0]{\catcode `\\12\catcode `\$12\catcode
  `\&12\catcode `\#12\catcode `\^12\catcode `\_12\catcode `\%12\relax}%
\providecommand \@@startlink[1]{}%
\providecommand \@@endlink[0]{}%
\providecommand \url  [0]{\begingroup\@sanitize@url \@url }%
\providecommand \@url [1]{\endgroup\@href {#1}{\urlprefix }}%
\providecommand \urlprefix  [0]{URL }%
\providecommand \Eprint [0]{\href }%
\providecommand \doibase [0]{https://doi.org/}%
\providecommand \selectlanguage [0]{\@gobble}%
\providecommand \bibinfo  [0]{\@secondoftwo}%
\providecommand \bibfield  [0]{\@secondoftwo}%
\providecommand \translation [1]{[#1]}%
\providecommand \BibitemOpen [0]{}%
\providecommand \bibitemStop [0]{}%
\providecommand \bibitemNoStop [0]{.\EOS\space}%
\providecommand \EOS [0]{\spacefactor3000\relax}%
\providecommand \BibitemShut  [1]{\csname bibitem#1\endcsname}%
\let\auto@bib@innerbib\@empty
\bibitem [{\citenamefont {Fulde}\ and\ \citenamefont
  {Ferrell}(1964)}]{Fulde:PRev135.1964}%
  \BibitemOpen
  \bibfield  {author} {\bibinfo {author} {\bibfnamefont {P.}~\bibnamefont
  {Fulde}}\ and\ \bibinfo {author} {\bibfnamefont {R.~A.}\ \bibnamefont
  {Ferrell}},\ }\bibfield  {title} {\bibinfo {title} {Superconductivity in a
  strong spin-exchange field},\ }\href
  {https://doi.org/10.1103/PhysRev.135.A550} {\bibfield  {journal} {\bibinfo
  {journal} {Phys. Rev.}\ }\textbf {\bibinfo {volume} {135}},\ \bibinfo {pages}
  {A550} (\bibinfo {year} {1964})}\BibitemShut {NoStop}%
\bibitem [{\citenamefont {Larkin}\ and\ \citenamefont
  {Ovchinnikov}(1964)}]{Larkin:JETP20.1965}%
  \BibitemOpen
  \bibfield  {author} {\bibinfo {author} {\bibfnamefont {A.~I.}\ \bibnamefont
  {Larkin}}\ and\ \bibinfo {author} {\bibfnamefont {Y.~N.}\ \bibnamefont
  {Ovchinnikov}},\ }\bibfield  {title} {\bibinfo {title} {Nonuniform state of
  superconductors},\ }\href@noop {} {\bibfield  {journal} {\bibinfo  {journal}
  {Zh. Eksp. Teor. Fiz.}\ }\textbf {\bibinfo {volume} {47}},\ \bibinfo {pages}
  {1136} (\bibinfo {year} {1964})},\ \bibinfo {note} {[Sov. Phys. JETP,
  \textbf{20}, 762 (1965)]}\BibitemShut {NoStop}%
\bibitem [{\citenamefont {Matsuda}\ and\ \citenamefont
  {Shimahara}(2007)}]{Matsuda:JPSJ76.2007}%
  \BibitemOpen
  \bibfield  {author} {\bibinfo {author} {\bibfnamefont {Y.}~\bibnamefont
  {Matsuda}}\ and\ \bibinfo {author} {\bibfnamefont {H.}~\bibnamefont
  {Shimahara}},\ }\bibfield  {title} {\bibinfo {title}
  {{Fulde--Ferrell--Larkin--Ovchinnikov} state in heavy fermion
  superconductors},\ }\href {https://doi.org/10.1143/JPSJ.76.051005} {\bibfield
   {journal} {\bibinfo  {journal} {J. Phys. Soc. Jpn.}\ }\textbf {\bibinfo
  {volume} {76}},\ \bibinfo {pages} {051005} (\bibinfo {year}
  {2007})}\BibitemShut {NoStop}%
\bibitem [{\citenamefont {Beyer}\ and\ \citenamefont
  {Wosnitza}(2013)}]{BeyerLTP13}%
  \BibitemOpen
  \bibfield  {author} {\bibinfo {author} {\bibfnamefont {R.}~\bibnamefont
  {Beyer}}\ and\ \bibinfo {author} {\bibfnamefont {J.}~\bibnamefont
  {Wosnitza}},\ }\bibfield  {title} {\bibinfo {title} {Emerging evidence for
  {FFLO} states in layered organic superconductors (review article)},\ }\href
  {https://doi.org/10.1063/1.4794996} {\bibfield  {journal} {\bibinfo
  {journal} {Low Temp. Phys.}\ }\textbf {\bibinfo {volume} {39}},\ \bibinfo
  {pages} {225} (\bibinfo {year} {2013})}\BibitemShut {NoStop}%
\bibitem [{\citenamefont {Wosnitza}(2018)}]{WosnitzaAnnPhys18}%
  \BibitemOpen
  \bibfield  {author} {\bibinfo {author} {\bibfnamefont {J.}~\bibnamefont
  {Wosnitza}},\ }\bibfield  {title} {\bibinfo {title} {{FFLO} states in layered
  organic superconductors},\ }\href {https://doi.org/10.1002/andp.201700282}
  {\bibfield  {journal} {\bibinfo  {journal} {Ann. Phys.}\ }\textbf {\bibinfo
  {volume} {530}},\ \bibinfo {pages} {1700282} (\bibinfo {year}
  {2018})}\BibitemShut {NoStop}%
\bibitem [{\citenamefont {Takada}\ and\ \citenamefont
  {Izuyama}(1969)}]{TakadaPrThPhys1969}%
  \BibitemOpen
  \bibfield  {author} {\bibinfo {author} {\bibfnamefont {S.}~\bibnamefont
  {Takada}}\ and\ \bibinfo {author} {\bibfnamefont {T.}~\bibnamefont
  {Izuyama}},\ }\bibfield  {title} {\bibinfo {title} {Superconductivity in a
  molecular field. {I}},\ }\href@noop {} {\bibfield  {journal} {\bibinfo
  {journal} {Prog. Theor. Phys.}\ }\textbf {\bibinfo {volume} {41}},\ \bibinfo
  {pages} {635} (\bibinfo {year} {1969})}\BibitemShut {NoStop}%
\bibitem [{\citenamefont {Machida}\ and\ \citenamefont
  {Nakanishi}(1984)}]{MachidaPhysRevB.30.122}%
  \BibitemOpen
  \bibfield  {author} {\bibinfo {author} {\bibfnamefont {K.}~\bibnamefont
  {Machida}}\ and\ \bibinfo {author} {\bibfnamefont {H.}~\bibnamefont
  {Nakanishi}},\ }\bibfield  {title} {\bibinfo {title} {Superconductivity under
  a ferromagnetic molecular field},\ }\href
  {https://doi.org/10.1103/PhysRevB.30.122} {\bibfield  {journal} {\bibinfo
  {journal} {Phys. Rev. B}\ }\textbf {\bibinfo {volume} {30}},\ \bibinfo
  {pages} {122} (\bibinfo {year} {1984})}\BibitemShut {NoStop}%
\bibitem [{\citenamefont {Shimahara}(1994)}]{ShimaharaPhysRevB.50.12760}%
  \BibitemOpen
  \bibfield  {author} {\bibinfo {author} {\bibfnamefont {H.}~\bibnamefont
  {Shimahara}},\ }\bibfield  {title} {\bibinfo {title} {{Fulde-Ferrell} state
  in quasi-two-dimensional superconductors},\ }\href
  {https://doi.org/10.1103/PhysRevB.50.12760} {\bibfield  {journal} {\bibinfo
  {journal} {Phys. Rev. B}\ }\textbf {\bibinfo {volume} {50}},\ \bibinfo
  {pages} {12760} (\bibinfo {year} {1994})}\BibitemShut {NoStop}%
\bibitem [{\citenamefont {Burkhardt}\ and\ \citenamefont
  {Rainer}(1994)}]{BurkhardtAnnPhys94}%
  \BibitemOpen
  \bibfield  {author} {\bibinfo {author} {\bibfnamefont {H.}~\bibnamefont
  {Burkhardt}}\ and\ \bibinfo {author} {\bibfnamefont {D.}~\bibnamefont
  {Rainer}},\ }\bibfield  {title} {\bibinfo {title}
  {{Fulde-Ferrell-Larkin-Ovchinnikov} state in layered superconductors},\
  }\href {https://doi.org/10.1002/andp.19945060305} {\bibfield  {journal}
  {\bibinfo  {journal} {Ann. Phys.}\ }\textbf {\bibinfo {volume} {506}},\
  \bibinfo {pages} {181} (\bibinfo {year} {1994})}\BibitemShut {NoStop}%
\bibitem [{\citenamefont {Dupuis}\ \emph {et~al.}(1993)\citenamefont {Dupuis},
  \citenamefont {Montambaux},\ and\ \citenamefont {S\'a~de
  Melo}}]{Dupuis:PhysRevLett.70.2613}%
  \BibitemOpen
  \bibfield  {author} {\bibinfo {author} {\bibfnamefont {N.}~\bibnamefont
  {Dupuis}}, \bibinfo {author} {\bibfnamefont {G.}~\bibnamefont {Montambaux}},\
  and\ \bibinfo {author} {\bibfnamefont {C.~A.~R.}\ \bibnamefont {S\'a~de
  Melo}},\ }\bibfield  {title} {\bibinfo {title} {Quasi-one-dimensional
  superconductors in strong magnetic field},\ }\href
  {https://doi.org/10.1103/PhysRevLett.70.2613} {\bibfield  {journal} {\bibinfo
   {journal} {Phys. Rev. Lett.}\ }\textbf {\bibinfo {volume} {70}},\ \bibinfo
  {pages} {2613} (\bibinfo {year} {1993})}\BibitemShut {NoStop}%
\bibitem [{\citenamefont {Dupuis}(1995)}]{Dupuis:PhysRevB.51.9074}%
  \BibitemOpen
  \bibfield  {author} {\bibinfo {author} {\bibfnamefont {N.}~\bibnamefont
  {Dupuis}},\ }\bibfield  {title} {\bibinfo {title}
  {{Larkin-Ovchinnikov-Fulde-Ferrell} state in quasi-one-dimensional
  superconductors},\ }\href {https://doi.org/10.1103/PhysRevB.51.9074}
  {\bibfield  {journal} {\bibinfo  {journal} {Phys. Rev. B}\ }\textbf {\bibinfo
  {volume} {51}},\ \bibinfo {pages} {9074} (\bibinfo {year}
  {1995})}\BibitemShut {NoStop}%
\bibitem [{\citenamefont {Dupuis}\ and\ \citenamefont
  {Montambaux}(1994)}]{Dupuis:PhysRevB.49.8993}%
  \BibitemOpen
  \bibfield  {author} {\bibinfo {author} {\bibfnamefont {N.}~\bibnamefont
  {Dupuis}}\ and\ \bibinfo {author} {\bibfnamefont {G.}~\bibnamefont
  {Montambaux}},\ }\bibfield  {title} {\bibinfo {title} {Superconductivity of
  quasi-one-dimensional conductors in a high magnetic field},\ }\href
  {https://doi.org/10.1103/PhysRevB.49.8993} {\bibfield  {journal} {\bibinfo
  {journal} {Phys. Rev. B}\ }\textbf {\bibinfo {volume} {49}},\ \bibinfo
  {pages} {8993} (\bibinfo {year} {1994})}\BibitemShut {NoStop}%
\bibitem [{\citenamefont {Croitoru}\ and\ \citenamefont
  {Buzdin}(2014)}]{CroitoruPhysRevB.89.224506}%
  \BibitemOpen
  \bibfield  {author} {\bibinfo {author} {\bibfnamefont {M.~D.}\ \bibnamefont
  {Croitoru}}\ and\ \bibinfo {author} {\bibfnamefont {A.~I.}\ \bibnamefont
  {Buzdin}},\ }\bibfield  {title} {\bibinfo {title} {Peculiarities of the
  orbital effect in the {Fulde-Ferrell-Larkin-Ovchinnikov} state in
  quasi-one-dimensional superconductors},\ }\href
  {https://doi.org/10.1103/PhysRevB.89.224506} {\bibfield  {journal} {\bibinfo
  {journal} {Phys. Rev. B}\ }\textbf {\bibinfo {volume} {89}},\ \bibinfo
  {pages} {224506} (\bibinfo {year} {2014})}\BibitemShut {NoStop}%
\bibitem [{\citenamefont {Gruenberg}\ and\ \citenamefont
  {Gunther}(1966)}]{Gruenberg:PRL16.1966}%
  \BibitemOpen
  \bibfield  {author} {\bibinfo {author} {\bibfnamefont {L.~W.}\ \bibnamefont
  {Gruenberg}}\ and\ \bibinfo {author} {\bibfnamefont {L.}~\bibnamefont
  {Gunther}},\ }\bibfield  {title} {\bibinfo {title} {{Fulde-Ferrell} effect in
  type-{II} superconductors},\ }\href
  {https://doi.org/10.1103/PhysRevLett.16.996} {\bibfield  {journal} {\bibinfo
  {journal} {Phys. Rev. Lett.}\ }\textbf {\bibinfo {volume} {16}},\ \bibinfo
  {pages} {996} (\bibinfo {year} {1966})}\BibitemShut {NoStop}%
\bibitem [{Note1()}]{Note1}%
  \BibitemOpen
  \bibinfo {note} {In this paper, we generalize this consideration for a
  \protect \emph {layered superconductor} in the magnetic field perpendicular
  to the layers and found that in this case, the critical Maki's parameter is
  even larger, for strong anisotropy it is $4.76$.}\BibitemShut {Stop}%
\bibitem [{\citenamefont {Houzet}\ and\ \citenamefont
  {Buzdin}(2001)}]{HouzetPhysRevB01}%
  \BibitemOpen
  \bibfield  {author} {\bibinfo {author} {\bibfnamefont {M.}~\bibnamefont
  {Houzet}}\ and\ \bibinfo {author} {\bibfnamefont {A.}~\bibnamefont
  {Buzdin}},\ }\bibfield  {title} {\bibinfo {title} {Structure of the vortex
  lattice in the {Fulde-Ferrell-Larkin-Ovchinnikov} state},\ }\href
  {https://doi.org/10.1103/PhysRevB.63.184521} {\bibfield  {journal} {\bibinfo
  {journal} {Phys. Rev. B}\ }\textbf {\bibinfo {volume} {63}},\ \bibinfo
  {pages} {184521} (\bibinfo {year} {2001})}\BibitemShut {NoStop}%
\bibitem [{\citenamefont {Houzet}\ and\ \citenamefont
  {Mineev}(2006)}]{HouzetPhysRevB2006}%
  \BibitemOpen
  \bibfield  {author} {\bibinfo {author} {\bibfnamefont {M.}~\bibnamefont
  {Houzet}}\ and\ \bibinfo {author} {\bibfnamefont {V.~P.}\ \bibnamefont
  {Mineev}},\ }\bibfield  {title} {\bibinfo {title} {Interplay of paramagnetic,
  orbital, and impurity effects on the phase transition of a normal metal to
  the superconducting state},\ }\href
  {https://doi.org/10.1103/PhysRevB.74.144522} {\bibfield  {journal} {\bibinfo
  {journal} {Phys. Rev. B}\ }\textbf {\bibinfo {volume} {74}},\ \bibinfo
  {pages} {144522} (\bibinfo {year} {2006})}\BibitemShut {NoStop}%
\bibitem [{\citenamefont {Maniv}\ and\ \citenamefont
  {Zhuravlev}(2008)}]{ManivPhysRevB2008}%
  \BibitemOpen
  \bibfield  {author} {\bibinfo {author} {\bibfnamefont {T.}~\bibnamefont
  {Maniv}}\ and\ \bibinfo {author} {\bibfnamefont {V.}~\bibnamefont
  {Zhuravlev}},\ }\bibfield  {title} {\bibinfo {title} {Dimensionality-driven
  changeover to first-order superconducting phase transitions in the {Pauli}
  paramagnetic limit},\ }\href {https://doi.org/10.1103/PhysRevB.77.134511}
  {\bibfield  {journal} {\bibinfo  {journal} {Phys. Rev. B}\ }\textbf {\bibinfo
  {volume} {77}},\ \bibinfo {pages} {134511} (\bibinfo {year}
  {2008})}\BibitemShut {NoStop}%
\bibitem [{\citenamefont {Zhuravlev}\ and\ \citenamefont
  {Maniv}(2009)}]{Zhuravlev:PRB80.2009}%
  \BibitemOpen
  \bibfield  {author} {\bibinfo {author} {\bibfnamefont {V.}~\bibnamefont
  {Zhuravlev}}\ and\ \bibinfo {author} {\bibfnamefont {T.}~\bibnamefont
  {Maniv}},\ }\bibfield  {title} {\bibinfo {title} {Nonperturbative theory of
  type-{II} superconductivity in the presence of a strong pauli paramagnetic
  effect},\ }\href {https://doi.org/10.1103/PhysRevB.80.174520} {\bibfield
  {journal} {\bibinfo  {journal} {Phys. Rev. B}\ }\textbf {\bibinfo {volume}
  {80}},\ \bibinfo {pages} {174520} (\bibinfo {year} {2009})}\BibitemShut
  {NoStop}%
\bibitem [{\citenamefont {Shimahara}(2009)}]{Shimahara:PRB80.2009}%
  \BibitemOpen
  \bibfield  {author} {\bibinfo {author} {\bibfnamefont {H.}~\bibnamefont
  {Shimahara}},\ }\bibfield  {title} {\bibinfo {title} {Transition from the
  vortex state to the {Fulde-Ferrell-Larkin-Ovchinnikov} state in
  quasi-two-dimensional superconductors},\ }\href
  {https://doi.org/10.1103/PhysRevB.80.214512} {\bibfield  {journal} {\bibinfo
  {journal} {Phys. Rev. B}\ }\textbf {\bibinfo {volume} {80}},\ \bibinfo
  {pages} {214512} (\bibinfo {year} {2009})}\BibitemShut {NoStop}%
\bibitem [{\citenamefont {Paglione}\ and\ \citenamefont
  {Greene}(2010)}]{Paglione:NatPhys6.2010}%
  \BibitemOpen
  \bibfield  {author} {\bibinfo {author} {\bibfnamefont {J.}~\bibnamefont
  {Paglione}}\ and\ \bibinfo {author} {\bibfnamefont {R.~L.}\ \bibnamefont
  {Greene}},\ }\bibfield  {title} {\bibinfo {title} {High-temperature
  superconductivity in iron-based materials},\ }\href
  {http://dx.doi.org/10.1038/nphys1759} {\bibfield  {journal} {\bibinfo
  {journal} {Nat. Phys.}\ }\textbf {\bibinfo {volume} {6}},\ \bibinfo {pages}
  {645} (\bibinfo {year} {2010})}\BibitemShut {NoStop}%
\bibitem [{\citenamefont {Stewart}(2011)}]{StewartRevModPhys.83.1589}%
  \BibitemOpen
  \bibfield  {author} {\bibinfo {author} {\bibfnamefont {G.~R.}\ \bibnamefont
  {Stewart}},\ }\bibfield  {title} {\bibinfo {title} {Superconductivity in iron
  compounds},\ }\href {https://doi.org/10.1103/RevModPhys.83.1589} {\bibfield
  {journal} {\bibinfo  {journal} {Rev. Mod. Phys.}\ }\textbf {\bibinfo {volume}
  {83}},\ \bibinfo {pages} {1589} (\bibinfo {year} {2011})}\BibitemShut
  {NoStop}%
\bibitem [{\citenamefont {Hosono}\ and\ \citenamefont
  {Kuroki}(2015)}]{Hosono:PhysC514.2015}%
  \BibitemOpen
  \bibfield  {author} {\bibinfo {author} {\bibfnamefont {H.}~\bibnamefont
  {Hosono}}\ and\ \bibinfo {author} {\bibfnamefont {K.}~\bibnamefont
  {Kuroki}},\ }\bibfield  {title} {\bibinfo {title} {Iron-based
  superconductors: Current status of materials and pairing mechanism},\ }\href
  {https://doi.org/http://dx.doi.org/10.1016/j.physc.2015.02.020} {\bibfield
  {journal} {\bibinfo  {journal} {Physica C}\ }\textbf {\bibinfo {volume}
  {514}},\ \bibinfo {pages} {399 } (\bibinfo {year} {2015})}\BibitemShut
  {NoStop}%
\bibitem [{\citenamefont {Si}\ \emph {et~al.}(2016)\citenamefont {Si},
  \citenamefont {Yu},\ and\ \citenamefont {Abrahams}}]{SiNatRevMat16}%
  \BibitemOpen
  \bibfield  {author} {\bibinfo {author} {\bibfnamefont {Q.}~\bibnamefont
  {Si}}, \bibinfo {author} {\bibfnamefont {R.}~\bibnamefont {Yu}},\ and\
  \bibinfo {author} {\bibfnamefont {E.}~\bibnamefont {Abrahams}},\ }\bibfield
  {title} {\bibinfo {title} {High-temperature superconductivity in iron
  pnictides and chalcogenides},\ }\href
  {http://dx.doi.org/10.1038/natrevmats.2016.17} {\bibfield  {journal}
  {\bibinfo  {journal} {Nat. Rev. Mat.}\ }\textbf {\bibinfo {volume} {1}},\
  \bibinfo {pages} {16017} (\bibinfo {year} {2016})}\BibitemShut {NoStop}%
\bibitem [{\citenamefont {Gurevich}(2010)}]{Gurevich:PRB82.2010}%
  \BibitemOpen
  \bibfield  {author} {\bibinfo {author} {\bibfnamefont {A.}~\bibnamefont
  {Gurevich}},\ }\bibfield  {title} {\bibinfo {title} {Upper critical field and
  the {Fulde-Ferrel-Larkin-Ovchinnikov} transition in multiband
  superconductors},\ }\href {https://doi.org/10.1103/PhysRevB.82.184504}
  {\bibfield  {journal} {\bibinfo  {journal} {Phys. Rev. B}\ }\textbf {\bibinfo
  {volume} {82}},\ \bibinfo {pages} {184504} (\bibinfo {year}
  {2010})}\BibitemShut {NoStop}%
\bibitem [{\citenamefont {Gurevich}(2011)}]{Gurevich:RPP74.2011}%
  \BibitemOpen
  \bibfield  {author} {\bibinfo {author} {\bibfnamefont {A.}~\bibnamefont
  {Gurevich}},\ }\bibfield  {title} {\bibinfo {title} {Iron-based
  superconductors at high magnetic fields},\ }\href
  {http://stacks.iop.org/0034-4885/74/i=12/a=124501} {\bibfield  {journal}
  {\bibinfo  {journal} {Rep. Prog. Phys.}\ }\textbf {\bibinfo {volume} {74}},\
  \bibinfo {pages} {124501} (\bibinfo {year} {2011})}\BibitemShut {NoStop}%
\bibitem [{\citenamefont {Adachi}\ and\ \citenamefont
  {Ikeda}(2015)}]{AdachiJPSJ15}%
  \BibitemOpen
  \bibfield  {author} {\bibinfo {author} {\bibfnamefont {K.}~\bibnamefont
  {Adachi}}\ and\ \bibinfo {author} {\bibfnamefont {R.}~\bibnamefont {Ikeda}},\
  }\bibfield  {title} {\bibinfo {title} {Possible field--temperature phase
  diagrams of two-band superconductors with paramagnetic pair-breaking},\
  }\href {https://doi.org/10.7566/JPSJ.84.064712} {\bibfield  {journal}
  {\bibinfo  {journal} {J. Phys. Soc. Jpn.}\ }\textbf {\bibinfo {volume}
  {84}},\ \bibinfo {pages} {064712} (\bibinfo {year} {2015})}\BibitemShut
  {NoStop}%
\bibitem [{\citenamefont {Ptok}\ and\ \citenamefont
  {Crivelli}(2013)}]{Ptok:JLowT172.2013}%
  \BibitemOpen
  \bibfield  {author} {\bibinfo {author} {\bibfnamefont {A.}~\bibnamefont
  {Ptok}}\ and\ \bibinfo {author} {\bibfnamefont {D.}~\bibnamefont
  {Crivelli}},\ }\bibfield  {title} {\bibinfo {title} {The
  {Fulde--Ferrell--Larkin--Ovchinnikov} state in pnictides},\ }\href
  {https://doi.org/10.1007/s10909-013-0871-0} {\bibfield  {journal} {\bibinfo
  {journal} {J. Low Temp. Phys}\ }\textbf {\bibinfo {volume} {172}},\ \bibinfo
  {pages} {226} (\bibinfo {year} {2013})}\BibitemShut {NoStop}%
\bibitem [{\citenamefont {Ptok}(2014)}]{Ptok:EPJB87.2014}%
  \BibitemOpen
  \bibfield  {author} {\bibinfo {author} {\bibfnamefont {A.}~\bibnamefont
  {Ptok}},\ }\bibfield  {title} {\bibinfo {title} {Influence of s{$_\pm$}
  symmetry on unconventional superconductivity in pnictides above the {Pauli}
  limit -- two-band model study},\ }\href
  {https://doi.org/10.1140/epjb/e2013-41007-2} {\bibfield  {journal} {\bibinfo
  {journal} {Eur. Phys. J. B}\ }\textbf {\bibinfo {volume} {87}},\ \bibinfo
  {pages} {2} (\bibinfo {year} {2014})}\BibitemShut {NoStop}%
\bibitem [{\citenamefont {Ptok}(2015)}]{Ptok:JPhysCM27.2015}%
  \BibitemOpen
  \bibfield  {author} {\bibinfo {author} {\bibfnamefont {A.}~\bibnamefont
  {Ptok}},\ }\bibfield  {title} {\bibinfo {title} {Multiple phase transitions
  in {Pauli}-limited iron-based superconductors},\ }\href
  {http://stacks.iop.org/0953-8984/27/i=48/a=482001} {\bibfield  {journal}
  {\bibinfo  {journal} {J. Phys. Condens. Matter}\ }\textbf {\bibinfo {volume}
  {27}},\ \bibinfo {pages} {482001} (\bibinfo {year} {2015})}\BibitemShut
  {NoStop}%
\bibitem [{\citenamefont {Ptok}\ \emph {et~al.}(2017)\citenamefont {Ptok},
  \citenamefont {Kapcia}, \citenamefont {Piekarz},\ and\ \citenamefont
  {Ole{\'s}}}]{Ptok:NJPhys19.2017}%
  \BibitemOpen
  \bibfield  {author} {\bibinfo {author} {\bibfnamefont {A.}~\bibnamefont
  {Ptok}}, \bibinfo {author} {\bibfnamefont {K.~J.}\ \bibnamefont {Kapcia}},
  \bibinfo {author} {\bibfnamefont {P.}~\bibnamefont {Piekarz}},\ and\ \bibinfo
  {author} {\bibfnamefont {A.~M.}\ \bibnamefont {Ole{\'s}}},\ }\bibfield
  {title} {\bibinfo {title} {The ab initio study of unconventional
  superconductivity in {CeCoIn$_5$} and {FeSe}},\ }\href
  {http://stacks.iop.org/1367-2630/19/i=6/a=063039} {\bibfield  {journal}
  {\bibinfo  {journal} {New J. Phys.}\ }\textbf {\bibinfo {volume} {19}},\
  \bibinfo {pages} {063039} (\bibinfo {year} {2017})}\BibitemShut {NoStop}%
\bibitem [{\citenamefont {Rajagopal}\ and\ \citenamefont
  {Vasudevan}(1966)}]{Rajagopal:PLett23.1966}%
  \BibitemOpen
  \bibfield  {author} {\bibinfo {author} {\bibfnamefont {A.}~\bibnamefont
  {Rajagopal}}\ and\ \bibinfo {author} {\bibfnamefont {R.}~\bibnamefont
  {Vasudevan}},\ }\bibfield  {title} {\bibinfo {title} {{De Haas-Van Alphen}
  oscillations in the critical temperature of type ii superconductors},\ }\href
  {https://doi.org/http://dx.doi.org/10.1016/0031-9163(66)90396-9} {\bibfield
  {journal} {\bibinfo  {journal} {Phys. Lett.}\ }\textbf {\bibinfo {volume}
  {23}},\ \bibinfo {pages} {539 } (\bibinfo {year} {1966})}\BibitemShut
  {NoStop}%
\bibitem [{\citenamefont {Gruenberg}\ and\ \citenamefont
  {Gunther}(1968)}]{Gruenberg:PRev176.1968}%
  \BibitemOpen
  \bibfield  {author} {\bibinfo {author} {\bibfnamefont {L.~W.}\ \bibnamefont
  {Gruenberg}}\ and\ \bibinfo {author} {\bibfnamefont {L.}~\bibnamefont
  {Gunther}},\ }\bibfield  {title} {\bibinfo {title} {Effect of orbital
  quantization on the critical field of type-{II} superconductors},\ }\href
  {https://doi.org/10.1103/PhysRev.176.606} {\bibfield  {journal} {\bibinfo
  {journal} {Phys. Rev.}\ }\textbf {\bibinfo {volume} {176}},\ \bibinfo {pages}
  {606} (\bibinfo {year} {1968})}\BibitemShut {NoStop}%
\bibitem [{\citenamefont {Te\ifmmode \check{s}\else
  \v{s}\fi{}anovi\ifmmode~\acute{c}\else \'{c}\fi{}}\ and\ \citenamefont
  {Rasolt}(1989)}]{Tesanovic:PRB39.1989}%
  \BibitemOpen
  \bibfield  {author} {\bibinfo {author} {\bibfnamefont {Z.}~\bibnamefont
  {Te\ifmmode \check{s}\else \v{s}\fi{}anovi\ifmmode~\acute{c}\else
  \'{c}\fi{}}}\ and\ \bibinfo {author} {\bibfnamefont {M.}~\bibnamefont
  {Rasolt}},\ }\bibfield  {title} {\bibinfo {title} {New type of
  superconductivity in very high magnetic fields},\ }\href
  {https://doi.org/10.1103/PhysRevB.39.2718} {\bibfield  {journal} {\bibinfo
  {journal} {Phys. Rev. B}\ }\textbf {\bibinfo {volume} {39}},\ \bibinfo
  {pages} {2718} (\bibinfo {year} {1989})}\BibitemShut {NoStop}%
\bibitem [{\citenamefont {Rieck}\ \emph {et~al.}(1990)\citenamefont {Rieck},
  \citenamefont {Scharnberg},\ and\ \citenamefont
  {Klemm}}]{Reick:PhysicaC170.1990}%
  \BibitemOpen
  \bibfield  {author} {\bibinfo {author} {\bibfnamefont {C.}~\bibnamefont
  {Rieck}}, \bibinfo {author} {\bibfnamefont {K.}~\bibnamefont {Scharnberg}},\
  and\ \bibinfo {author} {\bibfnamefont {R.}~\bibnamefont {Klemm}},\ }\bibfield
   {title} {\bibinfo {title} {Re-entrant superconductivity due to landau level
  quantization?},\ }\href
  {https://doi.org/http://dx.doi.org/10.1016/0921-4534(90)90311-2} {\bibfield
  {journal} {\bibinfo  {journal} {Physica C}\ }\textbf {\bibinfo {volume}
  {170}},\ \bibinfo {pages} {195 } (\bibinfo {year} {1990})}\BibitemShut
  {NoStop}%
\bibitem [{\citenamefont {MacDonald}\ \emph {et~al.}(1992)\citenamefont
  {MacDonald}, \citenamefont {Akera},\ and\ \citenamefont
  {Norman}}]{MacDonald:PRB45.1992}%
  \BibitemOpen
  \bibfield  {author} {\bibinfo {author} {\bibfnamefont {A.~H.}\ \bibnamefont
  {MacDonald}}, \bibinfo {author} {\bibfnamefont {H.}~\bibnamefont {Akera}},\
  and\ \bibinfo {author} {\bibfnamefont {M.~R.}\ \bibnamefont {Norman}},\
  }\bibfield  {title} {\bibinfo {title} {Landau quantization and
  particle-particle ladder sums in a magnetic field},\ }\href
  {https://doi.org/10.1103/PhysRevB.45.10147} {\bibfield  {journal} {\bibinfo
  {journal} {Phys. Rev. B}\ }\textbf {\bibinfo {volume} {45}},\ \bibinfo
  {pages} {10147} (\bibinfo {year} {1992})}\BibitemShut {NoStop}%
\bibitem [{\citenamefont {Maniv}\ \emph {et~al.}(1992)\citenamefont {Maniv},
  \citenamefont {Rom}, \citenamefont {Vagner},\ and\ \citenamefont
  {Wyder}}]{Maniv:PRB46.1992}%
  \BibitemOpen
  \bibfield  {author} {\bibinfo {author} {\bibfnamefont {T.}~\bibnamefont
  {Maniv}}, \bibinfo {author} {\bibfnamefont {A.~I.}\ \bibnamefont {Rom}},
  \bibinfo {author} {\bibfnamefont {I.~D.}\ \bibnamefont {Vagner}},\ and\
  \bibinfo {author} {\bibfnamefont {P.}~\bibnamefont {Wyder}},\ }\bibfield
  {title} {\bibinfo {title} {de {Haas}- van {Alphen} effect in the
  superconducting state of a two-dimensional metal},\ }\href
  {https://doi.org/10.1103/PhysRevB.46.8360} {\bibfield  {journal} {\bibinfo
  {journal} {Phys. Rev. B}\ }\textbf {\bibinfo {volume} {46}},\ \bibinfo
  {pages} {8360} (\bibinfo {year} {1992})}\BibitemShut {NoStop}%
\bibitem [{\citenamefont {Song}\ and\ \citenamefont
  {Koshelev}(2018)}]{SongHc2LifLay}%
  \BibitemOpen
  \bibfield  {author} {\bibinfo {author} {\bibfnamefont {K.~W.}\ \bibnamefont
  {Song}}\ and\ \bibinfo {author} {\bibfnamefont {A.~E.}\ \bibnamefont
  {Koshelev}},\ }\bibfield  {title} {\bibinfo {title} {Interplay between
  orbital-quantization effects and the {Fulde-Ferrell-Larkin-Ovchinnikov}
  instability in multiple-band layered superconductors},\ }\href
  {https://doi.org/10.1103/PhysRevB.97.224520} {\bibfield  {journal} {\bibinfo
  {journal} {Phys. Rev. B}\ }\textbf {\bibinfo {volume} {97}},\ \bibinfo
  {pages} {224520} (\bibinfo {year} {2018})}\BibitemShut {NoStop}%
\bibitem [{\citenamefont {Wosnitza}(1996)}]{WosnitzaFSLowDSC1996}%
  \BibitemOpen
  \bibfield  {author} {\bibinfo {author} {\bibfnamefont {J.}~\bibnamefont
  {Wosnitza}},\ }\href@noop {} {\emph {\bibinfo {title} {Fermi surfaces of
  Low-Dimensional Organic Metals and Superconductors}}}\ (\bibinfo  {publisher}
  {Springer, Berlin},\ \bibinfo {year} {1996})\BibitemShut {NoStop}%
\bibitem [{Note2()}]{Note2}%
  \BibitemOpen
  \bibinfo {note} {In the technical part, we use a natural system of units in
  which $k_B=1$ and $\hbar =1$.}\BibitemShut {Stop}%
\bibitem [{\citenamefont {Rajagopal}\ and\ \citenamefont
  {Ryan}(1991)}]{Rajagopal:PRB44.1991}%
  \BibitemOpen
  \bibfield  {author} {\bibinfo {author} {\bibfnamefont {A.~K.}\ \bibnamefont
  {Rajagopal}}\ and\ \bibinfo {author} {\bibfnamefont {J.~C.}\ \bibnamefont
  {Ryan}},\ }\bibfield  {title} {\bibinfo {title} {Quantum-state
  representations in a strong quantizing magnetic field: Pairing theory of
  superconductivity},\ }\href {https://doi.org/10.1103/PhysRevB.44.10280}
  {\bibfield  {journal} {\bibinfo  {journal} {Phys. Rev. B}\ }\textbf {\bibinfo
  {volume} {44}},\ \bibinfo {pages} {10280} (\bibinfo {year}
  {1991})}\BibitemShut {NoStop}%
\bibitem [{\citenamefont {Maniv}\ \emph {et~al.}(2001)\citenamefont {Maniv},
  \citenamefont {Zhuravlev}, \citenamefont {Vagner},\ and\ \citenamefont
  {Wyder}}]{Maniv:RMP73.2001}%
  \BibitemOpen
  \bibfield  {author} {\bibinfo {author} {\bibfnamefont {T.}~\bibnamefont
  {Maniv}}, \bibinfo {author} {\bibfnamefont {V.}~\bibnamefont {Zhuravlev}},
  \bibinfo {author} {\bibfnamefont {I.}~\bibnamefont {Vagner}},\ and\ \bibinfo
  {author} {\bibfnamefont {P.}~\bibnamefont {Wyder}},\ }\bibfield  {title}
  {\bibinfo {title} {Vortex states and quantum magnetic oscillations in
  conventional type-{II} superconductors},\ }\href
  {https://doi.org/10.1103/RevModPhys.73.867} {\bibfield  {journal} {\bibinfo
  {journal} {Rev. Mod. Phys.}\ }\textbf {\bibinfo {volume} {73}},\ \bibinfo
  {pages} {867} (\bibinfo {year} {2001})}\BibitemShut {NoStop}%
\bibitem [{\citenamefont {Helfand}\ and\ \citenamefont
  {Werthamer}(1966)}]{Helfand:PRev.1966}%
  \BibitemOpen
  \bibfield  {author} {\bibinfo {author} {\bibfnamefont {E.}~\bibnamefont
  {Helfand}}\ and\ \bibinfo {author} {\bibfnamefont {N.~R.}\ \bibnamefont
  {Werthamer}},\ }\bibfield  {title} {\bibinfo {title} {Temperature and purity
  dependence of the superconducting critical field, ${H}_{c2}$. {II}},\ }\href
  {https://doi.org/10.1103/PhysRev.147.288} {\bibfield  {journal} {\bibinfo
  {journal} {Phys. Rev.}\ }\textbf {\bibinfo {volume} {147}},\ \bibinfo {pages}
  {288} (\bibinfo {year} {1966})}\BibitemShut {NoStop}%
\bibitem [{\citenamefont {Werthamer}\ \emph {et~al.}(1966)\citenamefont
  {Werthamer}, \citenamefont {Helfand},\ and\ \citenamefont
  {Hohenberg}}]{Werthamer:PRev147.1966}%
  \BibitemOpen
  \bibfield  {author} {\bibinfo {author} {\bibfnamefont {N.~R.}\ \bibnamefont
  {Werthamer}}, \bibinfo {author} {\bibfnamefont {E.}~\bibnamefont {Helfand}},\
  and\ \bibinfo {author} {\bibfnamefont {P.~C.}\ \bibnamefont {Hohenberg}},\
  }\bibfield  {title} {\bibinfo {title} {Temperature and purity dependence of
  the superconducting critical field, ${H}_{c2}$. {III}. electron spin and
  spin-orbit effects},\ }\href {https://doi.org/10.1103/PhysRev.147.295}
  {\bibfield  {journal} {\bibinfo  {journal} {Phys. Rev.}\ }\textbf {\bibinfo
  {volume} {147}},\ \bibinfo {pages} {295} (\bibinfo {year}
  {1966})}\BibitemShut {NoStop}%
\bibitem [{\citenamefont {Kogan}\ and\ \citenamefont
  {Prozorov}(2012)}]{Kogan:RPP75.2012}%
  \BibitemOpen
  \bibfield  {author} {\bibinfo {author} {\bibfnamefont {V.~G.}\ \bibnamefont
  {Kogan}}\ and\ \bibinfo {author} {\bibfnamefont {R.}~\bibnamefont
  {Prozorov}},\ }\bibfield  {title} {\bibinfo {title} {Orbital upper critical
  field and its anisotropy of clean one- and two-band superconductors},\ }\href
  {http://stacks.iop.org/0034-4885/75/i=11/a=114502} {\bibfield  {journal}
  {\bibinfo  {journal} {Rep. Prog. Phys.}\ }\textbf {\bibinfo {volume} {75}},\
  \bibinfo {pages} {114502} (\bibinfo {year} {2012})}\BibitemShut {NoStop}%
\bibitem [{\citenamefont {Shoenberg}(1984)}]{ShoenbergMagnOscBook}%
  \BibitemOpen
  \bibfield  {author} {\bibinfo {author} {\bibfnamefont {D.}~\bibnamefont
  {Shoenberg}},\ }\href@noop {} {\emph {\bibinfo {title} {Magnetic oscillations
  in metals}}}\ (\bibinfo  {publisher} {Cambridge University Press,
  Cambridge},\ \bibinfo {year} {1984})\BibitemShut {NoStop}%
\bibitem [{Note3()}]{Note3}%
  \BibitemOpen
  \bibinfo {note} {The ratio $(\mu \pm 2t_z)/\omega _c$ in Eqs.\ \protect
  \textup {\hbox {\mathsurround \z@ \protect \normalfont (\ignorespaces \ref
  {eq:JQ1}\unskip \@@italiccorr )}} and \protect \textup {\hbox {\mathsurround
  \z@ \protect \normalfont (\ignorespaces \ref {eq:JQ2}\unskip \@@italiccorr
  )}} can be rewritten in a more common form as $F_\pm /H$, where $F_\pm
  =(c/2\pi e)A_\pm $ is the de Haas-van Alphen frequency and $A_\pm $ is the
  area of the corresponding extremal Fermi-surface cross section, see, e.g.,
  Ref.\ \cite {ShoenbergMagnOscBook}}\BibitemShut {NoStop}%
\bibitem [{\citenamefont {Gvozdikov}\ \emph {et~al.}(2003)\citenamefont
  {Gvozdikov}, \citenamefont {Jansen}, \citenamefont {Pesin}, \citenamefont
  {Vagner},\ and\ \citenamefont {Wyder}}]{Gvozdikov:PRB68.2003}%
  \BibitemOpen
  \bibfield  {author} {\bibinfo {author} {\bibfnamefont {V.~M.}\ \bibnamefont
  {Gvozdikov}}, \bibinfo {author} {\bibfnamefont {A.~G.~M.}\ \bibnamefont
  {Jansen}}, \bibinfo {author} {\bibfnamefont {D.~A.}\ \bibnamefont {Pesin}},
  \bibinfo {author} {\bibfnamefont {I.~D.}\ \bibnamefont {Vagner}},\ and\
  \bibinfo {author} {\bibfnamefont {P.}~\bibnamefont {Wyder}},\ }\bibfield
  {title} {\bibinfo {title} {Quantum magnetic oscillations of the chemical
  potential in superlattices and layered conductors},\ }\href
  {https://doi.org/10.1103/PhysRevB.68.155107} {\bibfield  {journal} {\bibinfo
  {journal} {Phys. Rev. B}\ }\textbf {\bibinfo {volume} {68}},\ \bibinfo
  {pages} {155107} (\bibinfo {year} {2003})}\BibitemShut {NoStop}%
\bibitem [{\citenamefont {Singleton}\ and\ \citenamefont
  {Mielke}(2002)}]{Singleton:ContemPhys43.2002}%
  \BibitemOpen
  \bibfield  {author} {\bibinfo {author} {\bibfnamefont {J.}~\bibnamefont
  {Singleton}}\ and\ \bibinfo {author} {\bibfnamefont {C.}~\bibnamefont
  {Mielke}},\ }\bibfield  {title} {\bibinfo {title} {Quasi-two-dimensional
  organic superconductors: A review},\ }\href
  {https://doi.org/10.1080/00107510110108681} {\bibfield  {journal} {\bibinfo
  {journal} {Contemporary Physics}\ }\textbf {\bibinfo {volume} {43}},\
  \bibinfo {pages} {63} (\bibinfo {year} {2002})}\BibitemShut {NoStop}%
\bibitem [{\citenamefont {Wosnitza}(2007)}]{Wosnitza:JLowTPhys146.2007}%
  \BibitemOpen
  \bibfield  {author} {\bibinfo {author} {\bibfnamefont {J.}~\bibnamefont
  {Wosnitza}},\ }\bibfield  {title} {\bibinfo {title} {Quasi-two-dimensional
  organic superconductors},\ }\href {https://doi.org/10.1007/s10909-006-9282-9}
  {\bibfield  {journal} {\bibinfo  {journal} {J. Low Temp. Phys.}\ }\textbf
  {\bibinfo {volume} {146}},\ \bibinfo {pages} {641} (\bibinfo {year}
  {2007})}\BibitemShut {NoStop}%
\bibitem [{\citenamefont {Murata}\ \emph {et~al.}(1988)\citenamefont {Murata},
  \citenamefont {Honda}, \citenamefont {Anzai}, \citenamefont {Tokumoto},
  \citenamefont {Takahashi}, \citenamefont {Kinoshita}, \citenamefont
  {Ishiguro}, \citenamefont {Toyota}, \citenamefont {Sasaki},\ and\
  \citenamefont {Muto}}]{Murata:SynMetals27.1988}%
  \BibitemOpen
  \bibfield  {author} {\bibinfo {author} {\bibfnamefont {K.}~\bibnamefont
  {Murata}}, \bibinfo {author} {\bibfnamefont {Y.}~\bibnamefont {Honda}},
  \bibinfo {author} {\bibfnamefont {H.}~\bibnamefont {Anzai}}, \bibinfo
  {author} {\bibfnamefont {M.}~\bibnamefont {Tokumoto}}, \bibinfo {author}
  {\bibfnamefont {K.}~\bibnamefont {Takahashi}}, \bibinfo {author}
  {\bibfnamefont {N.}~\bibnamefont {Kinoshita}}, \bibinfo {author}
  {\bibfnamefont {T.}~\bibnamefont {Ishiguro}}, \bibinfo {author}
  {\bibfnamefont {N.}~\bibnamefont {Toyota}}, \bibinfo {author} {\bibfnamefont
  {T.}~\bibnamefont {Sasaki}},\ and\ \bibinfo {author} {\bibfnamefont
  {Y.}~\bibnamefont {Muto}},\ }\bibfield  {title} {\bibinfo {title} {Transport
  properties of {$\kappa$-(BEDT-TTF)$_2$Cu(NCS)$_2$; H$_{C2}$}, its anisotropy
  and their pressure dependence},\ }\href
  {https://doi.org/https://doi.org/10.1016/0379-6779(88)90421-3} {\bibfield
  {journal} {\bibinfo  {journal} {Synthetic Metals}\ }\textbf {\bibinfo
  {volume} {27}},\ \bibinfo {pages} {A341 } (\bibinfo {year} {1988})},\
  \bibinfo {note} {proceedings of the International Conference on Science and
  Technology of Synthetic Metals}\BibitemShut {NoStop}%
\bibitem [{\citenamefont {Sasaki}\ \emph {et~al.}(2003)\citenamefont {Sasaki},
  \citenamefont {Fukuda}, \citenamefont {Yoneyama},\ and\ \citenamefont
  {Kobayashi}}]{SasakiPhysRevB.67.144521}%
  \BibitemOpen
  \bibfield  {author} {\bibinfo {author} {\bibfnamefont {T.}~\bibnamefont
  {Sasaki}}, \bibinfo {author} {\bibfnamefont {T.}~\bibnamefont {Fukuda}},
  \bibinfo {author} {\bibfnamefont {N.}~\bibnamefont {Yoneyama}},\ and\
  \bibinfo {author} {\bibfnamefont {N.}~\bibnamefont {Kobayashi}},\ }\bibfield
  {title} {\bibinfo {title} {{Shubnikov--de Haas effect in the quantum vortex
  liquid state of the organic superconductor
  $\ensuremath{\kappa}\ensuremath{-}(\mathrm{BEDT}\ensuremath{-}\mathrm{TTF}{)}_{2}\mathrm{Cu}(\mathrm{NCS}{)}_{2}$}},\
  }\href {https://doi.org/10.1103/PhysRevB.67.144521} {\bibfield  {journal}
  {\bibinfo  {journal} {Phys. Rev. B}\ }\textbf {\bibinfo {volume} {67}},\
  \bibinfo {pages} {144521} (\bibinfo {year} {2003})}\BibitemShut {NoStop}%
\bibitem [{\citenamefont {Meyer}\ \emph {et~al.}(1995)\citenamefont {Meyer},
  \citenamefont {Steep}, \citenamefont {Biberacher}, \citenamefont {Christ},
  \citenamefont {Lerf}, \citenamefont {Jansen}, \citenamefont {Joss},
  \citenamefont {Wyder},\ and\ \citenamefont {Andres}}]{MeyerEPL95}%
  \BibitemOpen
  \bibfield  {author} {\bibinfo {author} {\bibfnamefont {F.~A.}\ \bibnamefont
  {Meyer}}, \bibinfo {author} {\bibfnamefont {E.}~\bibnamefont {Steep}},
  \bibinfo {author} {\bibfnamefont {W.}~\bibnamefont {Biberacher}}, \bibinfo
  {author} {\bibfnamefont {P.}~\bibnamefont {Christ}}, \bibinfo {author}
  {\bibfnamefont {A.}~\bibnamefont {Lerf}}, \bibinfo {author} {\bibfnamefont
  {A.~G.~M.}\ \bibnamefont {Jansen}}, \bibinfo {author} {\bibfnamefont
  {W.}~\bibnamefont {Joss}}, \bibinfo {author} {\bibfnamefont {P.}~\bibnamefont
  {Wyder}},\ and\ \bibinfo {author} {\bibfnamefont {K.}~\bibnamefont
  {Andres}},\ }\bibfield  {title} {\bibinfo {title} {High-field de {Haas}-van
  {Alphen} studies of {$\kappa$-(BEDT-TTF)$_2$Cu(NCS)$_2$}},\ }\href
  {http://stacks.iop.org/0295-5075/32/i=8/a=011} {\bibfield  {journal}
  {\bibinfo  {journal} {Europhys. Lett.}\ }\textbf {\bibinfo {volume} {32}},\
  \bibinfo {pages} {681} (\bibinfo {year} {1995})}\BibitemShut {NoStop}%
\bibitem [{\citenamefont {Singleton}\ \emph {et~al.}(2002)\citenamefont
  {Singleton}, \citenamefont {Goddard}, \citenamefont {Ardavan}, \citenamefont
  {Harrison}, \citenamefont {Blundell}, \citenamefont {Schlueter},\ and\
  \citenamefont {Kini}}]{Singleton:PRL88.2002}%
  \BibitemOpen
  \bibfield  {author} {\bibinfo {author} {\bibfnamefont {J.}~\bibnamefont
  {Singleton}}, \bibinfo {author} {\bibfnamefont {P.~A.}\ \bibnamefont
  {Goddard}}, \bibinfo {author} {\bibfnamefont {A.}~\bibnamefont {Ardavan}},
  \bibinfo {author} {\bibfnamefont {N.}~\bibnamefont {Harrison}}, \bibinfo
  {author} {\bibfnamefont {S.~J.}\ \bibnamefont {Blundell}}, \bibinfo {author}
  {\bibfnamefont {J.~A.}\ \bibnamefont {Schlueter}},\ and\ \bibinfo {author}
  {\bibfnamefont {A.~M.}\ \bibnamefont {Kini}},\ }\bibfield  {title} {\bibinfo
  {title} {Test for interlayer coherence in a quasi-two-dimensional
  superconductor},\ }\href {https://doi.org/10.1103/PhysRevLett.88.037001}
  {\bibfield  {journal} {\bibinfo  {journal} {Phys. Rev. Lett.}\ }\textbf
  {\bibinfo {volume} {88}},\ \bibinfo {pages} {037001} (\bibinfo {year}
  {2002})}\BibitemShut {NoStop}%
\bibitem [{\citenamefont {Schmalian}(1998)}]{SchmalianPhysRevLett.81.4232}%
  \BibitemOpen
  \bibfield  {author} {\bibinfo {author} {\bibfnamefont {J.}~\bibnamefont
  {Schmalian}},\ }\bibfield  {title} {\bibinfo {title} {Pairing due to spin
  fluctuations in layered organic superconductors},\ }\href
  {https://doi.org/10.1103/PhysRevLett.81.4232} {\bibfield  {journal} {\bibinfo
   {journal} {Phys. Rev. Lett.}\ }\textbf {\bibinfo {volume} {81}},\ \bibinfo
  {pages} {4232} (\bibinfo {year} {1998})}\BibitemShut {NoStop}%
\bibitem [{\citenamefont {Kuroki}(2006)}]{KurokiJPSJ06}%
  \BibitemOpen
  \bibfield  {author} {\bibinfo {author} {\bibfnamefont {K.}~\bibnamefont
  {Kuroki}},\ }\bibfield  {title} {\bibinfo {title} {Pairing symmetry
  competition in organic superconductors},\ }\href
  {https://doi.org/10.1143/JPSJ.75.051013} {\bibfield  {journal} {\bibinfo
  {journal} {Journal of the Physical Society of Japan}\ }\textbf {\bibinfo
  {volume} {75}},\ \bibinfo {pages} {051013} (\bibinfo {year}
  {2006})}\BibitemShut {NoStop}%
\bibitem [{\citenamefont {De~Soto}\ \emph {et~al.}(1995)\citenamefont
  {De~Soto}, \citenamefont {Slichter}, \citenamefont {Kini}, \citenamefont
  {Wang}, \citenamefont {Geiser},\ and\ \citenamefont
  {Williams}}]{DeSotoPhysRevB.52.10364}%
  \BibitemOpen
  \bibfield  {author} {\bibinfo {author} {\bibfnamefont {S.~M.}\ \bibnamefont
  {De~Soto}}, \bibinfo {author} {\bibfnamefont {C.~P.}\ \bibnamefont
  {Slichter}}, \bibinfo {author} {\bibfnamefont {A.~M.}\ \bibnamefont {Kini}},
  \bibinfo {author} {\bibfnamefont {H.~H.}\ \bibnamefont {Wang}}, \bibinfo
  {author} {\bibfnamefont {U.}~\bibnamefont {Geiser}},\ and\ \bibinfo {author}
  {\bibfnamefont {J.~M.}\ \bibnamefont {Williams}},\ }\bibfield  {title}
  {\bibinfo {title} {{$^{13}\mathrm{C}$ {NMR} studies of the normal and
  superconducting states of the organic superconductor
  {\ensuremath{\kappa}-(ET${)}_{2}$Cu[N(CN${)}_{2}$]Br}}},\ }\href
  {https://doi.org/10.1103/PhysRevB.52.10364} {\bibfield  {journal} {\bibinfo
  {journal} {Phys. Rev. B}\ }\textbf {\bibinfo {volume} {52}},\ \bibinfo
  {pages} {10364} (\bibinfo {year} {1995})}\BibitemShut {NoStop}%
\bibitem [{\citenamefont {Carrington}\ \emph {et~al.}(1999)\citenamefont
  {Carrington}, \citenamefont {Bonalde}, \citenamefont {Prozorov},
  \citenamefont {Giannetta}, \citenamefont {Kini}, \citenamefont {Schlueter},
  \citenamefont {Wang}, \citenamefont {Geiser},\ and\ \citenamefont
  {Williams}}]{CarringtonPhysRevLett.83.4172}%
  \BibitemOpen
  \bibfield  {author} {\bibinfo {author} {\bibfnamefont {A.}~\bibnamefont
  {Carrington}}, \bibinfo {author} {\bibfnamefont {I.~J.}\ \bibnamefont
  {Bonalde}}, \bibinfo {author} {\bibfnamefont {R.}~\bibnamefont {Prozorov}},
  \bibinfo {author} {\bibfnamefont {R.~W.}\ \bibnamefont {Giannetta}}, \bibinfo
  {author} {\bibfnamefont {A.~M.}\ \bibnamefont {Kini}}, \bibinfo {author}
  {\bibfnamefont {J.}~\bibnamefont {Schlueter}}, \bibinfo {author}
  {\bibfnamefont {H.~H.}\ \bibnamefont {Wang}}, \bibinfo {author}
  {\bibfnamefont {U.}~\bibnamefont {Geiser}},\ and\ \bibinfo {author}
  {\bibfnamefont {J.~M.}\ \bibnamefont {Williams}},\ }\bibfield  {title}
  {\bibinfo {title} {Low-temperature penetration depth of
  {$\mathit{\ensuremath{\kappa}}\ensuremath{-}(\mathrm{ET}{)}_{2}\mathrm{Cu}[N(\mathrm{CN}{)}_{2}]\mathrm{Br}$
  and
  $\mathit{\ensuremath{\kappa}}\ensuremath{-}(\mathrm{ET}{)}_{2}\mathrm{Cu}(\mathrm{NCS}{)}_{2}$}},\
  }\href {https://doi.org/10.1103/PhysRevLett.83.4172} {\bibfield  {journal}
  {\bibinfo  {journal} {Phys. Rev. Lett.}\ }\textbf {\bibinfo {volume} {83}},\
  \bibinfo {pages} {4172} (\bibinfo {year} {1999})}\BibitemShut {NoStop}%
\bibitem [{\citenamefont {Milbradt}\ \emph {et~al.}(2013)\citenamefont
  {Milbradt}, \citenamefont {Bardin}, \citenamefont {Truncik}, \citenamefont
  {Huttema}, \citenamefont {Jacko}, \citenamefont {Burn}, \citenamefont {Lo},
  \citenamefont {Powell},\ and\ \citenamefont
  {Broun}}]{MilbradtPhysRevB.88.064501}%
  \BibitemOpen
  \bibfield  {author} {\bibinfo {author} {\bibfnamefont {S.}~\bibnamefont
  {Milbradt}}, \bibinfo {author} {\bibfnamefont {A.~A.}\ \bibnamefont
  {Bardin}}, \bibinfo {author} {\bibfnamefont {C.~J.~S.}\ \bibnamefont
  {Truncik}}, \bibinfo {author} {\bibfnamefont {W.~A.}\ \bibnamefont
  {Huttema}}, \bibinfo {author} {\bibfnamefont {A.~C.}\ \bibnamefont {Jacko}},
  \bibinfo {author} {\bibfnamefont {P.~L.}\ \bibnamefont {Burn}}, \bibinfo
  {author} {\bibfnamefont {S.-C.}\ \bibnamefont {Lo}}, \bibinfo {author}
  {\bibfnamefont {B.~J.}\ \bibnamefont {Powell}},\ and\ \bibinfo {author}
  {\bibfnamefont {D.~M.}\ \bibnamefont {Broun}},\ }\bibfield  {title} {\bibinfo
  {title} {In-plane superfluid density and microwave conductivity of the
  organic superconductor
  {$\ensuremath{\kappa}$-(BEDT-TTF)${}_{2}$Cu[N(CN)${}_{2}$]Br}: Evidence for
  $d$-wave pairing and resilient quasiparticles},\ }\href
  {https://doi.org/10.1103/PhysRevB.88.064501} {\bibfield  {journal} {\bibinfo
  {journal} {Phys. Rev. B}\ }\textbf {\bibinfo {volume} {88}},\ \bibinfo
  {pages} {064501} (\bibinfo {year} {2013})}\BibitemShut {NoStop}%
\bibitem [{\citenamefont {Taylor}\ \emph {et~al.}(2007)\citenamefont {Taylor},
  \citenamefont {Carrington},\ and\ \citenamefont
  {Schlueter}}]{TaylorPhysRevLett.99.057001}%
  \BibitemOpen
  \bibfield  {author} {\bibinfo {author} {\bibfnamefont {O.~J.}\ \bibnamefont
  {Taylor}}, \bibinfo {author} {\bibfnamefont {A.}~\bibnamefont {Carrington}},\
  and\ \bibinfo {author} {\bibfnamefont {J.~A.}\ \bibnamefont {Schlueter}},\
  }\bibfield  {title} {\bibinfo {title} {Specific-heat measurements of the gap
  structure of the organic superconductors
  $\ensuremath{\kappa}\mathrm{\text{\ensuremath{-}}}(\mathrm{ET}{)}_{2}\mathrm{Cu}[\mathrm{N}(\mathrm{CN}{)}_{2}]\mathrm{Br}$
  and
  $\ensuremath{\kappa}\mathrm{\text{\ensuremath{-}}}(\mathrm{ET}{)}_{2}\mathrm{Cu}(\mathrm{NCS}{)}_{2}$},\
  }\href {https://doi.org/10.1103/PhysRevLett.99.057001} {\bibfield  {journal}
  {\bibinfo  {journal} {Phys. Rev. Lett.}\ }\textbf {\bibinfo {volume} {99}},\
  \bibinfo {pages} {057001} (\bibinfo {year} {2007})}\BibitemShut {NoStop}%
\bibitem [{\citenamefont {Malone}\ \emph {et~al.}(2010)\citenamefont {Malone},
  \citenamefont {Taylor}, \citenamefont {Schlueter},\ and\ \citenamefont
  {Carrington}}]{MalonePhysRevB.82.014522}%
  \BibitemOpen
  \bibfield  {author} {\bibinfo {author} {\bibfnamefont {L.}~\bibnamefont
  {Malone}}, \bibinfo {author} {\bibfnamefont {O.~J.}\ \bibnamefont {Taylor}},
  \bibinfo {author} {\bibfnamefont {J.~A.}\ \bibnamefont {Schlueter}},\ and\
  \bibinfo {author} {\bibfnamefont {A.}~\bibnamefont {Carrington}},\ }\bibfield
   {title} {\bibinfo {title} {Location of gap nodes in the organic
  superconductors
  $\ensuremath{\kappa}\text{\ensuremath{-}}{(\text{ET})}_{2}\text{Cu}{(\text{NCS})}_{2}$
  and
  $\ensuremath{\kappa}\text{\ensuremath{-}}{(\text{ET})}_{2}\text{Cu}[\text{N}{(\text{CN})}_{2}]\text{Br}$
  determined by magnetocalorimetry},\ }\href
  {https://doi.org/10.1103/PhysRevB.82.014522} {\bibfield  {journal} {\bibinfo
  {journal} {Phys. Rev. B}\ }\textbf {\bibinfo {volume} {82}},\ \bibinfo
  {pages} {014522} (\bibinfo {year} {2010})}\BibitemShut {NoStop}%
\bibitem [{\citenamefont {Izawa}\ \emph {et~al.}(2001)\citenamefont {Izawa},
  \citenamefont {Yamaguchi}, \citenamefont {Sasaki},\ and\ \citenamefont
  {Matsuda}}]{IzawaPhysRevLett.88.027002}%
  \BibitemOpen
  \bibfield  {author} {\bibinfo {author} {\bibfnamefont {K.}~\bibnamefont
  {Izawa}}, \bibinfo {author} {\bibfnamefont {H.}~\bibnamefont {Yamaguchi}},
  \bibinfo {author} {\bibfnamefont {T.}~\bibnamefont {Sasaki}},\ and\ \bibinfo
  {author} {\bibfnamefont {Y.}~\bibnamefont {Matsuda}},\ }\bibfield  {title}
  {\bibinfo {title} {Superconducting gap structure of
  $\mathit{\ensuremath{\kappa}}\ensuremath{-}(\mathrm{BEDT}\ensuremath{-}\mathrm{TTF}{)}_{2}\mathrm{Cu}(\mathrm{NCS}{)}_{2}$
  probed by thermal conductivity tensor},\ }\href
  {https://doi.org/10.1103/PhysRevLett.88.027002} {\bibfield  {journal}
  {\bibinfo  {journal} {Phys. Rev. Lett.}\ }\textbf {\bibinfo {volume} {88}},\
  \bibinfo {pages} {027002} (\bibinfo {year} {2001})}\BibitemShut {NoStop}%
\bibitem [{\citenamefont {Uji}\ \emph {et~al.}(2018)\citenamefont {Uji},
  \citenamefont {Fujii}, \citenamefont {Sugiura}, \citenamefont {Terashima},
  \citenamefont {Isono},\ and\ \citenamefont {Yamada}}]{UjiPhysRevB2018}%
  \BibitemOpen
  \bibfield  {author} {\bibinfo {author} {\bibfnamefont {S.}~\bibnamefont
  {Uji}}, \bibinfo {author} {\bibfnamefont {Y.}~\bibnamefont {Fujii}}, \bibinfo
  {author} {\bibfnamefont {S.}~\bibnamefont {Sugiura}}, \bibinfo {author}
  {\bibfnamefont {T.}~\bibnamefont {Terashima}}, \bibinfo {author}
  {\bibfnamefont {T.}~\bibnamefont {Isono}},\ and\ \bibinfo {author}
  {\bibfnamefont {J.}~\bibnamefont {Yamada}},\ }\bibfield  {title} {\bibinfo
  {title} {Quantum vortex melting and phase diagram in the layered organic
  superconductor
  {$\ensuremath{\kappa}$-(BEDT-TTF)${}_{2}{\mathrm{Cu}(\mathrm{NCS})}_{2}$}},\
  }\href {https://doi.org/10.1103/PhysRevB.97.024505} {\bibfield  {journal}
  {\bibinfo  {journal} {Phys. Rev. B}\ }\textbf {\bibinfo {volume} {97}},\
  \bibinfo {pages} {024505} (\bibinfo {year} {2018})}\BibitemShut {NoStop}%
\bibitem [{\citenamefont {Medvedev}\ \emph {et~al.}(2009)\citenamefont
  {Medvedev}, \citenamefont {McQueen}, \citenamefont {Troyan}, \citenamefont
  {Palasyuk}, \citenamefont {Eremets}, \citenamefont {Cava}, \citenamefont
  {Naghavi}, \citenamefont {Casper}, \citenamefont {Ksenofontov}, \citenamefont
  {Wortmann},\ and\ \citenamefont {Felser}}]{Medvedev:NatMat8.2009}%
  \BibitemOpen
  \bibfield  {author} {\bibinfo {author} {\bibfnamefont {S.}~\bibnamefont
  {Medvedev}}, \bibinfo {author} {\bibfnamefont {T.~M.}\ \bibnamefont
  {McQueen}}, \bibinfo {author} {\bibfnamefont {I.~A.}\ \bibnamefont {Troyan}},
  \bibinfo {author} {\bibfnamefont {T.}~\bibnamefont {Palasyuk}}, \bibinfo
  {author} {\bibfnamefont {M.~I.}\ \bibnamefont {Eremets}}, \bibinfo {author}
  {\bibfnamefont {R.~J.}\ \bibnamefont {Cava}}, \bibinfo {author}
  {\bibfnamefont {S.}~\bibnamefont {Naghavi}}, \bibinfo {author} {\bibfnamefont
  {F.}~\bibnamefont {Casper}}, \bibinfo {author} {\bibfnamefont
  {V.}~\bibnamefont {Ksenofontov}}, \bibinfo {author} {\bibfnamefont
  {G.}~\bibnamefont {Wortmann}},\ and\ \bibinfo {author} {\bibfnamefont
  {C.}~\bibnamefont {Felser}},\ }\bibfield  {title} {\bibinfo {title}
  {Electronic and magnetic phase diagram of $\beta$-{Fe$_{1.01}$Se} with
  superconductivity at 36.7 {K} under pressure},\ }\href
  {http://dx.doi.org/10.1038/nmat2491} {\bibfield  {journal} {\bibinfo
  {journal} {Nat. Mater.}\ }\textbf {\bibinfo {volume} {8}},\ \bibinfo {pages}
  {630 EP } (\bibinfo {year} {2009})}\BibitemShut {NoStop}%
\bibitem [{\citenamefont {Kasahara}\ \emph {et~al.}(2014)\citenamefont
  {Kasahara}, \citenamefont {Watashige}, \citenamefont {Hanaguri},
  \citenamefont {Kohsaka}, \citenamefont {Yamashita}, \citenamefont
  {Shimoyama}, \citenamefont {Mizukami}, \citenamefont {Endo}, \citenamefont
  {Ikeda}, \citenamefont {Aoyama}, \citenamefont {Terashima}, \citenamefont
  {Uji}, \citenamefont {Wolf}, \citenamefont {von L{\"o}hneysen}, \citenamefont
  {Shibauchi},\ and\ \citenamefont {Matsuda}}]{Kasahara:PNAS111.2014}%
  \BibitemOpen
  \bibfield  {author} {\bibinfo {author} {\bibfnamefont {S.}~\bibnamefont
  {Kasahara}}, \bibinfo {author} {\bibfnamefont {T.}~\bibnamefont {Watashige}},
  \bibinfo {author} {\bibfnamefont {T.}~\bibnamefont {Hanaguri}}, \bibinfo
  {author} {\bibfnamefont {Y.}~\bibnamefont {Kohsaka}}, \bibinfo {author}
  {\bibfnamefont {T.}~\bibnamefont {Yamashita}}, \bibinfo {author}
  {\bibfnamefont {Y.}~\bibnamefont {Shimoyama}}, \bibinfo {author}
  {\bibfnamefont {Y.}~\bibnamefont {Mizukami}}, \bibinfo {author}
  {\bibfnamefont {R.}~\bibnamefont {Endo}}, \bibinfo {author} {\bibfnamefont
  {H.}~\bibnamefont {Ikeda}}, \bibinfo {author} {\bibfnamefont
  {K.}~\bibnamefont {Aoyama}}, \bibinfo {author} {\bibfnamefont
  {T.}~\bibnamefont {Terashima}}, \bibinfo {author} {\bibfnamefont
  {S.}~\bibnamefont {Uji}}, \bibinfo {author} {\bibfnamefont {T.}~\bibnamefont
  {Wolf}}, \bibinfo {author} {\bibfnamefont {H.}~\bibnamefont {von
  L{\"o}hneysen}}, \bibinfo {author} {\bibfnamefont {T.}~\bibnamefont
  {Shibauchi}},\ and\ \bibinfo {author} {\bibfnamefont {Y.}~\bibnamefont
  {Matsuda}},\ }\bibfield  {title} {\bibinfo {title} {Field-induced
  superconducting phase of {FeSe} in the {BCS-BEC} cross-over},\ }\href
  {https://doi.org/10.1073/pnas.1413477111} {\bibfield  {journal} {\bibinfo
  {journal} {Proc. Natl. Acad. Sci. USA}\ }\textbf {\bibinfo {volume} {111}},\
  \bibinfo {pages} {16309} (\bibinfo {year} {2014})}\BibitemShut {NoStop}%
\bibitem [{\citenamefont {Terashima}\ \emph {et~al.}(2014)\citenamefont
  {Terashima}, \citenamefont {Kikugawa}, \citenamefont {Kiswandhi},
  \citenamefont {Choi}, \citenamefont {Brooks}, \citenamefont {Kasahara},
  \citenamefont {Watashige}, \citenamefont {Ikeda}, \citenamefont {Shibauchi},
  \citenamefont {Matsuda}, \citenamefont {Wolf}, \citenamefont {B\"ohmer},
  \citenamefont {Hardy}, \citenamefont {Meingast}, \citenamefont {L\"ohneysen},
  \citenamefont {Suzuki}, \citenamefont {Arita},\ and\ \citenamefont
  {Uji}}]{Terashima:PRB90.2014}%
  \BibitemOpen
  \bibfield  {author} {\bibinfo {author} {\bibfnamefont {T.}~\bibnamefont
  {Terashima}}, \bibinfo {author} {\bibfnamefont {N.}~\bibnamefont {Kikugawa}},
  \bibinfo {author} {\bibfnamefont {A.}~\bibnamefont {Kiswandhi}}, \bibinfo
  {author} {\bibfnamefont {E.-S.}\ \bibnamefont {Choi}}, \bibinfo {author}
  {\bibfnamefont {J.~S.}\ \bibnamefont {Brooks}}, \bibinfo {author}
  {\bibfnamefont {S.}~\bibnamefont {Kasahara}}, \bibinfo {author}
  {\bibfnamefont {T.}~\bibnamefont {Watashige}}, \bibinfo {author}
  {\bibfnamefont {H.}~\bibnamefont {Ikeda}}, \bibinfo {author} {\bibfnamefont
  {T.}~\bibnamefont {Shibauchi}}, \bibinfo {author} {\bibfnamefont
  {Y.}~\bibnamefont {Matsuda}}, \bibinfo {author} {\bibfnamefont
  {T.}~\bibnamefont {Wolf}}, \bibinfo {author} {\bibfnamefont {A.~E.}\
  \bibnamefont {B\"ohmer}}, \bibinfo {author} {\bibfnamefont {F.}~\bibnamefont
  {Hardy}}, \bibinfo {author} {\bibfnamefont {C.}~\bibnamefont {Meingast}},
  \bibinfo {author} {\bibfnamefont {H.~v.}\ \bibnamefont {L\"ohneysen}},
  \bibinfo {author} {\bibfnamefont {M.-T.}\ \bibnamefont {Suzuki}}, \bibinfo
  {author} {\bibfnamefont {R.}~\bibnamefont {Arita}},\ and\ \bibinfo {author}
  {\bibfnamefont {S.}~\bibnamefont {Uji}},\ }\bibfield  {title} {\bibinfo
  {title} {Anomalous fermi surface in {FeSe} seen by {Shubnikov}-de {Haas}
  oscillation measurements},\ }\href
  {https://doi.org/10.1103/PhysRevB.90.144517} {\bibfield  {journal} {\bibinfo
  {journal} {Phys. Rev. B}\ }\textbf {\bibinfo {volume} {90}},\ \bibinfo
  {pages} {144517} (\bibinfo {year} {2014})}\BibitemShut {NoStop}%
\bibitem [{\citenamefont {Watson}\ \emph {et~al.}(2015)\citenamefont {Watson},
  \citenamefont {Kim}, \citenamefont {Haghighirad}, \citenamefont {Davies},
  \citenamefont {McCollam}, \citenamefont {Narayanan}, \citenamefont {Blake},
  \citenamefont {Chen}, \citenamefont {Ghannadzadeh}, \citenamefont
  {Schofield}, \citenamefont {Hoesch}, \citenamefont {Meingast}, \citenamefont
  {Wolf},\ and\ \citenamefont {Coldea}}]{Watson:PRB91.2015}%
  \BibitemOpen
  \bibfield  {author} {\bibinfo {author} {\bibfnamefont {M.~D.}\ \bibnamefont
  {Watson}}, \bibinfo {author} {\bibfnamefont {T.~K.}\ \bibnamefont {Kim}},
  \bibinfo {author} {\bibfnamefont {A.~A.}\ \bibnamefont {Haghighirad}},
  \bibinfo {author} {\bibfnamefont {N.~R.}\ \bibnamefont {Davies}}, \bibinfo
  {author} {\bibfnamefont {A.}~\bibnamefont {McCollam}}, \bibinfo {author}
  {\bibfnamefont {A.}~\bibnamefont {Narayanan}}, \bibinfo {author}
  {\bibfnamefont {S.~F.}\ \bibnamefont {Blake}}, \bibinfo {author}
  {\bibfnamefont {Y.~L.}\ \bibnamefont {Chen}}, \bibinfo {author}
  {\bibfnamefont {S.}~\bibnamefont {Ghannadzadeh}}, \bibinfo {author}
  {\bibfnamefont {A.~J.}\ \bibnamefont {Schofield}}, \bibinfo {author}
  {\bibfnamefont {M.}~\bibnamefont {Hoesch}}, \bibinfo {author} {\bibfnamefont
  {C.}~\bibnamefont {Meingast}}, \bibinfo {author} {\bibfnamefont
  {T.}~\bibnamefont {Wolf}},\ and\ \bibinfo {author} {\bibfnamefont {A.~I.}\
  \bibnamefont {Coldea}},\ }\bibfield  {title} {\bibinfo {title} {Emergence of
  the nematic electronic state in {FeSe}},\ }\href
  {https://doi.org/10.1103/PhysRevB.91.155106} {\bibfield  {journal} {\bibinfo
  {journal} {Phys. Rev. B}\ }\textbf {\bibinfo {volume} {91}},\ \bibinfo
  {pages} {155106} (\bibinfo {year} {2015})}\BibitemShut {NoStop}%
\bibitem [{\citenamefont {Audouard}\ \emph {et~al.}(2015)\citenamefont
  {Audouard}, \citenamefont {Duc}, \citenamefont {Drigo}, \citenamefont
  {Toulemonde}, \citenamefont {Karlsson}, \citenamefont {Strobel},\ and\
  \citenamefont {Sulpice}}]{AudouardEPL2015}%
  \BibitemOpen
  \bibfield  {author} {\bibinfo {author} {\bibfnamefont {A.}~\bibnamefont
  {Audouard}}, \bibinfo {author} {\bibfnamefont {F.}~\bibnamefont {Duc}},
  \bibinfo {author} {\bibfnamefont {L.}~\bibnamefont {Drigo}}, \bibinfo
  {author} {\bibfnamefont {P.}~\bibnamefont {Toulemonde}}, \bibinfo {author}
  {\bibfnamefont {S.}~\bibnamefont {Karlsson}}, \bibinfo {author}
  {\bibfnamefont {P.}~\bibnamefont {Strobel}},\ and\ \bibinfo {author}
  {\bibfnamefont {A.}~\bibnamefont {Sulpice}},\ }\bibfield  {title} {\bibinfo
  {title} {Quantum oscillations and upper critical magnetic field of the
  iron-based superconductor {FeSe}},\ }\href
  {http://stacks.iop.org/0295-5075/109/i=2/a=27003} {\bibfield  {journal}
  {\bibinfo  {journal} {Europhys. Lett.}\ }\textbf {\bibinfo {volume} {109}},\
  \bibinfo {pages} {27003} (\bibinfo {year} {2015})}\BibitemShut {NoStop}%
\bibitem [{\citenamefont {Shimojima}\ \emph {et~al.}(2014)\citenamefont
  {Shimojima}, \citenamefont {Suzuki}, \citenamefont {Sonobe}, \citenamefont
  {Nakamura}, \citenamefont {Sakano}, \citenamefont {Omachi}, \citenamefont
  {Yoshioka}, \citenamefont {Kuwata-Gonokami}, \citenamefont {Ono},
  \citenamefont {Kumigashira}, \citenamefont {B\"ohmer}, \citenamefont {Hardy},
  \citenamefont {Wolf}, \citenamefont {Meingast}, \citenamefont {L\"ohneysen},
  \citenamefont {Ikeda},\ and\ \citenamefont
  {Ishizaka}}]{Shimojima:PRB90.2014}%
  \BibitemOpen
  \bibfield  {author} {\bibinfo {author} {\bibfnamefont {T.}~\bibnamefont
  {Shimojima}}, \bibinfo {author} {\bibfnamefont {Y.}~\bibnamefont {Suzuki}},
  \bibinfo {author} {\bibfnamefont {T.}~\bibnamefont {Sonobe}}, \bibinfo
  {author} {\bibfnamefont {A.}~\bibnamefont {Nakamura}}, \bibinfo {author}
  {\bibfnamefont {M.}~\bibnamefont {Sakano}}, \bibinfo {author} {\bibfnamefont
  {J.}~\bibnamefont {Omachi}}, \bibinfo {author} {\bibfnamefont
  {K.}~\bibnamefont {Yoshioka}}, \bibinfo {author} {\bibfnamefont
  {M.}~\bibnamefont {Kuwata-Gonokami}}, \bibinfo {author} {\bibfnamefont
  {K.}~\bibnamefont {Ono}}, \bibinfo {author} {\bibfnamefont {H.}~\bibnamefont
  {Kumigashira}}, \bibinfo {author} {\bibfnamefont {A.~E.}\ \bibnamefont
  {B\"ohmer}}, \bibinfo {author} {\bibfnamefont {F.}~\bibnamefont {Hardy}},
  \bibinfo {author} {\bibfnamefont {T.}~\bibnamefont {Wolf}}, \bibinfo {author}
  {\bibfnamefont {C.}~\bibnamefont {Meingast}}, \bibinfo {author}
  {\bibfnamefont {H.~v.}\ \bibnamefont {L\"ohneysen}}, \bibinfo {author}
  {\bibfnamefont {H.}~\bibnamefont {Ikeda}},\ and\ \bibinfo {author}
  {\bibfnamefont {K.}~\bibnamefont {Ishizaka}},\ }\bibfield  {title} {\bibinfo
  {title} {Lifting of \textit{xz} / \textit{yz} orbital degeneracy at the
  structural transition in detwinned {FeSe}},\ }\href
  {https://doi.org/10.1103/PhysRevB.90.121111} {\bibfield  {journal} {\bibinfo
  {journal} {Phys. Rev. B}\ }\textbf {\bibinfo {volume} {90}},\ \bibinfo
  {pages} {121111} (\bibinfo {year} {2014})}\BibitemShut {NoStop}%
\bibitem [{\citenamefont {Nakayama}\ \emph {et~al.}(2014)\citenamefont
  {Nakayama}, \citenamefont {Miyata}, \citenamefont {Phan}, \citenamefont
  {Sato}, \citenamefont {Tanabe}, \citenamefont {Urata}, \citenamefont
  {Tanigaki},\ and\ \citenamefont {Takahashi}}]{Nakayama:PRL113.2014}%
  \BibitemOpen
  \bibfield  {author} {\bibinfo {author} {\bibfnamefont {K.}~\bibnamefont
  {Nakayama}}, \bibinfo {author} {\bibfnamefont {Y.}~\bibnamefont {Miyata}},
  \bibinfo {author} {\bibfnamefont {G.~N.}\ \bibnamefont {Phan}}, \bibinfo
  {author} {\bibfnamefont {T.}~\bibnamefont {Sato}}, \bibinfo {author}
  {\bibfnamefont {Y.}~\bibnamefont {Tanabe}}, \bibinfo {author} {\bibfnamefont
  {T.}~\bibnamefont {Urata}}, \bibinfo {author} {\bibfnamefont
  {K.}~\bibnamefont {Tanigaki}},\ and\ \bibinfo {author} {\bibfnamefont
  {T.}~\bibnamefont {Takahashi}},\ }\bibfield  {title} {\bibinfo {title}
  {Reconstruction of band structure induced by electronic nematicity in an
  {FeSe} superconductor},\ }\href
  {https://doi.org/10.1103/PhysRevLett.113.237001} {\bibfield  {journal}
  {\bibinfo  {journal} {Phys. Rev. Lett.}\ }\textbf {\bibinfo {volume} {113}},\
  \bibinfo {pages} {237001} (\bibinfo {year} {2014})}\BibitemShut {NoStop}%
\bibitem [{\citenamefont {Fedorov}\ \emph {et~al.}(2016)\citenamefont
  {Fedorov}, \citenamefont {Yaresko}, \citenamefont {Kim}, \citenamefont
  {Kushnirenko}, \citenamefont {Haubold}, \citenamefont {Wolf}, \citenamefont
  {Hoesch}, \citenamefont {Gr{\"u}neis}, \citenamefont {B{\"u}chner},\ and\
  \citenamefont {Borisenko}}]{FedorovSciRep2016}%
  \BibitemOpen
  \bibfield  {author} {\bibinfo {author} {\bibfnamefont {A.}~\bibnamefont
  {Fedorov}}, \bibinfo {author} {\bibfnamefont {A.}~\bibnamefont {Yaresko}},
  \bibinfo {author} {\bibfnamefont {T.~K.}\ \bibnamefont {Kim}}, \bibinfo
  {author} {\bibfnamefont {Y.}~\bibnamefont {Kushnirenko}}, \bibinfo {author}
  {\bibfnamefont {E.}~\bibnamefont {Haubold}}, \bibinfo {author} {\bibfnamefont
  {T.}~\bibnamefont {Wolf}}, \bibinfo {author} {\bibfnamefont {M.}~\bibnamefont
  {Hoesch}}, \bibinfo {author} {\bibfnamefont {A.}~\bibnamefont {Gr{\"u}neis}},
  \bibinfo {author} {\bibfnamefont {B.}~\bibnamefont {B{\"u}chner}},\ and\
  \bibinfo {author} {\bibfnamefont {S.~V.}\ \bibnamefont {Borisenko}},\
  }\bibfield  {title} {\bibinfo {title} {Effect of nematic ordering on
  electronic structure of {FeSe}},\ }\href
  {http://dx.doi.org/10.1038/srep36834} {\bibfield  {journal} {\bibinfo
  {journal} {Sci. Rep.}\ }\textbf {\bibinfo {volume} {6}},\ \bibinfo {pages}
  {36834} (\bibinfo {year} {2016})}\BibitemShut {NoStop}%
\bibitem [{\citenamefont {Watashige}\ \emph {et~al.}(2017)\citenamefont
  {Watashige}, \citenamefont {Arsenijevi{\'c}}, \citenamefont {Yamashita},
  \citenamefont {Terazawa}, \citenamefont {Onishi}, \citenamefont {Opherden},
  \citenamefont {Kasahara}, \citenamefont {Tokiwa}, \citenamefont {Kasahara},
  \citenamefont {Shibauchi}, \citenamefont {von L{\"o}hneysen}, \citenamefont
  {Wosnitza},\ and\ \citenamefont {Matsuda}}]{Watashige:JPSJ86.2017}%
  \BibitemOpen
  \bibfield  {author} {\bibinfo {author} {\bibfnamefont {T.}~\bibnamefont
  {Watashige}}, \bibinfo {author} {\bibfnamefont {S.}~\bibnamefont
  {Arsenijevi{\'c}}}, \bibinfo {author} {\bibfnamefont {T.}~\bibnamefont
  {Yamashita}}, \bibinfo {author} {\bibfnamefont {D.}~\bibnamefont {Terazawa}},
  \bibinfo {author} {\bibfnamefont {T.}~\bibnamefont {Onishi}}, \bibinfo
  {author} {\bibfnamefont {L.}~\bibnamefont {Opherden}}, \bibinfo {author}
  {\bibfnamefont {S.}~\bibnamefont {Kasahara}}, \bibinfo {author}
  {\bibfnamefont {Y.}~\bibnamefont {Tokiwa}}, \bibinfo {author} {\bibfnamefont
  {Y.}~\bibnamefont {Kasahara}}, \bibinfo {author} {\bibfnamefont
  {T.}~\bibnamefont {Shibauchi}}, \bibinfo {author} {\bibfnamefont
  {H.}~\bibnamefont {von L{\"o}hneysen}}, \bibinfo {author} {\bibfnamefont
  {J.}~\bibnamefont {Wosnitza}},\ and\ \bibinfo {author} {\bibfnamefont
  {Y.}~\bibnamefont {Matsuda}},\ }\bibfield  {title} {\bibinfo {title}
  {Quasiparticle excitations in the superconducting state of {FeSe} probed by
  thermal hall conductivity in the vicinity of the {BCS--BEC} crossover},\
  }\href {https://doi.org/10.7566/JPSJ.86.014707} {\bibfield  {journal}
  {\bibinfo  {journal} {J. Phys. Soc. Jpn.}\ }\textbf {\bibinfo {volume}
  {86}},\ \bibinfo {pages} {014707} (\bibinfo {year} {2017})}\BibitemShut
  {NoStop}%
\bibitem [{\citenamefont {Mineev}(2000)}]{Mineev:PhilMagB80.2000}%
  \BibitemOpen
  \bibfield  {author} {\bibinfo {author} {\bibfnamefont {V.~P.}\ \bibnamefont
  {Mineev}},\ }\bibfield  {title} {\bibinfo {title} {Phase transition into the
  superconducting mixed state and the de {Haas}-van {Alphen} effect},\ }\href
  {https://doi.org/10.1080/13642810008208594} {\bibfield  {journal} {\bibinfo
  {journal} {Philos. Mag. B}\ }\textbf {\bibinfo {volume} {80}},\ \bibinfo
  {pages} {307} (\bibinfo {year} {2000})}\BibitemShut {NoStop}%
\bibitem [{\citenamefont {Champel}\ and\ \citenamefont
  {Mineev}(2001)}]{Champel:PhilMagB81.2001}%
  \BibitemOpen
  \bibfield  {author} {\bibinfo {author} {\bibfnamefont {T.}~\bibnamefont
  {Champel}}\ and\ \bibinfo {author} {\bibfnamefont {V.~P.}\ \bibnamefont
  {Mineev}},\ }\bibfield  {title} {\bibinfo {title} {de {Haas}--van {Alphen}
  effect in two- and quasi-two-dimensional metals and superconductors},\ }\href
  {https://doi.org/10.1080/13642810108216525} {\bibfield  {journal} {\bibinfo
  {journal} {Philos. Mag. B}\ }\textbf {\bibinfo {volume} {81}},\ \bibinfo
  {pages} {55} (\bibinfo {year} {2001})}\BibitemShut {NoStop}%
\end{thebibliography}%

\end{document}